\documentclass[aps,prl,reprint,showpacs,superscriptaddress]{revtex4-2}

\usepackage{comment}
\usepackage[pdftex]{graphicx}
\usepackage[pdftex]{epsfig}
\usepackage{epstopdf}
\usepackage{color}
\usepackage{amssymb,amsmath}
\usepackage{graphicx}
\usepackage{wrapfig}
\usepackage{dcolumn}
\usepackage{multirow}
\usepackage{tabularx}
\usepackage{tabularray}
\usepackage{xltabular}
\usepackage[hypertexnames=false,colorlinks,urlcolor=blue,citecolor=blue]{hyperref}
\usepackage{siunitx}
\usepackage[dvipsnames]{xcolor}
\usepackage[capitalise]{cleveref}
\usepackage{float}
\usepackage[caption=false]{subfig}
\usepackage{bm}
\usepackage{quantikz}
\usepackage[export]{adjustbox}

\usepackage{mathtools}

\newtheorem{theorem}{Theorem}
\usepackage{amsmath,amsfonts}
\usepackage{mathrsfs}
\usepackage{quantikz}
\usepackage[scr=rsfso]{mathalpha}

\DeclareSIUnit\au{\text{at.\,u.}}
\DeclareSIUnit\hartree{\text{Ha}}
\DeclareSIUnit\angstrom{\text{\AA}}

\renewcommand{\Re}{\mathrm{Re}}

\newcommand{\abs}{\mathscr{A}}
\newcommand{\Apot}{\hat{{\bf A}}}
\newcommand{\aan}[2]{\hat{a}_{#1,#2}}
\newcommand{\acr}[2]{\hat{a}^{\dagger}_{#1,#2}}
\newcommand{\Bfld}{\hat{{\bf B}}}

\newcommand{\uvec}{{\bf u}}
\newcommand{\nel}{\eta}
\newcommand{\nn}{\zeta}
\newcommand{\bg}{{\bf g}}
\newcommand{\hel}{\hat{H}_{\textrm{el}}}
\newcommand{\helph}{\hat{H}_{\textrm{el-ph}}}
\newcommand{\hpf}{\hat{H}_{\textrm{PF}}}
\newcommand{\hph}{\hat{H}_{\textrm{ph}}}
\newcommand{\id}{\hat{I}}
\newcommand{\bk}{{\bf k}}
\newcommand{\opacity}{\kappa}
\newcommand{\maxocc}{\Lambda}

\newcommand{\ngridel}{N_\textrm{G}}
\newcommand{\ngridph}{N_{\gamma}}
\newcommand{\bpo}[1]{\hat{\bf p}_{#1}}
\newcommand{\br}{{\bf r}}
\newcommand{\bro}[1]{\hat{\bf r}_{#1}}
\newcommand{\bR}{{\bf R}}
\newcommand{\spin}[1]{\hat{\bm{\sigma}}_{#1}}
\newcommand{\tee}{t}
\newcommand{\tke}{\hat{T}_{\textrm{e}}}
\newcommand{\trans}{\mathscr{T}}
\newcommand{\helvec}[2]{{\bf u}_{#1,#2}}
\renewcommand{\vee}{\hat{V}_{\textrm{ee}}}
\newcommand{\vei}{\hat{V}_{\textrm{ei}}}

\newcommand{\om}[1]{\omega(#1)}
\newcommand{\nion}{\zeta}

\newcommand{\npf}{N_\textrm{pf}}
\newcommand{\ns}{N_\textrm{s}}

\newcommand{\rws}{R_\textrm{ws}}
\newcommand{\wigner}[6]{\left(\begin{array}{ccc} #1 & #2 & #3\\ #4 & #5 & #6 \end{array}\right)}
\newcommand{\ereg}{\epsilon_\mathrm{reg}}
\newcommand{\eden}{\mathrm{n}}

\newcommand{\papertitle}{An approach for calculating astrophysical opacities on quantum computers}

\begin{document}

\title{\papertitle}
\author{Shivesh Pathak}
\author{Alina Kononov}
\author{Andrew D. Baczewski}
\affiliation{Quantum Algorithms and Applications Collaboratory, Sandia National Laboratories, Albuquerque NM, USA}

\begin{abstract}
We propose a quantum algorithmic protocol for calculating astrophysical opacities.
Our implementation uses first- and second-quantized representations of interacting electronic and photonic subsystems and Hamiltonian simulation via the interaction picture.
Inferring opacity from momentum-resolved measurements of the photonic register yields a direct relationship between qubit count and spectral range/resolution.
Logical resource estimates for the classically challenging problem of solar iron opacity are comparable to another high-energy-density physics problem (Rubin \emph{et al.}, PNAS 121(3) (2024)).
\end{abstract}

\maketitle

\textit{Introduction}---Opacity quantifies the extent to which a medium absorbs and scatters light as a function of frequency~\cite{huebner2014opacity}.
Atomistic models of opacity play a central role in larger scale radiation-hydrodynamic models of the structure and evolution of stars~\cite{rogers1994astrophysical}, inertial fusion implosions~\cite{hu2014first,betti2016inertial}, and high-energy-density plasmas created in laboratories around the world~\cite{bailey2009experimental,heeter2017conceptual,zastrau2021high}.
Opacity models are particularly valuable when it is difficult to create and measure the medium in a laboratory, and the opacity of iron in the solar interior is exemplary in this regard.
Over the years, discrepancies between high-precision laboratory astrophysics experiments and painstaking theoretical analyses have attracted intense interest~\cite{bailey2015higher,nagayama2019systematic,perry2020progress,bailey2025oxygen,loisel2025first}.
While this Letter does not address these discrepancies directly, we propose a protocol by which future quantum computers might predict astrophysical (and other) opacities.

Classical models of opacity typically rely on configuration interaction expansions of the electronic wavefunction for a single ion with semiclassical coupling to light~\cite{seaton1995opacity,blancard2011solar,colgan2013light,hansen2007hybrid, hansen2023self}. 
These methods scale unfavorably with the number of relevant configurations, which rapidly increases with atomic number under partial ionization in the warm or hot dense matter regimes.
In practice, various truncations degrade the description of both the equilibrium electronic structure and the electronic transitions responsible for opacity~\cite{hansen2007hybrid}.
The semiclassical treatment of light further neglects multiphoton effects, which a classical computer could capture at even greater expense~\cite{more2017opacity,kruse2019two,kruse2021two}.

A quantum computer could achieve greater accuracy than a classical computer through its ability to more efficiently represent the Hilbert space of a plasma coupled to an electromagnetic field.
The prospect of practical quantum advantage for solving electronic structure problems on quantum computers has spurred wide-ranging efforts to develop quantum algorithms for simulating molecules and materials~\cite{cao2019quantum,bauer2020quantum,mcardle2020quantum} along with detailed estimates of the number of qubits and quantum operations required for specific classically challenging problems~\cite{reiher2017elucidating,lee2021even,von2021quantum,su2021fault,rubin2023fault,caesura2025faster,low2025fast}.
At high temperatures, accurately sampling thermal states introduces additional complications~\cite{bobrow2026resource}, but certain prospects for quantum advantage are expected to improve in warm and hot dense matter~\cite{babbush2023quantum}.
Despite the high value proposition offered by simulations of experimentally challenging regimes, proposals for quantum applications in high-energy-density physics~\cite{rubin2024quantum} remain scarce.

As a foundational starting point, we consider a simplified opacity model that neglects relativistic electronic effects, inter-ion interactions, and ionic motion.
While these simplifications discard important spectral line splitting and broadening mechanisms, our model retains exact electron-electron interactions and captures many-body and multi-photon effects, aspects that may contribute to discrepancies between theory and experiment but remain exceedingly challenging or infeasible for classical computations.
In particular, our quantum protocol uses the Pauli-Fierz Hamiltonian~\cite{hiroshima2001ground, spohn2004dynamics,mukhopadhyay2024quantum} for a semi-relativistic description of electron-photon coupling near a single plasma ion.
App.~\ref{app:definition_of_the_hamiltonian} describes the Hamiltonian and its Hilbert space, including comments about incorporating the neglected physics~\cite{sm}.
Hartree atomic units are used throughout, unless otherwise indicated.

\begin{figure}
    \centering
        \begin{quantikz}[column sep=0.335cm]
        \lstick{\ket{0}$_\textrm{el}$} & \gate{\hat{U}_\textrm{i}^\textrm{el}} & \gate[2]{
        e^{-i\hpf \tee}} & & \rstick[2]{$\ns$} \\
        \lstick{\ket{0}$_\textrm{ph}$} & \gate{\hat{U}_\textrm{i}^\textrm{ph}} & & \meter{} &  \\
        \end{quantikz}
    \caption{
    Quantum circuit for the opacity protocol.
    First, $\hat{U}_\textrm{i}^\textrm{el}$ and $\hat{U}_\textrm{i}^\textrm{ph}$ initialize the electronic (el) and photonic (ph) registers into a Slater determinant and a maximally occupied Gaussian wave packet, respectively.
    Then, $e^{-i\hpf \tee}$ implements Hamiltonian simulation using the interaction picture algorithm~\cite{low2019hamiltonian}.
    Finally, momentum-resolved measurements of the photonic occupation numbers are related to opacity.
    Repeating the circuit $\ns$ times resolves the variance of the observable and thermal effects in the electronic distribution.
    }
    \label{fig:circuit}
\end{figure}
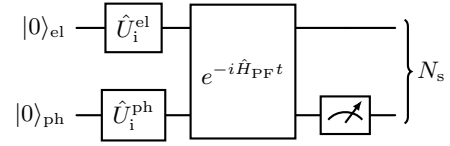

\begin{figure}
    \centering
    \includegraphics[width=\columnwidth]{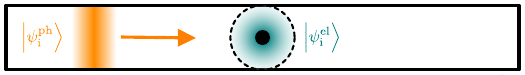}
    \includegraphics[width=\columnwidth]{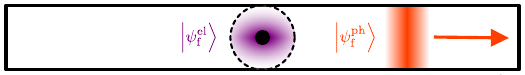}
    \caption{
    Simulation cell schematic.
    The central black point and dashed circle represent the nucleus and the atom sphere of radius $\rws$, respectively.
    The outer rectangle corresponds to the simulation cell of volume $\Omega$.
    The photons begin outside the atom sphere, with momentum oriented toward it (top panel).
    During the simulation, light passes through the atom sphere and back into the surrounding vacuum (bottom panel).
    To keep the atom visible, the simulation cell is not drawn to scale for the conditions considered in our resource estimates.
    }
    \label{fig:schematic}
\end{figure}

\begin{figure}
    \centering
    \includegraphics[width=\columnwidth]{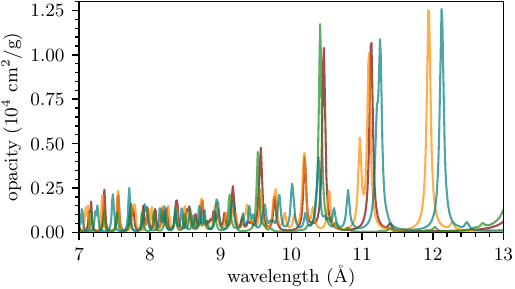}
    \caption{
    Representative spectra targeted by the opacity simulation protocol.
    Each color shows exemplary features sampled by an individual shot.
    Repeating the simulation for different initial electronic configurations drawn from a thermal ensemble will provide the average opacity.
    }
    \label{fig:opacity_example}
\end{figure}

\textit{Quantum algorithmic protocol}---Fig.~\ref{fig:circuit} summarizes our proposed quantum protocol~\footnote{Not shown are other ancilla registers required to implement Hamiltonian simulation.}. 
Two quantum registers encode digital representations of first-quantized electrons~\cite{Kassal_2008}~and second-quantized photons~\cite{mcardle2019digital,sawaya2020resource}.
As shown in Fig.~\ref{fig:schematic}, the simulated electrons lie within an atom sphere of radius $\rws$ determined by the plasma density, while the photons begin as a wave packet outside of the atom sphere.
After appropriately initializing each subsystem, the composite system evolves according to its Hamiltonian as the photons propagate forward, interact with the electrons, and re-emerge outside the atom sphere.
Ultimately, measuring the photonic register determines the number of photons transmitted through the atom and thus the atom's opacity, producing a spectrum like the ones in Fig.~\ref{fig:opacity_example}.
This procedure is repeated $\ns$ times to reduce statistical errors and estimate the thermal ensemble average over different pure initial states.

The electronic register, encoded in first quantization, consists of $\nel\log_2 N_G$ qubits.
Here $\nel$ is the number of electrons in the simulation and $N_G$ is the total number of real-space grid points.
For each shot, the electronic register is initialized into a Slater determinant state via an efficient protocol using Givens rotations~\cite{rubin2024quantum, babbush2023quantum}.
Over many shots, the Slater determinants recover the appropriate thermal distribution of the electrons~\footnote{We note that one could also average over correlated initial states, given access to appropriate state preparation circuits and weights.}.

The photonic register, encoded in second quantization, consists of $\ngridph \log_2 \maxocc$ qubits.
Here $\ngridph$ is the number of photonic modes and $\maxocc$ is the maximum photonic occupancy.
The photonic register is initialized identically for each shot, as a Gaussian wave packet with maximum photonic occupancy.
App.~\ref{subapp:initial-state_prep} describes further details of the state-preparation circuits 
($\hat{U}_\textrm{i}^{\textrm{el}}, $ $\hat{U}_\textrm{i}^{\textrm{ph}}$)~\cite{sm}.

Dynamical simulation is performed according to the atomic Pauli-Fierz Hamiltonian:
\begin{align}
    \hpf = &\underbrace{\sum_{i=1}^{\nel} \left[\frac{\bpo{i}^2}{2} - \frac{Z}{|\bro{i}|} + \frac{1}{2}\sum_{j \neq i=1}^{\nel} \frac{1}{|\bro{i}-\bro{j}|} \right]}_{\hel}+\nonumber\\
    &\underbrace{\sum_{\mu = \pm 1} \sum_{\bk} \omega(\bk) \acr{\bk}{\mu}\aan{\bk}{\mu}}_{\hph} + \label{eq:HPF_AA}\\
    & \underbrace{\sum_{i = 1}^{\nel} \frac{1}{2c}\left[\frac{1}{c}\Apot^2(\bro{i}) -\lbrace \bpo{i},\Apot(\bro{i})\rbrace - \spin{i} \cdot \Bfld(\bro{{i}}) \right]}_{\helph}. \nonumber
\end{align}
Here $\bro{i}$, $\bpo{i}$, and $\spin{i}$ are the first-quantized electron position, momentum and spin operators for electron $i$, and $\acr{\bk}{\mu}, \aan{\bk}{\mu}$ are the second-quantized creation and annihilation operators for the photons in a mode with momentum~$\bk$ and polarization $\mu$.
The nuclear charge is $Z$ and the free-field dispersion relation is $\omega(\bk) = c|\bk|$.
The magnetic vector potential takes the form
\begin{equation}
    \Apot(\bro{i}) = \Omega^{-1/2} \sum_{\mu = \pm 1,\ \bk} \ \frac{ \helvec{\bk}{\mu}}{\sqrt{2 \omega(\bk)}}\Big[\acr{\bk}{\mu} e^{-i\bk \cdot \bro{i}} + 
    \mathrm{h.c.}
    \Big].
\end{equation}
$\Omega$ is the simulation cell volume, $\helvec{\bk}{\mu}$ is the polarization unit vector for polarization $\mu$ at momentum $\bk$, and h.c.\ denotes Hermitian conjugate terms.
The magnetic field operator is $\Bfld(\bro{{i}}) = \nabla_{\bro{i}} \times \Apot(\bro{i})$.
The justification for and construction of this Hamiltonian can be found in App.~\ref{app:definition_of_the_hamiltonian}~\cite{sm}.

The Hamiltonian simulation of $\hpf$ is conducted in the interaction picture~\cite{low2019hamiltonian}.
This algorithm involves repeated application of an electronic Hamiltonian simulation circuit ($\exp(i\hel \tau)$), a photonic Hamiltonian simulation circuit ($\exp(i\hph \tau)$), and a block-encoding of the electron-photon interaction term ($\helph$), where $\tau$ is the discrete time step \cite{low2019hamiltonian}.
The electronic simulation can be carried out using higher-order Trotter product formulas for real-space electronic simulation \cite{low2023complexity,babbush2023quantum}, which were shown to be more efficient than quantum signal processing for comparable conditions in Ref.~\cite{rubin2024quantum}. 
The photonic simulation is fast-forwarded, and the block encoding of the interaction term is efficiently implemented by using the sparsity of the operator.
A detailed discussion of the implementation of each circuit and overheads for the interaction picture algorithm can be found in App.~\ref{subapp:Hamiltonian_simulation_via_interaction}~\cite{sm}.

A simple relationship gives the opacity $\opacity$ from the photon transmission probability $\trans_\mathrm{atom}$,
\begin{equation}
\opacity(\bk) = - \rho_i^{1/3} \ln(\trans_\mathrm{atom}(\bk))/\rho,
\end{equation}
where $\rho$ ($\rho_i$) is the mass (nuclear) density of the plasma.
The transmission probability $\trans_\mathrm{atom}(\bk)$ is related to the difference between the final and initial photon number in a given momentum mode, which can be read off directly by measuring the photonic register.
Accumulating results over $\ns$ shots allows resolving statistical uncertainties in the photon number and sampling over the thermal ensemble.
App.~\ref{subapp:observable_estimation} proves the relation between opacity and atomic transmission probability and estimates $\ns$~\cite{sm}.

\begin{figure}
    \centering
    \includegraphics[width=\columnwidth]{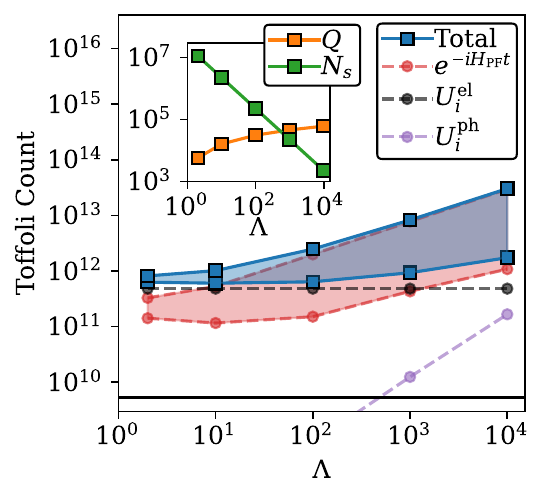}
    \caption{Logical resource estimates for the opacity simulation of solar iron using the interaction picture algorithm with pseudized K-shell electrons. 
    Total Toffoli counts are shown for a single shot of the full simulation as a function of the maximum photonic occupancy ($\maxocc$). 
    The total simulation cost is decomposed into contributions from the simulation cost of the Pauli-Fierz Hamiltonian ($e^{-i\hpf t}$) and the state preparation costs for the electronic ($\hat{U}_\mathrm{i}^\mathrm{el}$) and photonic ($\hat{U}_\mathrm{i}^\mathrm{ph}$) subsystems. 
    Shading indicates ranges between pessimistic and optimistic bounds involved in the Hamiltonian simulation cost.
    The horizontal solid black line shows the simulation cost of only the electronic part of the Hamiltonian ($e^{-i\hel t}$) for reference. 
    The inset shows total qubit counts $(Q)$ and the number of shots required $(\ns)$, including observable estimation and thermal sampling overheads.
    \label{fig:full_cost}}
\end{figure}

\textit{Resource estimates}---
We present logical resource estimates for using our protocol to calculate the opacity of iron at a temperature of \SI{2.26e6}{\kelvin} and electron density of \SI{4e22}{\centi\meter}$^{-3}$. 
These conditions occur near the radiative-convective boundary in the sun and they correspond to the largest discrepancies between classical calculations and experimental data reported in prior work~\cite{bailey2015higher}.
In Fig.~\ref{fig:full_cost}, we provide a relatively coarse description of the resource requirements in terms of the number of logical qubits ($Q$), Toffoli gates, and circuit repetitions ($\ns$) needed to execute the proposed simulation protocol as a function of the maximum photonic occupancy ($\maxocc$).

Our estimates depend on the physical and numerical parameters summarized in Table~\ref{tab:parameters}.
The atomic number of iron determines $\nel=26$ for a single average atom.
Several of the other parameters are informed by iron opacity experiments \cite{bailey2015higher} as discussed in App.~\ref{subapp:parameters_experimental}~\cite{sm}, where the \mbox{7\,--\,\SI{13}{\angstrom}} wavelength range and $\sim \SI{0.1}{\angstrom}$ feature width define our target spectral range and resolution.
The simulation cell volume is $\Omega =(2\rws)^2 L$, where the plasma density gives $\rws$ and the spectral resolution determines the cell length $L$.
The grid spacing $\Delta_r$ required to accurately resolve the iron atom's electronic structure is estimated through classical average-atom calculations as described in App.~\ref{subapp:parameters_numerical}~\cite{sm}.
Then, the number of electronic grid points is $\ngridel=\Omega/\Delta_r^3$.

\begin{table}
    \centering
    \bgroup
    \def\arraystretch{1.5}
    \begin{tabular}{c|c|c|c|c|c}
        $\nel$ & $\Omega$ & $\Delta_r$ & $\ngridel$ & $\ngridph$ & \tee \\\hline
          26 & $1.3\times 10^{7}$ & $10^{-2}$ & $1.3\times 10^{13}$ & 4500 & 0.42 
    \end{tabular}
    \caption{Parameters determining the quantum resources required to predict iron opacity at solar conditions using the present protocol.    
    }
    \label{tab:parameters}
    \egroup
\end{table}

The number of photonic modes, $\ngridph$, is determined by the spectral range and resolution, with a four-fold overhead to adequately capture the tails of the Gaussian wave packet.
Finally, the simulation time $t$ is determined by the distance the photonic wave packet must travel to pass through the atom, including enough spatial separation between the photonic and electronic subsystems in both the initial and final states to avoid biasing computed absorption rates.
Notably, wider spectral ranges lead to shorter $t$ because a more spatially localized photonic wave packet can begin closer to the atom.
A detailed discussion of the simulation parameters can be found in App.~\ref{app:parameters_for_solar_iron}, and
a detailed discussion of the impact of other simulation parameters on the resources requirements is presented in App.~\ref{app:resources-int}~\cite{sm}.

In Fig.~\ref{fig:full_cost}, the state preparation costs, $Q$, and $\ns$ all have straightforward scalings with $\maxocc$.
The electronic state preparation cost is independent of $\maxocc$ and the photonic state preparation cost scales as $\mathcal{O}(\maxocc \log \maxocc)$, with further details in App.~\ref{subapp:initial-state_prep}~\cite{sm}.
The number of logical qubits scales logarithmically with $\maxocc$ due to the second-quantized encoding of the photons.
The number of shots scales inversely with $\maxocc$, since the probability of measuring a photon absorption event scales linearly with the number of photons in the simulation. 
A detailed statistical treatment of the number of shots required to estimate the opacity spectrum is found in App.~\ref{subapp:observable_estimation}~\cite{sm}.

The scaling of the Toffoli count for implementing $e^{-i\hpf t}$ is less straightforward because of the distribution of momenta in the photonic state.
Each photon with momentum $\bk$ in the simulation contributes a cost scaling as $|\bk|^{-1/2}$ arising from the norm of $\Apot$.
The minimum relevant momentum, $\min|\bk|$, determines a bound for the simulation cost.
The targeted wavelength range corresponds to a minimum momentum of $k_{\textrm{min}}^{\textrm{targ}} \approx 0.3$.
However, the initial Gaussian wave packet must carry small, but non-zero, weight at $\Delta_k \approx 2\times 10^{-3} \ll k_{\textrm{min}}^{\textrm{targ}}$ to localize the initial photonic wave packet outside of the atom sphere.

Thus, Fig.~\ref{fig:full_cost} shows a range of Hamiltonian simulation costs depending on the treatment of $\min |\bk|$.
The upper curve corresponds to the pessimistic strict bound wherein $\min|\bk| = \Delta_k$, and the lower curve corresponds to the more optimistic bound wherein $\min|\bk| = k_{\textrm{min}}^{\textrm{targ}}$.
The $\min|\bk| = \Delta_k$ case is a worst-case scenario based on operator norms, while the $\min|\bk| = k_{\textrm{min}}^{\textrm{targ}}$ case represents a state-dependent estimate informed by the initial photonic distribution.
In the initial Gaussian wave packet, the probability of occupying a mode with $|\bk| = \Delta_k$ is about $10^{3}$ times lower than for $|\bk| = k_{\textrm{min}}^{\textrm{targ}}$, indicating that the true cost of the simulation certainly lies closer to the lower estimate.
App.~\ref{subapp:parameters_experimental} and \ref{app:resources-int} contain further details about the photonic distribution and its implications for the cost estimates, respectively~\cite{sm}.

The total Toffoli count is a sum of the state preparation and Hamiltonian simulation costs.
The upper curve, corresponding to the worst-case bound, is dominated by the Hamiltonian simulation cost for $\maxocc \geq 10^1$, resulting in the $\maxocc^{1/2}$ scaling of the Toffoli count arising from the norm of $\helph$.
The lower curve, however, is dominated by the constant electronic state preparation cost for $\maxocc < 10^3$.
In this case, the Hamiltonian simulation cost begins to outweigh the state preparation cost only for $\maxocc > 10^3$, where the Toffoli count exhibits a transition from a nearly constant value to the asymptotic $\maxocc^{1/2}$ scaling.

We expect the lower estimate to more accurately reflect the total simulation cost, where the best choice of $\maxocc$ is between $10^3$ and $ 10^4$.
This range of $\maxocc$ reduces the number of shots required by 2\,--\,3 orders of magnitude without appreciably increasing the Toffoli counts per shot relative to smaller $\maxocc$.
In this region, the state preparation and Hamiltonian simulation costs become comparable contributions to the total Toffoli count, indicating that further resource reductions for opacity simulation require improving \emph{both} electronic structure simulation algorithms~\cite{low2025fast,berry2025quantum} and electronic structure state preparation algorithms.
The total Toffoli cost here is about 2 orders of magnitude larger than the pure electronic structure Hamiltonian simulation cost for the same simulation time, shown as the solid horizontal line in Fig.~\ref{fig:full_cost}.

The resource estimates reported in Fig.~\ref{fig:full_cost} use a pseudization scheme for the K-shell electrons, which do not participate in the transitions relevant to the iron opacity simulation at the thermodynamic conditions of interest.
App.~\ref{app:pseudization} provides the implementation details of our pseudization scheme~\cite{sm}.
Since only the K-shell electrons are pseudized, the overhead incurred from computing non-local contributions to the pseudopotential are negligible compared to the electronic structure simulation primitive.
We find in App.~\ref{app:resources-int} that the K-shell pseudization reduces the total Toffoli count by up to 3 orders of magnitude.


\textit{Conclusions}---
Our protocol frames opacity estimation as the direct simulation of an atomistic photon transmission measurement.
We treat the atom in first quantization to facilitate the use of large basis sets and the representation of highly excited configurations relevant to extreme thermodynamic conditions.
We treat the photons in second quantization such that measurement outcomes in the computational basis are easily related to opacity, and light-matter interaction is rendered in terms of simple kinematic arithmetic.
This is a natural setting for interaction-picture simulation because the free-field dynamics of the photons is fast forwardable, and the short simulation time is determined by the time required for a wave packet to pass through the simulated matter at the speed of light.
The intensity of the illuminating field and spectral resolution are free parameters by which we can adjust the computational overhead per shot, with our quantum resource estimates describing the overhead of estimating many spectral features at once.

This is conceptually similar to simulation methods in high-energy physics~\cite{jordan2012quantum,rhodes2024exponential,hardy2026scattering}.
We also expect that this approach can be translated into other applications involving light-matter interaction (e.g., optical or UV spectra of molecules or condensed phases) where larger cross sections could reduce the resource requirements.
While a naive approach would require $\mathcal{O}(\abs^{-1/2})$ depth to estimate a cross section associated with probability $\abs$, explicit representation and simulation of the field driving that process boosts the probability of observing a relevant event by the intensity ($\maxocc^{1/2}$) of that field. 
For the case of a bosonic field, this approach incurs a $\mathcal{O}(\log\maxocc)$ qubit overhead, but an $\mathcal{O}(\maxocc^{1/2})$ depth overhead that might neutralize any resource advantage in an asymptotic setting\,---\,on top of a prospective $\mathcal{O}(\maxocc)$ state preparation cost.
However, constant-factor sweet spots for which the cost of implementing the interaction term is not the dominant cost of simulation (as in iron opacity) are likely to appear in other settings.

While it would remove multi-photon effects, the qubit overhead of the electromagnetic field could be reduced by (i) removing the photonic subsystem, (ii) implementing $\helph$ using classical fields with prescribed spacetime dependence, and (iii) estimating the opacity from an observable on the atomic subsystem~\footnote{For example, time-dependent multipole moment tensors.}.
The block encoding of the interaction term would still incur a cost comparable to an interaction with an explicit photonic subsystem, in both cases scaling with the intensity of the field.
However, the number of spectral features probed in a single shot would still require the same number of ancilla registers recording a response function of the atomic subsystem.
The overhead for a comparable implementation with a classical field would be $\mathcal{O}(1)$ qubit per wave number, relative to $\mathcal{O}(\log \maxocc)$ per mode in the fully quantum protocol.
A detailed comparison to resource estimates for other approaches to estimating response functions~\cite{roggero2019dynamic,kokcu2024linear,fomichev2025fast,kharazi2025efficient} is left to future work.
Finally, we note that a fully Trotterized approach to Hamiltonian simulation might perform competitively (see App.~\ref{app:trotterization}), though commutators involving the light-matter coupling will compound certain challenging features of many-body Trotterization~\cite{burgarth2024strong,fang2026trotter,fang2026trotterization}, especially when scaling the intensity to boost the signal from which opacity is inferred.

\textit{Acknowledgments}---We thank 
Evan Borras, 
Yale Fan, 
Stephanie Hansen, 
Lex Kemper, 
Milad Marvian, 
Taisuke Nagayama, 
Charles Starrett, 
and Shirley Temple 
for helpful technical discussions.
SP and AK were supported by the Sandia National Laboratories' Laboratory Directed Research and Development (LDRD) Project No.\ 233078.
ADB was supported by the National Nuclear Security Administration's Advanced Simulation and Computing program and the Department of Energy (DOE) Office
of Fusion Energy Sciences “Foundations for quantum simulation of warm dense
matter” project

This work was performed, in part, at the Center for Integrated Nanotechnologies, an Office of Science User Facility operated for the U.S. Department of Energy (DOE) Office of Science.
This article has been co-authored by employees of National Technology \& Engineering Solutions of Sandia, LLC under Contract No. DE-NA0003525 with the U.S. Department of Energy (DOE).
The employees own all right, title and interest in and to the article and are solely responsible for its contents. 
The United States Government retains and the publisher, by accepting the article for publication, acknowledges that the United States Government retains a non-exclusive, paid-up, irrevocable, world-wide license to publish or reproduce the published form of this article or allow others to do so, for United States Government purposes. 
The DOE will provide public access to these results of federally sponsored research in accordance with the \href{https://www.energy.gov/downloads/doe-public-access-plan}{DOE Public Access Plan}.

\let\oldaddcontentsline\addcontentsline
\renewcommand{\addcontentsline}[3]{}
\bibliography{main}
\let\addcontentsline\oldaddcontentsline

\clearpage
\widetext
\begin{center}
\textbf{\large Supplemental Materials: \papertitle}
\end{center}

\setcounter{section}{0}
\setcounter{page}{1}
\setcounter{secnumdepth}{3}
\appendix
\makeatletter

\tableofcontents
\clearpage

\section{Conventions and Notation}
\label{app:notation}

\renewcommand{\theequation}{A\arabic{equation}}
\renewcommand{\thefigure}{A\arabic{figure}}
\renewcommand{\thetable}{A\arabic{table}}
\setcounter{figure}{0}
\setcounter{table}{0}

Throughout, we use atomic units in which $\hbar = e = m_e = 4\pi\epsilon_0 = 1$ and $c \approx 137$.
Operators are indicated by hats, vectors are indicated by boldface, Hilbert spaces are indicated by a calligraphic font, and sets are indicated by blackboard bold.
Notation is largely defined at first use, but we provide a comprehensive list in \mbox{Table \ref{tab:notation}} for convenience.

\begin{longtblr}[
    caption = {Symbol names and definitions used throughout this work.},
    label = {tab:notation},
    ]{colspec={|l|X|}, rowhead=1}
    \hline\hline
    Symbol & Definition\\\hline
    $\Apot$ & The magnetic vector potential operator.\\
    $\abs(\bk)$ & The absorption probability for a photon with wave vector $\bk$ after it interacts with matter.\\
     $\aan{\bk}{\mu}$ & Photonic annihilation operator that acts on a mode with wave vector $\bk$ and helicity $\mu$.\\
     $\acr{\bk}{\mu}$ & Photonic creation operator that acts on a mode with wave vector $\bk$ and helicity $\mu$.\\
     $\alpha$ & Upper bound for an operator norm.\\
     $\Bfld$ & The magnetic field operator.\\
     $c$ & The speed of light, $\approx 137$ in Hartree atomic units.\\
     $\mathbb{C}^2$ & The two-dimensional vector space over the complex numbers.\\
     $\Delta$ & Grid spacing, with subscripts denoting the discretized quantity. $\Delta_r$, $\Delta_k$, and $\Delta_\lambda$ correspond to the real-space, momentum, and wavelength grids. \\
    $\epsilon$ & Total additive error in the Hamiltonian simulation. \\
    $\ereg$ & Regularization constant for the Coulomb interaction.\\
     $\mathcal{F}$ & The photonic Fock space, spanned by multi-photon states indexed by wave number and helicity.\\
     $\bg$ & An index for a uniform grid, $\bg \in \mathbb{Z}^3$.\\
     $\mathbb{G}_N$ & A uniform grid that is a subset of $\mathbb{Z}^3$ with $N^{1/3}$ points per dimension.\\
     $\nel$ & The total number of electrons. For an iron opacity simulation using an atomic model, $\nel\approx 26$.  \\ 
     $\hel$ & The contribution to the Pauli-Fierz Hamiltonian that governs the electronic subsystem. \\
     $\helph$ & The contribution to the Pauli-Fierz Hamiltonian that governs the coupling between the electronic and photonic subsystems. \\
     $\hpf$ & The Pauli-Fierz Hamiltonian.\\
     $\hph$ & The contribution to the Pauli-Fierz Hamiltonian that governs the photonic subsystem.\\
     $\theta$ & The polar angle in spherical coordinates.\\
     $\id$ & The identity. We use a typical overloaded notation in which the space that this operator acts on is not explicitly indicated and evident from context.\\
     $\bk$ & The wave vector associated with a photonic mode, $\bk \in \mathbb{R}^3$.    \\
     $\bar{\bk}$ & The average wave vector of a photonic wave packet.\\
     $\tilde{k}$ & The order of a Trotterized time-evolution algorithm.\\
     $\mathbb{K}$ & A set of occupied photonic modes, $\mathbb{K}\subset\mathbb{R}^3$.\\
     $\bk_\bg$ & The momentum vector corresponding to grid index $\bg$.\\
     $k_\mathrm{max}^\mathrm{targ}$ & The maximum wave number of interest targeted by the opacity calculation, $k_\mathrm{max}^\mathrm{targ}\in \mathbb{R^+}$.\\
     $k_\mathrm{min}^\mathrm{targ}$ & The minimum wave number of interest targeted by the opacity calculation, $k_\mathrm{min}^\mathrm{targ}\in \mathbb{R^+}$.\\
     $k_\mathrm{max}^\mathrm{sim}$ & The maximum wave number supported by the planar photonic wave packet within the quantum simulation, $k_\mathrm{max}^\mathrm{sim}\in\mathbb{R}^+$.\\
     $k_\mathrm{min}^\mathrm{sim}$ & The minimum wave number supported by the planar photonic wave packet within the quantum simulation, $k_\mathrm{min}\in\mathbb{R}$. Note that $k_\mathrm{min}^\mathrm{sim}$ can be negative.\\
     $\opacity$ & Opacity, the primary quantity of interest in this work.\\
     $\ell$ & The angular momentum quantum number of a single-electron atomic orbital.\\
     $\maxocc$ & The maximum occupancy of any given photonic mode.\\
     $\lambda$ & The wavelength of a photon.\\
     $\mathcal{L}^2(\mathbb{R}^3)$ & The space of square-integrable functions on $\mathbb{R}^3$.\\
     $m$ & The magnetic quantum number of a single-electron atomic orbital.\\
     $\mu$ & Index that distinguishes the helicity of a photonic mode, $\mu \in \lbrace -1,+1\rbrace$.\\
     $n$ & The principal quantum number of a single-electron atomic orbital.\\
     $\mathcal{N}$ & A normalization constant.\\
     $\ngridel$ & The number of grid points used to discretize the electronic Hilbert space.\\
     $\ngridph$ & The number of modes used to discretize the photonic Fock space.\\
     $\npf$ & The number of shots per feature. \\
     $\ns$ & The number of shots, or the number of times the quantum circuit is repeated.\\
     $N_\mathrm{wp}$ & The number of Gaussian wave packets in the initial photonic state.\\
     $\hat{n}_{\bk}$ & The helicity-unresolved photonic number operator $\hat{n}_{\bk}=\sum_\mu \hat{n}_{\bk,\mu}$. \\
     $\hat{n}_{\bk,\mu}$ & The helicity-resolved photonic number operator $\hat{n}_{\bk,\mu}=\acr{\bk}{\mu}\aan{\bk}{\mu}$. \\
     $\bpo{j}$ & The momentum operator for electron $j$.\\
     $Q$ & The number of logical qubits required for the simulation. Subscripts denote contributions from the system register and auxiliary qubits required to implement various subroutines.\\
     $\mathbb{R}^3$ & The three-dimensional vector space over the real numbers.    \\
     $r$   & The radial distance from the origin.\\
     $\br$ & A position vector. \\
     $\tilde{r}$ & The number of time steps in a Trotterized time-evolution scheme. \\
     $\bro{j}$ & The position operator for electron $j$.\\
     $\br_\bg$ & The position vector corresponding to grid index $\bg$.\\
     $\bR_{J}$ & The (classical) position of ion $J$, $\bR_{J} \in \mathbb{R}^3$. Within the atomic model, $\bR=0$.\\
     $\bar{\bR}$ & The average position of a photonic wave packet.\\
     $\rws$ & The Wigner-Seitz radius of the atom sphere.\\
     $R_n^\ell(r)$ & The radial wavefunction of a single-electron atomic orbital, indexed by principal quantum number $n$ and angular momentum quantum number $\ell$.\\
     $\rho_i$ & The ion number density of the plasma.\\
     $\spin{j}$ & The spin operator for electron $j$.\\
     $\sigma_k$ & The standard deviation of a photonic wave packet in momentum space.\\
     $\sigma_r$ & The standard deviation of a photonic wave packet in real space.\\
     $T$ & The number of logical T gates required for each shot of the simulation. Subscripts or parentheses denote T gate costs of specific subroutines.\\
     $\trans(\bk)$ & The transmission probability for a photon with wave vector $\bk$ after it interacts with matter.\\
     $\tee$ & Time, typically the total time over which the atom and photon field undergo Hamiltonian evolution.\\
     $\tke$ & The contribution to $\hel$ due to the kinetic energy of the electrons.  \\
     $\tau$ & The time step Hamiltonian simulation. \\
     $\uvec$ & A unit vector in $\mathbb{R}^3$.\\
     $\hat{U}_{\aan{\bk}{\mu}}$ & Unitary operator that decrements the occupancy of a photonic mode with wave vector $\bk$ and helicity $\mu$.\\
     $\hat{U}_{\acr{\bk}{\mu}}$ & Unitary operator that increments the occupancy of a photonic mode with wave vector $\bk$ and helicity $\mu$.\\
     $\helvec{\bk}{\mu}$ & The helicity vector of a photonic mode with wave vector $\bk$ and helicity $\mu$, $\helvec{\bk}{\mu} \in \mathbb{R}^3$.\\
     $\vee$ & The contribution to $\hel$ due to the instantaneous Coulomb interactions between electrons. \\
     $\vei$ & The contribution to $\hel$ due to the instantaneous Coulomb interactions between electrons and ions. \\
     $\varphi$ & The azimuthal angle in spherical coordinates.\\
     $\ket{\phi}$ & A single-particle electronic orbital obtained from a classical mean-field calculation.\\
     $\ket{\psi_\mathrm{i}}$ & The initial state prepared for the full system. Initial states for the electronic and photonic subsystems are indicated by el and ph superscripts, respectively.\\
     $\ket{\psi_\mathrm{f}}$ & The final state after evolving $\ket{\psi_\mathrm{i}}$ under $\hpf$ for time $t$.\\
     $Y_\ell^m(\theta,\varphi)$ & The spherical harmonic of degree $\ell$ and order $m$.\\
     $\om{\bk}$ & The angular frequency of a photon with wave vector $\bk$, equivalent to its energy up to a factor of $\hbar$ and related to its wave number ($|\bk|$) through the dispersion relation $\om{\bk}=c|\bk|$.\\
     $\Omega$ & The integration volume over which the photon field is defined.\\
     $\mathbb{Z}^3$ & The three-dimensional vector space over the integers.\\
     $Z_{J}$ & The nuclear charge of ion $J$. For an iron opacity simulation, $Z=26$.\\
     $\nion$ & The number of ions. For an atomic model, $\nion=1$.\\
     \hline\hline
\end{longtblr}

\clearpage
\section{Definition of the Hamiltonian and Hilbert space}
\label{app:definition_of_the_hamiltonian}

\renewcommand{\theequation}{B\arabic{equation}}
\renewcommand{\thefigure}{B\arabic{figure}}
\setcounter{figure}{0}

We first define the Pauli-Fierz Hamiltonian in the Born-Oppenheimer approximation, $\hat{H}_{\textrm{PF}}$.
We have relied on elements of the derivations and notations from Hiroshima and Spohn~\cite{hiroshima2001ground, spohn2004dynamics} and Mukhopadhyay \textit{et al.}~\cite{mukhopadhyay2024quantum}.
$\hat{H}_{\textrm{PF}}$ is a semi-relativistic approximation to quantum electrodynamics, accounting for both the electronic and photonic degrees of freedom, with the former treated non-relativistically and the latter fully relativistically.
It is defined as
\begin{equation}
    \hpf = \underbrace{\tke + \vee + \vei}_{\hel} + \underbrace{\sum_{\mu = \pm 1} \sum_{\bk} \omega(\bk) \acr{\bk}{\mu}\aan{\bk}{\mu}}_{\hph} + \underbrace{\sum_{i = 1}^{\nel} \frac{1}{2c}\left[\frac{1}{c}\Apot^2(\bro{i}) -\lbrace \bpo{i},\Apot(\bro{i})\rbrace - \spin{i} \cdot \Bfld(\bro{{i}}) \right]}_{\helph}. \label{eq:pauli-fierz_hamiltonian}
\end{equation}
$\hel$ describes spin-1/2 non-relativistic electrons that interact with each other through an instantaneous Coulomb force and $\hph$ describes the electromagnetic field.
$\helph$ accounts for the interaction between the electrons and this field through the lowest-order coupling from quantum electrodynamics (the dipole and spin-magnetic field couplings) in the Coulomb gauge.
This interaction will lead to scattering and absorption of an imposed electromagnetic field by the electrons.
Opacity quantifies the extent to which these processes occur as a function of the frequency of the photons involved.

The Hilbert space associated with $\hpf$ is
\begin{equation}
    \mathcal{H}_\textrm{PF} = \mathcal{A}\left[\left(\mathbb{C}^2 \otimes \mathcal{L}^{2}(\mathbb{R}^{3})\right)^{\otimes \nel}\right] \otimes \mathcal{F},
\end{equation}
where the first factor describes $\nel$ spin-1/2 electrons in first quantization and the second factor describes the electromagnetic field in second quantization.
$\mathcal{A}$ antisymmetrizes the $\eta$-fold tensor product of single-electron Hilbert spaces, each of which is comprised of its spin ($\mathbb{C}^2$) and translational ($\mathcal{L}^{2}(\mathbb{R}^3)$) degrees of freedom.
$\mathcal{F}$ is the Fock space spanned by multi-photon states generated by products of bosonic creation operators $\acr{\bk}{\mu}$ acting on the vacuum $|\emptyset\rangle$, where $\mu \in \lbrace -1,+1\rbrace$ indicates the helicity of the photon and $\bk \in \mathbb{R}^3$ indicates its wave vector.
$\hel$ and $\hph$ act non-trivially on the electronic and photonic factors, respectively. 
We elide any explicit notation to indicate their trivial actions on either complementary factor.
$\helph$ acts non-trivially on both factors.

The electronic Hilbert space is discretized on a uniform grid, and individual terms comprising $\hel$ are then
\begin{subequations}
\begin{align}
    \tke &= \sum_{i=1}^{\nel} \frac{\bpo{i}^2}{2} = \sum \limits_{i=1}^{\nel} \textrm{QFT}_i\left(\sum \limits_{\bg \in \mathbb{G}_{\ngridel}} \frac{|\bk_\bg|^2}{2}\ket{\bg}\!\!\bra{\bg}_i \right)\textrm{QFT}^{\dagger}_i, \\
    \vee &= \frac{1}{2}\sum_{i \neq j = 1}^{\nel}\frac{1}{|\bro{i}-\bro{j}|}= \frac{1}{2} \sum_{i \neq j = 1}^{\nel} \sum \limits_{\bg,\bg' \in \mathbb{G}_{\ngridel}} \frac{1}{|\br_\bg - \br_{\bg'}|}\ket{\bg}\!\!\bra{\bg}_{i}\ket{\bg'}\!\!\bra{\bg'}_j,~\textrm{and}\label{eq:vee}\\
    \vei &=  -\sum_{i = 1}^{\nel} \sum_{l = 1}^{\nn} \frac{Z_l}{|\bro{i}-\bR_l|} = -\sum_{i = 1}^{\nel} \sum_{l = 1}^{\nn}  \sum_{\bg \in \mathbb{G}_{\ngridel}} \frac{Z_l}{|\br_\bg - \bR_l|}\ket{\bg}\!\!\bra{\bg}_i. \label{eq:vei}   
\end{align}
\end{subequations}
Here $\nn$ is the number of nuclei, $Z_l$ is the charge of the $l$th nucleus, $\br_\bg$ are electronic positions, and $\bR_l$ are nuclear positions.
The electronic positions are associated with points on a uniform grid $\mathbb{G}_{\ngridel}$ such that
\begin{subequations}\label{eq:discretization}
\begin{align}
    \br_\bg &= \bg \left(\frac{\Omega}{\ngridel}\right)^{1/3}, \\
    \bk_\bg &= \bg \left(\frac{2\pi}{\Omega^{1/3}}\right),~\textrm{and} \\
    \bg & \in \mathbb{G}_{\ngridel} = \left[-\frac{\ngridel^{1/3}}{2},\frac{\ngridel^{1/3}}{2}\right]^{3},
\end{align}
\end{subequations}
where $\Omega$ is the volume of the domain discretized by the grid.
Representing an $\nel$-electron wavefunction on this grid requires $\nel \left(\log_2(\ngridel)+1\right)$ qubits, with the additional $\nel$ qubits accounting for the spin degree of freedom of each electron.
The choice of $\ngridel$ is determined by the target precision and the desired spectral resolution (see App.~\ref{app:parameters_for_solar_iron}).

The exact form of $\vee$ in Eq.~\eqref{eq:vee} is unbounded because of the divergence at $\br_\bg=\br_{\bg'}$, complicating simulation error bounds for the electronic subsystem~\cite{burgarth2023state,burgarth2024strong}.
Thus, our resource estimates assume a regularized version of the electron-electron interaction,
\begin{equation}
    \vee \rightarrow \frac{1}{2}\sum_{i \neq j = 1}^{\nel}\frac{1}{(|\bro{i}-\bro{j}|^2 + \ereg^2)^{1/2}}.
    \label{eq:vee_regularized}
\end{equation}
The regularization constant $\ereg$ is small compared to the grid spacing and it is determined by the target precision (see App.~\ref{app:parameters_for_solar_iron}).

The photonic Fock space is discretized into $\ngridph$ modes with definite wave vector and helicity.
The associated bosonic creation and annihilation operators satisfy the commutation relations
\begin{subequations}
\begin{align}
    &[\acr{\bk}{\mu}, \acr{\bk^\prime}{\mu^\prime}] = 0,\\
    &[\aan{\bk}{\mu}, \aan{\bk^\prime}{\mu^\prime}] = 0,~\text{and} \\
    &[\aan{\bk}{\mu}, \acr{\bk^\prime}{\mu^\prime}] = \delta_{\mu \mu^\prime} \delta_{\bk \bk^\prime}.
\end{align}
\end{subequations}
The occupation number of each mode is encoded in binary with a maximum occupancy less than $\maxocc$, requiring $\ngridph\log_2(\maxocc)$ qubits to represent the wavefunction of the photonic subsystem.
The choice of $\ngridph$ is determined by the set of spectral features that will be probed in any single round of our protocol (see App.~\ref{app:parameters_for_solar_iron}), and the choice of $\maxocc$ will influence the signal-to-noise ratio in the observable estimate and thus the number of shots required to achieve a certain target precision (see App.~\ref{subapp:observable_estimation}).

In $\helph$ the light-matter couplings are through the magnetic vector potential and magnetic field at each electronic position.
In our mixed first-/second-quantized discretization of $\mathcal{H}_{\textrm{PF}}$, these operators are 
\begin{subequations}\label{eq:aebfields}
\begin{align}
    \Apot(\bro{i}) &= \Omega^{-1/2} \sum_{\mu = \pm 1, \bk} \ \frac{1}{\sqrt{2 \omega(\bk)}} \helvec{\bk}{\mu}\Big[\acr{\bk}{\mu} e^{-i\bk \cdot \bro{i}} + \aan{\bk}{\mu}e^{i\bk \cdot \bro{i}}\Big]\nonumber\\ 
    &= \Omega^{-1/2} \sum \limits_{\bg \in \mathbb{G}_{\ngridel}}\sum_{\mu = \pm 1, \bk} \ \frac{1}{\sqrt{2 \omega(\bk)}} \helvec{\bk}{\mu}\Big[\acr{\bk}{\mu} e^{-i\bk \cdot \br_{\bg}} + \aan{\bk}{\mu}e^{i\bk \cdot \br_{\bg}}\Big] \ket{\bg}\!\!\bra{\bg}_i, \\
    \Bfld(\bro{i}) &= i \Omega^{-1/2} \sum_{\mu = \pm 1, \bk} \frac{1}{\sqrt{2 \omega(\bk)}} \bk \times \helvec{\bk}{\mu}\Big[\acr{\bk}{\mu} e^{-i\bk \cdot \bro{i}} - \aan{\bk}{\mu} e^{i\bk \cdot \bro{i}}\Big]\nonumber\\
    &= i \Omega^{-1/2}  \sum \limits_{\bg \in \mathbb{G}_{\ngridel}} \sum_{\mu = \pm 1, \bk} \frac{1}{\sqrt{2 \omega(\bk)}} \bk \times \helvec{\bk}{\mu}\Big[\acr{\bk}{\mu} e^{-i\bk \cdot \br_{\bg}} - \aan{\bk}{\mu} e^{i\bk \cdot \br_{\bg}}\Big] \ket{\bg}\!\!\bra{\bg}_i,
\end{align}
\end{subequations}
where $\omega(\bk) = c|\bk|$ and $\helvec{\bk}{\mu}$ are the helicity vectors.
The electronic spin in the magnetic-field coupling term $\spin{i} \cdot \Bfld(\bro{{i}})$ is simply 
\begin{equation}
    \spin{i} = \frac{1}{2}\left(\hat{X}_i \uvec_x + \hat{Y}_i \uvec_y + \hat{Z}_i \uvec_z  \right),
\end{equation}
where $\lbrace \hat{X}_i,\hat{Y}_i,\hat{Z}_i\rbrace$ are Pauli matrices acting on the $i$th electron's spin and $\lbrace \uvec_x,\uvec_y,\uvec_z\rbrace$ are the associated Cartesian unit vectors.
The electronic momenta in the dipole coupling term $\lbrace \bpo{i},\Apot(\bro{i})\rbrace$ are implemented as 
\begin{equation}
    \bpo{i} = \textrm{QFT}_i \left( \sum \limits_{\bg \in \mathbb{G}_{\ngridel}} \bk_{\bg} \ket{\bg}\!\!\bra{\bg}_i\right) \textrm{QFT}^\dagger_{i},
\end{equation}
where the QFT on the $i$th electronic register is defined as
\begin{equation}
    \textrm{QFT}_i = \frac{1}{\sqrt{\ngridel}}\sum \limits_{\bg, \bg' \in \mathbb{G}_{\ngridel}} e^{i\frac{2\pi}{\ngridel}{\bg'}\cdot{\bg}} \ket{\bg'}\!\!\bra{\bg}_i.
\end{equation}
Notably, we use commensurate electronic and photonic grids, \emph{i.e.}, the photonic momentum modes are a subset of the electronic momentum grid, $\bk\in\{\bk_\bg\}$.

The full Pauli-Fierz Hamiltonian in Eq.~\eqref{eq:pauli-fierz_hamiltonian} would suffice to model the opacity of a dense many-atom plasma and the protocol in this paper could be easily adapted to this fully generic setting.
However, the conditions typical of astrophysical systems allow us to leverage additional assumptions that reduce the cost of quantum simulation.
In particular, we assume that the plasma density is sufficiently low that interactions between neighboring ions can be neglected, allowing us to restrict the electronic degrees of freedom to those of a single average atom, \emph{i.e.}, $\nion=1$.
Such atomic models have long been used in astrophysical modeling, as well as broader high-energy density science applications~\cite{seaton1995opacity, blancard2011solar, colgan2013light, hansen2007hybrid, hansen2023self}. 
We refer to Ref.~\onlinecite{callow2022first} for a derivation of the average-atom model from first principles.
We will henceforth assume that our average atom occupies a sphere of Wigner-Seitz radius $\rws$ that is centered at the origin ($\bR=0$) and contained within the uniform cubic grid of volume $\Omega$. 
The number density of atoms is then $\rho_i=3/(4\pi \rws^3)$, and Eq.~\eqref{eq:pauli-fierz_hamiltonian} reduces to Eq.~\eqref{eq:HPF_AA} of the main text.

The average-atom model given by Eq.~\eqref{eq:HPF_AA} of the main text describes many-body interactions among electrons near the central ion exactly, overcoming approximations made in classical calculations like the exchange-correlation functional, truncation in configuration interaction treatments, and lifetime broadening.
However, it does not include other important phenomenology like electronic relativistic effects leading to spectral line splitting, which will be the subject of future work.
Various environmental effects on the average atom, \emph{e.g.}, Stark broadening due to plasma microfields or Doppler broadening due to ion motion, can be accounted for through the inclusion of an external potential, \emph{i.e.}, augmentation of the one-body term in Eq.~\eqref{eq:vei}.
We expect that including these environmental effects will not dramatically change the resource estimates in this manuscript because the cost of simulating $\hel$ is dominated by $\vee$, and costs associated with the Coulomb singularity in Eq.~\eqref{eq:vei} will further dominate any contributions due to smoother Stark or Doppler broadening terms.

\clearpage
\section{Quantum algorithmic protocol for opacity}
\label{app:quantum_alg_for_opacity}

\renewcommand{\theequation}{C\arabic{equation}}
\renewcommand{\thefigure}{C\arabic{figure}}
\setcounter{figure}{0}

In App.~\ref{app:definition_of_the_hamiltonian} we described how to encode the Hilbert space of an average-atom Pauli-Fierz Hamiltonian in the system register's $Q_\mathrm{sys} = \eta \left(\log_2(\ngridel)+1\right) + \ngridph\log_2(\maxocc)$ qubits.
In this Appendix, we describe our quantum algorithmic protocol for calculating the opacity of a plasma comprised of such atoms.
As shown in Fig.~\ref{fig:circuit} of the main text, the steps of the protocol consist of preparing an appropriate initial state for the atom and photonic field (App.~\ref{subapp:initial-state_prep}), evolving it over time using the interaction picture formalism (App.~\ref{subapp:Hamiltonian_simulation_via_interaction}), and estimating the opacity from measurements of the number of photons remaining in each mode after the evolution (App.~\ref{subapp:observable_estimation}).

As an alternative to the interaction picture formalism, we also considered a fully Trotterized time evolution algorithm.
However, we ultimately found lower resource estimates for interaction picture simulation.
App.~\ref{app:trotterization} provides a detailed discussion of Trotterized simulation costs.

\subsection{Initial-state preparation}
\label{subapp:initial-state_prep}

The system register is initialized in a product state between the electronic and photonic subsystems,
\begin{equation}
    \ket{\psi_\mathrm{i}} = \ket{\psi_\mathrm{i}^\mathrm{el}} \otimes \ket{\psi_\mathrm{i}^\mathrm{ph}}.
\end{equation}
The subsystems will become entangled through their evolution under $\hpf$, and measurements on the photonic register will probe the atom's contribution to the opacity of a thermal ensemble (see Fig.~\ref{fig:opacity_example} in the main text).
The pure initial electronic state $|\psi_\mathrm{i}^\mathrm{el}\rangle$ represents a sample from a thermal ensemble that is prepared as described in Section \ref{subapp:e_prep}.
Rather than representing a physical photonic field, the initial photonic state $|\psi_\mathrm{i}^\mathrm{ph}\rangle$ is chosen to efficiently probe the opacity over a specific range of interesting spectral features.
We describe the photonic state preparation algorithm in Section \ref{subapp:photon_prep}.

\subsubsection{Electronic state preparation}
\label{subapp:e_prep}
The initial electronic state is a Slater determinant state sampled according to Boltzmann statistics at the physically relevant temperature scale of 1.91\,--\,\SI{2.26}{\mega\kelvin}, or about \SI{10}{\hartree}.
This approach is currently the state-of-the-art in end-to-end resource estimates for finite-temperature quantum simulations \cite{rubin2024quantum}, and it relies on classical calculations to sample configurations expected to contribute to the opacity in the spectral range of interest.
Since each initial electronic state has a corresponding set of allowed transitions  (see Fig.~\ref{fig:opacity_example} in the main text), sampling from the thermal ensemble is equivalent to sampling opacity features.
Thus, the thermal sampling incurs no additional overheads in the number of required shots.

We use an efficient protocol to prepare the single Slater determinant wavefunctions~\cite{rubin2024quantum, babbush2023quantum}.
This method prepares the Slater determinant in second quantization and uses an additional $\nel$ auxiliary qubits to map the state to first quantization.
The Toffoli cost of preparing such a state is
\begin{equation}
    T_\mathrm{prep}^\mathrm{el} = \nel \ngridel T_\textrm{Givens} + \left[\ngridel(3\nel + \lceil \log_2(\nel + 1) \rceil - 2)\right] = \mathcal{O}(\nel \ngridel).
\end{equation}
The first term is the cost of preparing the second-quantized state, where $T_\textrm{Givens}$ is the cost of a Givens rotation. 
The second term is the cost of mapping this second-quantized state to first quantization.

If we were to naively carry out this procedure over the entire set of $10^{16}$ discretized grid points for an all-electron simulation of the iron atom, we would find $T_\mathrm{prep}^\mathrm{el} \sim 10^{17}$, which would overshadow the cost of 
interaction picture dynamics in Section~\ref{subapp:Hamiltonian_simulation_via_interaction} by multiple orders of magnitude.
Pseudizing the 1s electrons allows a coarser grid with $10^{13}$ points (see App.~\ref{subapp:parameters_numerical}) and reduces the naive electronic state preparation cost to $T_\mathrm{prep}^\mathrm{el} \sim 10^{14}$, which still exceeds time evolution costs unless $\maxocc\gtrsim 10^6$.
However, the initial electronic state is confined to the atom sphere, a small portion of the entire simulation cell (see Fig.~\ref{fig:schematic} of the main text).
This smaller volume of $4.4\times 10^3$ atomic units contains $4.4\times 10^{12}$ ($4.4\times 10^9$) grid points in the all-electron (1s pseudized) case, reducing the electronic state preparation cost to $T_\mathrm{prep}^\mathrm{el} \sim 10^{14}$ ($10^{11}$).

\subsubsection{Photonic state preparation}
\label{subapp:photon_prep}


The initial photonic state must be spatially localized outside of the atom and contain momentum modes corresponding to the targeted spectral range for the opacity.
In general, these requirements can be satisfied by a collection of $N_\mathrm{wp}$ Gaussian wave packets,
\begin{equation}
    \ket{\psi_\mathrm{i}^\mathrm{ph}} = \prod_{J=1}^{N_\mathrm{wp}} \left(\frac{1}{\mathcal{N}_J} \sum_{\bk \in \mathbb{K}_J} e^{-i\bk\cdot {\bf \bar{R}}_J} \, e^{-(k_\shortparallel-|\bar{\bk}_J|)^2/(2\sigma_k^\shortparallel)^2} \, e^{-(k_\perp)^2/2(\sigma_k^\perp)^2} \;U_{\acr{\bk}{\mu_J}}^{\maxocc_2} \right)^{\maxocc_1} |\emptyset\rangle.
    \label{eq:initial_photonic_state_general}
\end{equation}
Here, ${\bf \bar{R}}_J$, $\bar{\bk}_J$, and $\mu_J$ denote each wave packet's average initial position, average wave vector, and polarization.
Since the photons propagate toward the atom at the origin, 
${\bf \bar{R}}_J$ and $\bar{\bk}_J$ are antiparallel.
Each occupied momentum mode $\bk$ is decomposed into components that are parallel and perpendicular to $\bar{\bk}_J$, where $k_\shortparallel = (\bk\cdot\bar{\bk}_J)/|\bar{\bk}_J|$ and $k_\perp^2=|\bk|^2 - k_\shortparallel^2$.
The Gaussian weights involve $\sigma_k^\shortparallel$ and $\sigma_k^\perp$, which are standard deviations parallel and perpendicular to the average direction of motion, respectively.
$\mathcal{N}_J$ are normalization constants given by
\begin{equation}
    \mathcal{N}_J^2 = \sum_{\bk\in\mathbb{K}_J} e^{-(k_\shortparallel-|\bar{\bk}_J|)^2/2(\sigma_k^\shortparallel)^2} \, e^{-(k_\perp)^2/(\sigma_k^\perp)^2} .
\end{equation}
Provided the $\bar{\bk}_J$ are commensurate with the discrete momentum grid, $\mathcal{N}_J=\mathcal{N}$ are all the same.
Each wave packet contains $\maxocc=\maxocc_1 \maxocc_2$ photons consisting of $\maxocc_1$ independent groups of $\maxocc_2$ entangled photons, where each of the $\maxocc_2$ photons within a group occupies the same momentum mode.
$\mathbb{K}_J$ represents a restricted set of momentum modes occupied by each wave packet.
Finally, the unitary 
\begin{equation}
    \hat{U}_{\acr{\bk}{\mu_J}}^{\maxocc_2} \propto (\acr{\bk}{\mu_J})^{\maxocc_2}
\end{equation}
adds $\maxocc_2$ to the occupancy of momentum mode $\bk, \mu_J$.
The proportionality relationship is due to the fact that $\acr{\bk}{\mu_J}$ is not unitary.

Increasing the total number of photons by choosing large values of $N_\mathrm{wp}$, $\maxocc_1$, and $\maxocc_2$ reduces the number of times the algorithm must be repeated to detect sufficient absorption events for an accurate opacity estimate (see Section~\ref{subapp:observable_estimation}).
However, as discussed in App. \ref{app:parameters_for_solar_iron}, including multiple wave packets with distinct $\bar{\bk}_J$ requires more momentum modes and increases the size of the photonic register.
On the other hand, including multiple wave packets with the same $\bar{\bk}_J$ but different $\bar{\bR}_J$ requires a longer simulation time for all of the wave packets to pass through the atom.
To reduce resource requirements for a single shot, we proceed with a single wave packet with arbitrary helicity.
This choice further allows us to reduce the simulation cell volume and let $k_\perp=0$, $\sigma_k^\perp\rightarrow 0$ so that the wave packet becomes planar (see App. \ref{app:parameters_for_solar_iron}).
The initial photonic state then simplifies into
\begin{equation}
    \ket{\psi_\mathrm{i}^\mathrm{ph}} = \left(\frac{1}{\mathcal{N}} \sum_{\bk \in \mathbb{K}} e^{-i\bk\cdot {\bf \bar{R}}} \, e^{-|\bk-\bar{\bk}|^2/(2\sigma_{k})^2} \;U_{\acr{\bk}{\mu}}^{\maxocc_2} \right)^{\maxocc_1} |\emptyset\rangle,
    \label{eq:initial_photonic_state_planar}
\end{equation}
where $\bk \parallel \bar{\bk}$ for all $\bk \in \mathbb{K}$.
Given a uniform momentum grid with spacing $\Delta_k$ that spans $k_\mathrm{min}\leq |\bk|\leq k_\mathrm{max}$, we have $\mathbb{K}=\{\bk_j\}$ where $\bk_j = (k_\mathrm{min} + j\Delta_k) \bar{\bk}/|\bar{\bk}|$ and $j$ is an integer between $0$ and $\ngridph-1 = (k_\mathrm{max}-k_\mathrm{min})/\Delta_k $.

\begin{figure}
    \begin{quantikz}
        \lstick{$\ket{0}^{\otimes \log_2 N_{1q}}$} & \gate{\hat{U}_{1q}} & \gate[2]{\hat{U}_{1q\rightarrow 2q}^{\maxocc_2}} & \gate{\hat{U}_{1q}^\dagger} & \rstick{$\ket{0}^{\otimes \log_2 N_{1q}}$} \\
        \lstick{$\ket{\psi}_{2q}$} &&&&
    \end{quantikz}
    \caption{Circuit for adding $\maxocc_2$ entangled photons to the second-quantized Gaussian wave packet encoded in $\ket{\psi}_{2q}$.
    First, $\hat{U}_{1q}$ prepares a Gaussian state on an auxiliary first-quantized register.
    Then, $\hat{U}_{1q\rightarrow 2q}$ performs coherent arithmetic to increase the photon occupancies in $\ket{\psi}_{2q}$, thereby translating the first-quantized Gaussian state into second quantization.
    Finally, $\hat{U}_{1q}^\dagger$ uncomputes the first-quantized Gaussian state and amplitude amplification ensures a high probability of measuring $\ket{0}$ on the auxiliary register.
    }
    \label{fig:ph_state_prep}
\end{figure}
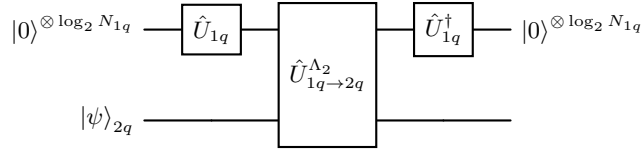

The photonic state in Eq.~\eqref{eq:initial_photonic_state_planar} can be prepared by repeating the circuit shown in Fig.~\ref{fig:ph_state_prep} $\maxocc_1$ times.
This state preparation protocol involves an auxiliary register with $\log_2 N_{1q}$ qubits, which is used to encode a first-quantized Gaussian wave packet prepared by $\hat{U}_{1q}$:
\begin{equation}
    \hat{U}_{1q} \ket{0} = \frac{1}{\mathcal{N}} \sum_{j=0}^{\ngridph -1} e^{-i\bk_j\cdot {\bf \bar{R}}} \, e^{-|\bk_j-\bar{\bk}|^2/(2\sigma_{k})^2} \ket{k_\mathrm{min}/\Delta_k +j}.
    \label{eq:1dgaussian}
\end{equation}
To implement $\hat{U}_{1q}$, we begin with the Gaussian state preparation method from Ref.~\cite{bagherimehrab2022nearly} to prepare
\begin{equation}
    \hat{U}_\mathrm{Gauss}\ket{0} = \frac{1}{\mathcal{N}} \sum_{j=0}^{\ngridph -1} e^{-|\bk_j - \bar{\bk}|^2/(2\sigma_k)^2} \,\ket{k_\mathrm{min}/\Delta_k + j}
\end{equation}
at $\mathcal{O}(\log \ngridph)$ cost.
The additional phase factor in Eq.~\eqref{eq:1dgaussian} can be applied by transforming the wave packet into real space, translating it, and transforming back into momentum space.
That is, we implement
\begin{equation}
    \hat{U}_{1q} = \textrm{QFT}^\dagger\; \mathcal{T}(|\bar{\bR}|) \; \textrm{QFT} \;  \hat{U}_\mathrm{Gauss},
\end{equation}
where $\mathcal{T}(|\bar{\bR}|)$ shifts the register by $|\bar{\bR}|/\Delta_r$.
Note that avoiding aliasing errors within the QFTs requires including the full momentum grid in the first-quantized register so that $N_{1q}=\ngridel^{1/3} \geq \ngridph$.
The translation can be accomplished using only Clifford gates, while the QFTs require $\mathcal{O}((\log N_{1q})^2)$ T gates.
The overall cost of $\hat{U}_{1q}$ is $\mathcal{O}((\log N_{1q})^2)$, a negligible contribution to the total state preparation cost.

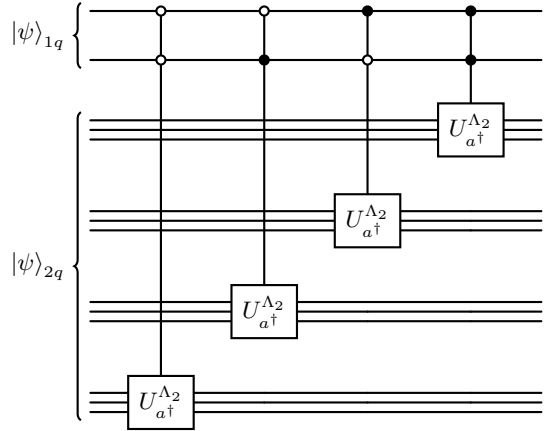
\begin{figure}[h]
    \begin{quantikz}[wire types = {q,q,b,b,b,b}, classical gap=0.05in]
     \lstick[2]{$\ket{\psi}_{1q}$} & \octrl{1} & \octrl{1} & \ctrl{1} & \ctrl{1} & \\
     & \octrl{4} & \ctrl{3} & \octrl{2} & \ctrl{1} & \\
     \lstick[4]{$\ket{\psi}_{2q}$} &&&& \gate{U_{a^\dagger}^{\maxocc_2}} & \\
     &&& \gate{U_{a^\dagger}^{\maxocc_2}} &&\\
     && \gate{U_{a^\dagger}^{\maxocc_2}}&&& \\
     & \gate{U_{a^\dagger}^{\maxocc_2}} &&&& \\
    \end{quantikz}
    \caption{Circuit implementing $U_{1q\rightarrow 2q}^{\maxocc_2}$ to add a photonic state encoded in first quantization within $\ket{\psi}_{1q}$ to the second-quantized system register $\ket{\psi}_{2q}$. 
    This small illustrative example corresponds to $N_{1q}=4$, $\maxocc=8$.
    For each relevant bit string in $\ket{\psi}_{1q}$, a controlled $U_{a^\dagger}^{\maxocc_2}$ operation adds $\maxocc_2$ to the occupation of the corresponding momentum mode in $\ket{\psi}_{2q}$.
    \label{fig:translation_subroutine}
    }
\end{figure}

Once the wave packet has been encoded in the auxiliary first-quantized register $\ket{\psi}_{1q}$, we transfer it into the second-quantized system register $\ket{\psi}_{2q}$ through the unitary
\begin{equation}
    \hat{U}_{1q\rightarrow 2q}^{\maxocc_2} = \sum_{j=0}^{\ngridph-1} \ket{k_\mathrm{min}/\Delta_k + j}\bra{k_\mathrm{min}/\Delta_k + j} \otimes U_{\acr{\bk_j}{\mu}}^{\maxocc_2}.
\end{equation}
The $\hat{U}_{1q\rightarrow 2q}^{\maxocc_2}$ subroutine consists of controlled arithmetic as illustrated in Fig.~\ref{fig:translation_subroutine}.
Using unary iteration \cite{babbush2018}, the $\ket{\psi}_{1q}$ controls contribute a T count of $4\ngridph - 4$.
There are also $\ngridph$ controlled applications of $\hat{U}_{a^\dagger}^{\maxocc_2}$, each of which involves $4\log_2\maxocc -4$ T gates to perform $\log_2\maxocc$-bit addition \cite{Gidney2018halvingcostof}.
Thus, the total T count for implementing $U_{1q\rightarrow 2q}^{\maxocc_2}$ is 
\begin{equation}
    T(U_{1q\rightarrow 2q}^{\maxocc_2}) = 4\ngridph\log_2\maxocc -4.
    \label{eq:U1q2q_cost}
\end{equation}

Finally, we use amplitude amplification to ensure a high probability that the uncomputed auxiliary register results in the $\ket{0}$ state.
The overlap with $\ket{0}$ on the first-quantized register at the end of the circuit shown in Fig.~\ref{fig:ph_state_prep} is
\begin{subequations}
\begin{align}
    \bra{0}_{1q} (\hat{U}_{1q}^\dagger\otimes I) \hat{U}_{1q\rightarrow 2q}^{\maxocc_2}(\hat{U}_{1q}\otimes I)\ket{0}_{1q}\ket{\psi}_{2q}
    &= \sum_{j''=j_\mathrm{min}}^{j_\mathrm{max}} c_{j''}^*\bra{j''} \left(\sum_{j'=j_\mathrm{min}}^{j_\mathrm{max}} \ket{j'}\bra{j'}\otimes \hat{U}_{\acr{\bk_{j'}}{\mu}}^{\maxocc_2}\right) \sum_{j=j_\mathrm{min}}^{j_\mathrm{max}} c_j \ket{j}\ket{\psi}_{2q} \\
    &= \sum_{j=j_\mathrm{min}}^{j_\mathrm{max}} |c_j|^2 \hat{U}_{\acr{\bk_j}{\mu}}^{\maxocc_2} \ket{\psi}_{2q},
\end{align}
\end{subequations}
where
\begin{equation}
    c_j = \frac{1}{\mathcal{N}} e^{-i\bk_j\cdot {\bar{\bR}}}e^{-|\bk_j - \bar{\bk}|^2/(2\sigma_k)^2} 
\end{equation}
and $j_\mathrm{min}=k_\mathrm{min}/\Delta_k$, $j_\mathrm{max}=k_\mathrm{max}/\Delta_k$ index the minimum and maximum occupied momentum modes.
Thus, the success probability of the circuit in Fig.~\ref{fig:ph_state_prep} is
\begin{equation}
    P = \sum_j |c_j|^4 = \frac{1}{\mathcal{N}^4} \sum_{j=j_\mathrm{min}}^{j_\mathrm{max}} e^{-|\bk_j - \bar{\bk}|^2/\sigma_k^2} \gtrsim \frac{\Delta_k}{2\sqrt{\pi}\sigma_k},
    \label{eq:ampampsuccess}
\end{equation}
where the approximate bound holds when $\Delta_k \ll \sigma_k$ (continuous limit) and approaches saturation as $\ngridph\rightarrow\infty$ (negligible truncation).
The required number of amplitude amplification iterations is then
\begin{equation}
    \frac{1}{\sqrt{P}} \lesssim (4\pi)^{1/4}\left(\frac{\sigma_k}{\Delta_k}\right)^{1/2}.
    \label{eq:ampamprounds}
\end{equation}
The simulation parameters discussed in App.~\ref{app:parameters_for_solar_iron} lead to $\sigma_k/\Delta_k \approx 473$ and a nearly saturated success probability such that $1/\sqrt{P}\approx 40.$

Ignoring contributions that scale sublinearly with $\ngridph$, Eqs.~\eqref{eq:U1q2q_cost} and Eq.~\eqref{eq:ampamprounds} give a total T count for photonic state preparation of approximately
\begin{equation}
    T_\mathrm{prep}^\mathrm{ph} \approx 4 (4\pi)^{1/4} \left(\frac{\sigma_k}{\Delta_k}\right)^{1/2} \ngridph \maxocc_1 \log_2\maxocc \approx 160 \ngridph \maxocc_1 \log_2\maxocc.
\end{equation}
Note that $\sigma_k/\Delta_k$ is proportional to $\ngridph$ such that the overall scaling is $T_\mathrm{prep}^\mathrm{ph}=\mathcal{O}(\ngridph^{3/2}\maxocc_1\log\maxocc)$.

While $\ngridph$, $\sigma_k$, and $\Delta_k$ are determined by the desired spectral range and resolution for the computed opacity (see App.~\ref{app:parameters_for_solar_iron}), the number of photons $\maxocc$ and their distribution across uncorrelated groups ($\maxocc_1$ and $\maxocc_2$) are essentially free parameters that can be optimized to minimize overall costs.
As discussed in App.~\ref{subapp:observable_estimation} on observable estimation, the number of repetitions of the algorithm $\ns$ generally decreases for large $\maxocc$ because the number of absorption events is proportional to the initial number of photons.
While $\maxocc_1=1$ minimizes photonic state preparation costs for fixed $\maxocc$, under this choice the fully entangled photons only sample a single momentum mode per shot such that $\ns$ saturates at $\ngridph$.
Also, $\maxocc$ should be chosen such that nonlinear optical effects --- which are believed to be absent in experimental conditions \cite{kruse2019two,kruse2021two} --- remain negligible.
In practice, ensuring that the simulation operates within the linear-response regime will require repeating the protocol with multiple different values of $\maxocc$.



\subsection{Hamiltonian simulation in the interaction picture}
\label{subapp:Hamiltonian_simulation_via_interaction}
The interaction picture simulation algorithm allows us to approximate Hamiltonian simulation of a general Hamiltonian $\hat{H} = \hat{A} + \hat{B}$ with asymptotic scaling only in the norm of $\hat{B}$ by making use of interaction picture dynamics
\begin{equation}
|\psi_I(t)\rangle = e^{-i\hat{H}_I(t)}|\psi_I(0)\rangle, \ |\psi_I(t)\rangle \equiv e^{i\hat{A}t}|\psi(t)\rangle, \ \hat{H}_I(t) \equiv e^{i\hat{A}t} \hat{B} e^{-i\hat{A}t}. 
\end{equation}
Note that while $||\hat{H}||=||\hat{A} + \hat{B}||$, the interaction picture Hamiltonian norm $||\hat{H}_I(t)|| = ||\hat{B}||$.
In the case where $||\hat{B}||\ll ||\hat{A}||$, and assuming that the non-interacting term $e^{-i\hat{A}t}$ can be implemented efficiently, the interaction picture simulation presents a significant advantage over the ordinary formalism.

This property is present in the opacity simulation problem since $\hat{B} = \helph$ has a small norm compared to $\hat{A} = \hel + \hph$.
We begin by specifying the resource estimates from the algorithm, restated from Lemma 6 in Ref.~\onlinecite{low2019hamiltonian} with modified notation and explicit overheads for the asymptotic expressions in blue:
\begin{theorem}[Lemma 6 \cite{low2019hamiltonian}]
Let $A, B \in \mathbb{C}^{2^{Q_\mathrm{sys}} \times 2^{Q_\mathrm{sys}}}$, let $\alpha_A$ and $\alpha_B$ be known constants such that $||A|| \leq \alpha_A$ and  $||B|| \leq \alpha_B$. 
Assume the existence of a unitary oracle that implements the Hamiltonian within the interaction picture, denoted HAM-T$\in \mathbb{C}^{2^{Q_\mathrm{sys} + Q_\mathrm{aux}}\times 2^{Q_\mathrm{sys} + Q_\mathrm{aux}}}$ which implicity depends on the time-step size $\tau \in \mathcal{O}(\alpha_B^{-1})$ \textcolor{blue}{$[\tau = 1/(2\alpha_B)$, Corollary 4 \cite{low2019hamiltonian}]} and number of time-steps $M \in \mathcal{O}(\frac{t}{\epsilon}(\alpha_A + \alpha_B))$ \textcolor{blue}{$[M \geq \frac{16t\ln 2 }{\epsilon}(2\alpha_A + \alpha_B)$, Eq.~21 and Lemma 4 \cite{low2019hamiltonian}]}, such that
\begin{equation}
    (\langle 0|a \otimes 1_s)\textrm{HAM-T}(|0\rangle_a \otimes 1_s) = \sum_{m = 0}^{M-1} |m\rangle \langle m|\otimes \frac{e^{iA\tau m/M}Be^{-iA\tau m/M}}{\alpha_B},
\end{equation}
for all $t \geq 2\alpha_B \tau$, the time-evolution operator $e^{-i(A+B)t}$ may be approximated to error $\epsilon$ with the following cost.
\begin{enumerate}
    \item Simulations of $e^{-iA\tau}: \mathcal{O}(\alpha_Bt)$, \textcolor{blue}{\Big[$2\alpha_Bt$\Big]}
    \item Queries to HAM-T: $\mathcal{O}(\alpha_B t \frac{\log(\alpha_B t/\epsilon)}{\log \log(\alpha_B t/\epsilon)})$, \textcolor{blue}{\Big[$2\alpha_BtK,\ K = \lceil \frac{-1 + 2\ln(2\alpha_Bt/\epsilon)}{\ln\ln(2\alpha_Bt/\epsilon)+1} \rceil$, Lemma 5 \cite{low2019hamiltonian}\Big]}
    \item Qubits: $Q_\mathrm{sys} + \mathcal{O}(Q_\mathrm{aux} + \log(\frac{t}{\epsilon}(\alpha_A + \alpha_B)))$,
     \textcolor{blue}{\Big[$Q_\mathrm{sys} + Q_\mathrm{aux} + \log_2M$\Big]}
    \item Primitive gates: $\mathcal{O}(\alpha_Bt(Q_\mathrm{aux} + \log(\frac{t}{\epsilon}(\alpha_A + \alpha_B)))\frac{\log(\alpha_B t/\epsilon)}{\log \log(\alpha_B t/\epsilon)}).$ \textcolor{blue}{\Big[$2\alpha_Bt(Q_\mathrm{aux} + \log_2M)K$\Big]}
\end{enumerate}
\label{thm:interaction-picture}
\end{theorem}
Here, $Q_\mathrm{sys}$ and $Q_\mathrm{aux}$ denote the system register size and the number of auxiliary qubits, respectively.
It turns out that an even more efficient simulation is possible in the case where $t < \tau = 1/(2\alpha_B)$.
In such a case, we find that Theorem~\ref{thm:interaction-picture} simplifies to
\begin{theorem}[Lemma 6, Theorem 3 \cite{low2019hamiltonian}]
Consider the conditions of Theorem~\ref{thm:interaction-picture} with the further condition that the simulation time $t < \tau = 1/(2\alpha_B)$.
In this case the interaction picture simulation can be carried out with the following costs
\begin{enumerate}
    \item Simulations of $e^{-iA\tau}: \mathcal{O}(1)$, \textcolor{blue}{\Big[$1$\Big]}
    \item Queries to HAM-T: $\mathcal{O}( \frac{\log(1/\epsilon)}{\log \log(1/\epsilon)})$, \textcolor{blue}{\Big[$K,\ K = \lceil \frac{-1 + 2\ln(1/\epsilon)}{\ln\ln(1/\epsilon)+1} \rceil$, Lemma 5 \cite{low2019hamiltonian}\Big]}
    \item Qubits: $Q_\mathrm{sys} + \mathcal{O}(Q_\mathrm{aux} + \log(\frac{t}{\epsilon}(\alpha_A + \alpha_B)))$,
     \textcolor{blue}{\Big[$Q_\mathrm{sys} + Q_\mathrm{aux} + \log_2M$\Big]}
    \item Primitive gates: $\mathcal{O}((Q_\mathrm{aux} + \log(\frac{t}{\epsilon}(\alpha_A + \alpha_B)))\frac{\log(1/\epsilon)}{\log \log(1/\epsilon)}).$ \textcolor{blue}{\Big[$(Q_\mathrm{aux} + \log_2M)K$\Big]}
\end{enumerate}
\label{thm:interaction-picture-reduced}
\end{theorem}

The next piece is to compute the complexity of HAM-T, which is given in the proof for Theorem 7. 
We do not go into detail, just re-iterating the relevant parts as theorem:
\begin{theorem}[From Theorem 7 \cite{low2019hamiltonian}]
Assume one has access to an exact block-encoding $O_B$ of $B$ with $Q_\mathrm{aux}$ extra qubits, and an implementation of $e^{-iAt}$. Constructing HAM-T to accuracy $\delta$ requires
\begin{enumerate}
    \item A controlled application of each unitary $e^{iA\tau/M}$, $e^{iA 2\tau/M}$...$e^{iA2^{\lceil \log_2 M\rceil}\tau/M}$, with error $\delta/\log_2M$,
    \item One application of $O_B$.
\end{enumerate}
For application within the interaction picture Hamiltonian, to ensure an additive error of $\epsilon$, we would require $\delta= \epsilon/(2\alpha_B t K)$.
\label{thm:ham-t}
\end{theorem}

From Theorem~\ref{thm:interaction-picture} we then can estimate the cost of the simulation algorithm for error $\epsilon$ and total simulation time $t$ as:
\begin{subequations}
\begin{align}
& T(t,\epsilon) = \frac{t}{\tau} \cdot  T_A\left(\tau, \frac{\epsilon}{t/\tau}\right) + \frac{tK}{\tau} \cdot \left( T_B\left(\frac{\epsilon}{tK/\tau} \right) + \sum_{i=0}^{\lceil \log_2M \rceil } T_A\left(\frac{2^i\tau}{M}, \frac{\epsilon}{tK \lceil \log_2M \rceil/\tau}\right)\right), \\
& Q(t,\epsilon) = Q_\mathrm{sys} + Q_\mathrm{aux} + \log_2M, \\
& \tau = \frac{1}{2\alpha_B}, \ M \geq \frac{16\ln 2 t(\alpha_A + \alpha_B)}{\epsilon}, \ K = \Bigg\lceil \frac{-1 + 2\ln(2\alpha_Bt/\epsilon)}{\ln\ln(2\alpha_Bt/\epsilon)+1} \Bigg\rceil.
\end{align}
\label{eq:interaction-simulation-cost}
\end{subequations}
Here, $T_{A}(t, \epsilon)$ and $T_{B}(\epsilon)$ represent the gate cost of implementing $e^{iAt}$ and $O_B$, respectively.
In the reduced case of Theorem~\ref{thm:interaction-picture-reduced}, the total cost is
\begin{subequations}
\begin{align}
& T(t,\epsilon) =  T_A\left(t, \epsilon \right) + K \cdot \left( T_B(\epsilon/K) +  \sum_{i=0}^{\lceil \log_2M \rceil } T_A\left(\frac{2^i t}{M}, \frac{\epsilon}{K \lceil \log_2M \rceil}\right)\right), \\
& Q(t,\epsilon) = Q_\mathrm{sys} + \log_2M, \\
& M \geq \frac{16\ln 2 t(\alpha_A + \alpha_B)}{\epsilon}, \ K = \Bigg\lceil \frac{-1 + 2\ln(1/\epsilon)}{\ln\ln(1/\epsilon)+1} \Bigg\rceil.
\end{align}
\label{eq:interaction-simulation-cost-reduced}
\end{subequations}
In the following we will construct expressions for $\alpha_A, \alpha_B, T_A(t,\epsilon), T_B(\epsilon)$ in terms of the basic physical parameters relevant to the solar opacity simulation given in Appendix~\ref{app:parameters_for_solar_iron}. 
The values of $t, \epsilon$ are also specified in the same section.

\subsubsection{Norms $\alpha_A, \alpha_B$}
To bound the norms 
\begin{subequations}
    \begin{align}
    ||\hat{A}|| &= ||\hel+\hph|| \leq ||\hel|| + ||\hph|| \leq \alpha_A,\\
    ||\hat{B}|| &= ||\helph|| \leq \alpha_B,
    \end{align}
\end{subequations}
we will consider the norms of the Hamiltonian terms $\hel, \hph, \helph$ individually.
First, to bound the norm of the photonic Hamiltonian, we note that each register can have at most $\maxocc$ occupancy and the energy of each mode is $\omega(\vec{k}) = c|\bk| \leq ck_\mathrm{max}^\mathrm{sim}$, where $k_\mathrm{max}^\mathrm{sim}$ is the maximum photonic momentum included in the simulation.
The total norm of the photonic term is then bounded by:
\begin{equation}
||\hph|| \leq \maxocc \ngridph ck_\mathrm{max}^\mathrm{sim}.
\label{eq:norm-ph}
\end{equation}
Next, the electronic term can be bounded by the difference of the largest and smallest eigenvalues, 
\begin{equation}
||\hel|| \leq \frac{3\pi^2\nel}{2\Delta_r^2} + \frac{\nel Z}{\Delta_r} + \frac{\nel^2}{\ereg},
\label{eq:norm-el}
\end{equation}
where the first term accounts for the kinetic energy, the second and third terms account for the 1- and 2-body potential energies, and we have used the discretization given in Eq.~\eqref{eq:discretization}.
Therefore the parameter $\alpha_A$ is
\begin{equation}
\alpha_A = \maxocc \ngridph ck_\mathrm{max}^\mathrm{sim} + \frac{3\pi^2\nel}{2\Delta_r^2} + \frac{\nel Z}{\Delta_r} + \frac{\nel^2}{\ereg}.
\label{eq:alphaA}
\end{equation}

To bound the norm for the interaction term $||\helph||$, we focus on the dipole coupling term:
\begin{equation}
\frac{1}{2c} \left|\left|\sum_{i=1}^\eta \{\bpo{i},\Apot(\bro{i})\}\right|\right| \leq \frac{\nel}{c} \left|\left|\bpo{i} \cdot \Apot(\bro{i})\right|\right| \leq \frac{\sqrt{3}\pi\nel}{c\Delta_r} ||\Apot(\bro{i})||,
\end{equation}
where the largest eigenvalue of the electronic momentum operator is $\sqrt{3}\pi/\Delta_r$.
To get the norm of the field operator, we look at the expression for the operator in Eq.~\eqref{eq:aebfields}.
There are $\ngridph$ terms in the sum, with each term having norm at most $2\sqrt{\maxocc/2\omega_\mathrm{min}}$, where the $2\sqrt{\maxocc}$ factor comes from the action of the raising and lowering operators while $\omega_\mathrm{min}$ is the minimum relevant photonic energy.
The full norm then has the expression
\begin{equation}
||\mathbf{A}(\bro{i})|| \leq 2\ngridph \sqrt{\frac{\maxocc}{2\Omega \omega_\mathrm{min}}}.
\end{equation}

Since the simulated photonic wave packet will have support over momentum modes near the origin (see Appendix~\ref{app:parameters_for_solar_iron}), a strict bound requires taking $\omega_\mathrm{min}=c\Delta_k$.
However, the probability amplitude of momentum modes near the origin is about $10^3$ times smaller than at the center of the wave packet (see Fig.~\ref{fig:photon_truncation}).
Thus, an effective $\omega_\mathrm{min}$ of $c k_\mathrm{min}^\mathrm{targ}$ is more representative of a state-dependent bound considering the predominantly occupied modes, where $k_\mathrm{min}^\mathrm{targ}$ is the minimum wave number within the targeted experimental spectral range.

Even for $\omega_\mathrm{min}=c\Delta_k$, the parameters of the solar opacity simulation lead to $||\Apot(\bro{i})||/c\ll||\bpo{i}||$ such that the $\Apot^2(\bro{i})$ term yields a negligible contribution to $||\helph||$.
We also neglect the contribution from the magnetic-field coupling term $\spin{i} \cdot \Bfld(\bro{i})$ because $||\Bfld(\bro{i})||\approx ||\Apot(\bro{i})||$ (see Eq.~\eqref{eq:aebfields} and note that $|\bk|<1$ for the spectral range of interest) while $||\spin{i}||\ll||\bpo{i}||$.
Thus, the complete expression for $\alpha_B$ is 
\begin{equation}
\alpha_B = \frac{2\sqrt{3}\pi\ngridph \nel}{c\Delta_r}\sqrt{\frac{\maxocc}{2\Omega \omega_\mathrm{min}}}.
\label{eq:alphaB}
\end{equation}

\subsubsection{The gate cost $T_A(t,\epsilon)$}
The operator $e^{-i\hat{A}t}$ is easy to implement since $\hat{A} = \hel + \hph$ and $[\hel, \hph]=0$, thereby $e^{-i\hat{A}t} = e^{-i\hel t} \,e^{-i\hph t}$.
The total gate cost for each term is determined below in Eqs.~\eqref{eq:ham-ph-resources} and \eqref{eq:ham-el-resources} and the cost of $T_A$ can then be expressed as
\begin{equation}
T_A(t,\epsilon) = T_{e^{-i\hph t}}(t,\epsilon) + T_{e^{-i\hel t}}(t,\epsilon) 
\label{eq:ham-A-resources}.
\end{equation}

First, we consider $T_{e^{-i\hph t}}(t,\epsilon)$.
Since $\hph$ is diagonal in the computational basis, we can fast-forward the implementation for its time-evolution unitary.
Consider a register for photonic mode $|\bk,\mu \rangle$, which contains $\lceil \log_2\maxocc \rceil$ qubits for the $\maxocc$ maximum occupancy.
On the qubits $j \in [1, \lceil 2\log_2\maxocc \rceil]$, we can apply the phase gate $P(2^j \omega(\bk)t)$.
The cumulative affect of the phase gates on the register is 
\begin{equation}
    \prod_j P(2^j \omega(\bk)t) \, |\bk, \mu\rangle = e^{i\omega(\bk) \sum_{j} 2^j b_j t}|\bk, \mu\rangle = e^{i\omega(\bk)\hat{n}_{\bk,\mu} t}|\bk, \mu\rangle,
\end{equation}
where $b_j$ is binary value of the $j$th qubit in the register.
Parallel application of these phase gates on all photonic registers applies $\exp(-it\hph)$ to the photonic state.
Each phase gate requires in the worst case $4\log_2(1/\epsilon)$ T gates~\cite{ross2016optimal} for accuracy $\epsilon$, though synthesis with a better constant prefactor is feasible with probabilistic methods~\cite{kliuchnikov2023shorter}.
The total T-gate cost is at most
\begin{equation}
T_{e^{-i\hph t}}(t,\epsilon) = 4\ngridph \lceil \log_2 \maxocc \rceil \log_2\left(\frac{\ngridph \lceil \log_2 \maxocc \rceil}{\epsilon}\right),
\label{eq:ham-ph-resources}
\end{equation}
with no ancilla qubits required.

Next, we consider $T_{e^{-i\hel t}}(t,\epsilon)$.
To implement $\exp(-it\hel)$, we make use of the optimized, first-quantized, Trotterized implementation of Hamiltonian simulation for electronic structure designed previously for stopping power calculations~\cite{rubin2024quantum}.
Therein, the simulation cost is dominated by the evaluation of square roots in the two-body Coulomb repulsion term.
The exponentiation of the Coulomb repulsion takes on the order of $2395 \nel (\nel - 1)/2$ Toffolis.
For a $\tilde{k}_\textrm{el}$th order expansion, one requires additionally $2\tilde{k}_\textrm{el} - 1$ such exponentials for the product formula.
Lastly, for the number of Trotter steps, an optimized expression was found via numerical simulation:
\begin{equation}
    \tilde{r}_\textrm{el} \cong t^{1 + 1/\tilde{k}_\textrm{el}}(||\tau||_1 + ||\nu||_{1, [\nel]})^{1 - 1/\tilde{k}_\textrm{el}}(\xi_{\tilde{k}_\textrm{el}} ||\tau||_1 ||\nu||_{1, [\nel]} \nel/\epsilon)^{1/\tilde{k}_\textrm{el}}, 
\end{equation}
wherein one has a constant $\xi_{\tilde{k}_\textrm{el}} \approx 10^{-2},\ 2 \times 10^{-4},\ 3 \times 10^{-6},\ 3 \times 10^{-8}$ for $\tilde{k}_\textrm{el} = 2, 4, 6, 8$, and  
\begin{equation}
    ||\tau||_1 \leq\frac{1}{2(\Delta_r^2 + \ereg^2)}, \ \  ||\nu||_{1, [\nel]} \leq \pi^{1/3} (3\nel/4)^{2/3} \frac{1}{(\Delta_r^2 + \ereg^2)^{1/2}}.
\end{equation}
The total resources required therein are:
\begin{equation}
    T_{e^{-i\hel t}}(t,\epsilon) = \tilde{r}_\textrm{el} \times 2395 (2\tilde{k}_\textrm{el}-1) \nel(\nel - 1)/2.
    \label{eq:ham-el-resources}
\end{equation}

\subsubsection{The gate cost $T_B(\epsilon)$}
Our objective is to block-encode the interaction Hamiltonian:
\begin{equation}
\helph = \frac{1}{c}\sum_{i=1}^{\nel} \bpo{i} \cdot \Apot(\bro{i}).
\end{equation}
We can write this operator using the expression for the field operator
\begin{equation}
\Apot(\bro{i}) = \Omega^{-1/2} \sum \limits_{\bg \in \mathbb{G}_{\ngridel}}\sum_{\mu = \pm 1, \bk} \ \frac{1}{\sqrt{2 \omega(\bk)}} \helvec{\bk}{\mu}\Big[\acr{\bk}{\mu} e^{-i\bk \cdot \br_{\bg}} + \aan{\bk}{\mu}e^{i\bk \cdot \br_{\bg}}\Big] \ket{\bg}\!\!\bra{\bg}_i,
\end{equation}
and we find:
\begin{subequations}
\begin{align}
\helph = &c^{-1}\Omega^{-1/2} \sum_{i=1}^{\nel} \sum_{\mathbf{p}} \mathbf{p} |\mathbf{p}\rangle \langle \mathbf{p}|_i \cdot \sum \limits_{\bg \in \mathbb{G}_{\ngridel}}\sum_{\mu = \pm 1, \bk} \ \frac{1}{\sqrt{2 \omega(\bk)}} \helvec{\bk}{\mu}\Big[\acr{\bk}{\mu} e^{-i\bk \cdot \br_{\bg}} + \aan{\bk}{\mu}e^{i\bk \cdot \br_{\bg}}\Big] \ket{\bg}\!\!\bra{\bg}_i \\
= & \frac{\Omega^{-1/2}}{c}\sum_{i=1}^{\nel} \sum_{\mathbf{g}} \sum_{\mathbf{p},\mu,\mathbf{k}}  \frac{ \mathbf{p} \cdot \helvec{\bk}{\mu}}{\sqrt{2 \omega(\bk)}}\Big[\acr{\bk}{\mu} e^{-i\bk \cdot \br_{\bg}} + \aan{\bk}{\mu}e^{i\bk \cdot \br_{\bg}}\Big] e^{-i\mathbf{p} \cdot \mathbf{r}_g}\ket{\mathbf{p}}\!\!\bra{\bg}_i \\
= & \frac{\Omega^{-1/2}}{c} \sum_{i=1}^{\nel} \sum_{\bpo{} \in \tilde{G}_{\nel}} \sum_{\mu = \pm 1,\mathbf{k}}  \frac{ \mathbf{p} \cdot \helvec{\bk}{\mu}}{\sqrt{2 \omega(\bk)}}\Big[\acr{\bk}{\mu} |\mathbf{p}\rangle \langle \mathbf{p}+\mathbf{k}|_i + \aan{\bk}{\mu}|\mathbf{p}\rangle \langle \mathbf{p}-\mathbf{k}|_i\Big].
\end{align}
\end{subequations}
To arrive at the third line we have carried out the sum over $\mathbf{g}$ in the second line.

We note that the above operator is sparse.
Starting with a given basis state, there are only $s = 4\nel\ngridph$ other basis states that it it couples to.
One can contrast this to the total number of basis states $2^{\nel \log_2 \ngridel}2^{\ngridph \log_2 \maxocc} = \ngridel^{\nel} \maxocc^{\ngridph}.$
As such, we will implement the block encoding using two layers: a simple LCU over the electron index $i$ and a sparse encoding for each electron operator $\mathcal{J}^{(i)}$ defined as such
\begin{equation}
\mathcal{U}(\helph) = \sum_{i=1}^{\nel} \mathcal{J}^{(i)} + h.c., \ \mathcal{J}^{(i)} \equiv \frac{\Omega^{-1/2}}{c} \sum_{\mathbf{p}} \sum_{\{\mu,\mathbf{k}\}} \frac{ \mathbf{p} \cdot \helvec{\bk}{\mu}}{\sqrt{2 \omega(\bk)}} \hat{a}_{\mathbf{k},\mu}|\mathbf{p}\rangle \langle \mathbf{p-k}|_i.
\end{equation}
Each electronic operator $\mathcal{J}^{(i)}$ is $2\ngridph$ sparse which is seen by the action on a computational basis state
\begin{equation}
\mathcal{J}^{(i)}|\mathbf{q}\rangle_i \otimes |\{\mathbf{k}_j,\mu_j, n_j\}\rangle = \frac{\Omega^{-1/2}}{c}\sum_{\mu,\mathbf{k}} \frac{ (\mathbf{q+k}) \cdot \helvec{\bk}{\mu}}{\sqrt{2 \omega(\bk)}} \Bigg(|\mathbf{q+k}\rangle_i \otimes |\{\mathbf{k}_j,\mu_j, n_j - \delta_{\mathbf{k},\mathbf{k}_j}\delta_{\mu,\mu_j}\}\rangle \Bigg) 
\label{eq:ji-action}
\end{equation}
Here $|\mathbf{q}\rangle_i$ is the electronic momentum basis state for electron $i$, and the photonic basis state is defined as:
\begin{equation}
 |\{\mathbf{k}_j,\mu_j, n_j\}\rangle \equiv \prod_{j=1} \left(\hat{a}_{\bk_j,\mu_j} \right)^{n_j}|0\rangle. 
\end{equation}

To construct the operator $\mathcal{J}^{(i)}$ we make use of a known efficient sparse block-encoding theorem
\begin{theorem}[Theorem 4.1 \cite{camps2023}]
Let $c(j,l)$ be a function that gives the row index of the $l$th (among a list of $s$) nonzero matrix elements in the $j$th column of an s-sparse matrix $A \in \mathbb{C}^{N \times N}$ with $N = 2^n$, where $s = 2^m$. If there exists a unitary $O_c$ such that
\begin{equation}
    O_c|l\rangle |j\rangle  = |l\rangle |c(j,l)\rangle
\end{equation}
and a unitary $O_A$ such that 
\begin{equation}
    O_A |0\rangle |l\rangle |j\rangle = \left(A_{c(j,l),j}|0\rangle + \sqrt{1 - A_{c(j,l),j}^2}|1\rangle \right)|l\rangle |j\rangle,
\end{equation}
then 
\begin{equation}
U_A = (I_2 \otimes D_s \otimes I_N)(I_2 \otimes O_c)O_A(I_2 \otimes D_s \otimes I_N)
\end{equation}
block encodes $A/s$. Here $D_s$ is called the diffusion operator and is defined as $D_s \equiv H^{\otimes m}.$
\end{theorem}
As such, we only need to specify the complexity of the primitive $O_c, O_A$ operators and a single call to each will permit us to block encode the operator $\mathcal{J}^{(i)}$. 

\paragraph{Unitary $O_c$}
In our case $l$ encodes $\mathbf{k}$ and $\mu$, so we will consider an auxiliary register which encodes $|\mu\rangle |\mathbf{k}\rangle$ in binary requiring $\log_2(2\ngridph)$ auxiliary qubits.
The action of $O_c$ is thereby
\begin{equation}
O_c |\mu\rangle|\mathbf{k}\rangle \Big(|\mathbf{q}\rangle_i \otimes |\{\mathbf{k}_j,\mu_j, n_j\}\rangle \Big) = |\mu\rangle |\mathbf{k}\rangle \Big(|\mathbf{q+k}\rangle_i \otimes |\{\mathbf{k}_j,\mu_j, n_j - \delta_{\mathbf{k},\mathbf{k}_j} \delta_{\mu,\mu_j}\}\rangle\Big).
\end{equation}
The action on the electronic part of the register is binary addition, since both the auxiliary $\bk$ register and $\mathbf{q}_i$ registers are stored in binary.
The action on the photonic part is trickier since it requires decrementation on the $\log_2\maxocc$ photonic register corresponding to momentum $\bk$.
Since the photonic momentum is stored classically, this requires us to carry out multi-controlled decrements on the $|\mu\rangle|\bk\rangle$ auxiliary register with a total $2\ngridph$ such calls.

\begin{figure}[h]
\begin{quantikz}
\lstick{\ket{\mu}} &  & \ctrl{1} & \octrl{1} & \ctrl{1} &  \octrl{1} &\\
\lstick{\ket{\mathbf{k}}} & \gate[2][1.7cm]{\textrm{add}} & \ctrl{2} & \ctrl{3} & \octrl{4} & \octrl{5} &\\
\lstick{\ket{\mathbf{q}_i}} & \gateinput{$\mathbf{q}_i$}\gateoutput{$\mathbf{q}_i\oplus \mathbf{k}$} & & & & &\\
\lstick[4]{\ket{\{\mathbf{k}_j,\mu_j, n_j\}}} &  &\gate{\textrm{inc}} & & & &\\
& & & \gate{\textrm{inc}} & & &\\
& & & & \gate{\textrm{inc}} & &\\
& & & & & \gate{\textrm{inc}} & 
\end{quantikz}
\caption{Quantum circuit for $O_c$ in the simplified case of $\ngridph = 2$. For larger $\ngridph$ the sequence of double controls can be optimized using the standard sawtooth reduction in~\cite{babbush2018}.}
\label{fig:oc}
\end{figure}
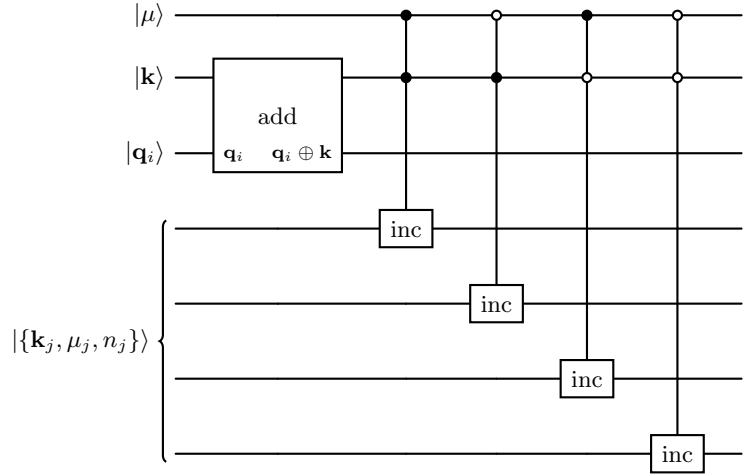

A circuit which implements $O_c$ is shown in Fig.~\ref{fig:oc} for the simple case of $\ngridph = 2$.
The multi-controlled logic can be generalized in the case of $\ngridph > 2$ and reduced to linear-T complexity using the standard sawtooth reduction in~\cite{babbush2018} requiring an additional $\log_2(2\ngridph)$ ancilla qubits.
The total cost of $O_c$ requires $2\log_2(2\ngridph)$ additional qubits, and the T cost is
\begin{equation}
T_{O_c} = T_\mathrm{add}(\log_2 \ngridel) + 2\ngridph T_\mathrm{inc}(\log_2\maxocc) + T_\mathrm{sawtooth}(\log_2 (2\ngridph)).
\end{equation}
The cost for these primitive gates are as follows: $T_\mathrm{sawtooth}(N) = T_\mathrm{add}(N) = 4N-4$ \cite{babbush2018}, \cite{Gidney2018halvingcostof} and $T_\mathrm{inc} = 8n+4$ \cite{rhodes2024exponential} using controlled signed-increment-decrement (SID) gates.
Hence the total gate and qubit costs for $O_c$ are
\begin{subequations}
\begin{align}
&T_{O_c} = \left\{4\log_2\ngridel - 4\right\} + 2\ngridph(8\log_2\maxocc + 4) +  \left\{4\log_2(2\ngridph) - 4\right\}.\\
& Q_{O_c} = 2\log_2(2\ngridph).
\end{align}
\label{eq:costs-oc}
\end{subequations}

\paragraph{Unitary $O_A$}
The action of $O_A$ in our notation will be 
\begin{equation}
O_A |0\rangle |\mu\rangle|\mathbf{k}\rangle \Big(|\mathbf{q}\rangle_i \otimes |\{\mathbf{k}_j,\mu_j, n_j\}\rangle \Big) = \Big(A_{\mathbf{q},\mathbf{k},\mu}|0\rangle  + \sqrt{1 - A_{\mathbf{q},\mathbf{k},\mu}^2}|1\rangle\Big) |\mu\rangle|\mathbf{k}\rangle \Big(|\mathbf{q}\rangle_i \otimes |\{\mathbf{k}_j,\mu_j, n_j\}\rangle \Big),
\end{equation}
where 
\begin{equation}
A_{\mathbf{q},\mathbf{k},\mu} \equiv \frac{\Omega^{-1/2}}{c}\frac{ (\mathbf{q+k}) \cdot \helvec{\bk}{\mu}}{\sqrt{2 \omega(\bk)}} . 
\end{equation}
We will implement this using QROM~\cite{babbush2018} and controlled phase kick-back.

A naive implementation of the QROM would be as
\begin{equation}
\textrm{QROM}|\mu\rangle |\mathbf{k}\rangle  \Big(|\mathbf{q}\rangle_i \otimes |\{\mathbf{k}_j,\mu_j, n_j\}\rangle \Big) |0^{\otimes l}\rangle  = |\mu\rangle |\mathbf{k}\rangle  \Big(|\mathbf{q}\rangle_i \otimes |\{\mathbf{k}_j,\mu_j, n_j\}\rangle \Big) |[A_{\mathbf{q},\mathbf{k},\mu}]_{bin}^l\rangle,
\end{equation}
loading in the binary representation of our coefficient in a register of size $l$.
The QROM thereby has $D = 2\ngridph \ngridel $ bitstrings with length $l$ each.
The issue here is that since we are working in first quantization we have permitted $\ngridel$ to be very large with the intent that it only appears logarithmically in any algorithm.
Since the T cost of QROM scales linearly in $D$, it will additionally scale linearly in $\ngridel$, which is not optimal.

Rather than storing the entire coefficient in QROM, consider instead storing on the part which depends on the photonic momentum
\begin{equation}
a_{\mathbf{q},\mu,x} \equiv \frac{\Omega^{-1/2}}{c} \frac{\mathbf{u}_{\mathbf{k},\mu}^x}{\sqrt{2\omega(\mathbf{k})}},
\end{equation}
where $x \in \{0,1,2\}$ are the three vector component, thereby having $3\times 2\ngridph$ unique coefficients in QROM.
We can then compute the required output via the following sequence of operations.
First, add the auxiliary momentum register to $\mathbf{q}_i$, then load in the set of coefficients from QROM, then carry out binary multiplication between the registers:
\begin{subequations}
\begin{align}
& {\textrm{add: }} |\mu\rangle |\mathbf{k}\rangle  \Big(|\mathbf{q+k}\rangle_i \otimes |\{\mathbf{k}_j,\mu_j, n_j\}\rangle \Big) |0^{\otimes 3l}\rangle  \\
& {\textrm{QROM: }} |\mu\rangle |\mathbf{k}\rangle  \Big(|\mathbf{q+k}\rangle_i \otimes |\{\mathbf{k}_j,\mu_j, n_j\}\rangle \Big) |a_{\mathbf{q},\mu,x}\rangle |a_{\mathbf{q},\mu,y}\rangle |a_{\mathbf{q},\mu,z}\rangle  \\
& {\textrm{mult: }} |\mu\rangle |\mathbf{k}\rangle  \Big(|\mathbf{q+k}\rangle_i \otimes |\{\mathbf{k}_j,\mu_j, n_j\}\rangle \Big) |(q_x+k_x)a_{\mathbf{q},\mu,x}\rangle |(q_y+k_y)a_{\mathbf{q},\mu,y}\rangle |(q_z+k_z)a_{\mathbf{q},\mu,z}\rangle  \\
& 2\times \textrm{add: } |\mu\rangle |\mathbf{k}\rangle  \Big(|\mathbf{q+k}\rangle_i \otimes |\{\mathbf{k}_j,\mu_j, n_j\}\rangle \Big) |(q_x+k_x)a_{\mathbf{q},\mu,x}\rangle |(q_y+k_y)a_{\mathbf{q},\mu,y}\rangle |(\mathbf{q+k}) \cdot \mathbf{a}_{\mathbf{q},\mu}\rangle. 
\end{align}
\end{subequations}
The last step adds the components in-place to the third register giving us our final result.
Note that all operations involving the electronic momentum are now carried out in binary, giving the expected $\log_2 \ngridel$ scaling we desire.

With the value $A_{\mathbf{q},\mathbf{k},\mu}$ stored in the register, one can carry out the desired rotation using controlled phase kickback
\cite{Kassal_2008}.
The idea behind phase kickback is that if you have a circuit which can execute:
\begin{equation}
\mathcal{A}|\mathbf{x},y\rangle = |\mathbf{x}, y\oplus A(\mathbf{x})\rangle, \
\end{equation}
then the action on the following state is as:
\begin{equation}
\mathcal{A}|\mathbf{x}\rangle |G\rangle = e^{-iA(\mathbf{x})}|\mathbf{x}\rangle |G\rangle, \ |G\rangle \equiv \sum_{y} \frac{e^{2\pi i y/M}}{\sqrt{M}} |y\rangle.
\end{equation}
Here the phase gradient state $|G\rangle$ can be prepared as 
\begin{equation}
|G\rangle = \bigotimes_{i=1}^l \left\{R_z\left(\frac{2\pi}{2^{i+1}} \right) H|0\rangle\right\}.
\end{equation}
If an additional control qubit is used then one finds
\begin{equation}
\left(|0\rangle \langle 0| + \mathcal{A}|1\rangle \langle 1|\right)|0\rangle |\mathbf{x}\rangle |G\rangle = (|0\rangle + e^{-iA(\mathbf{x})}|1\rangle)|\mathbf{x}\rangle|G\rangle.
\end{equation}
This is identical to the desired action of $O_A$ except that the rotation is $R_z$ whereas we desire an $R_y$ rotation: this can be accomplished with a $SH$ correction.

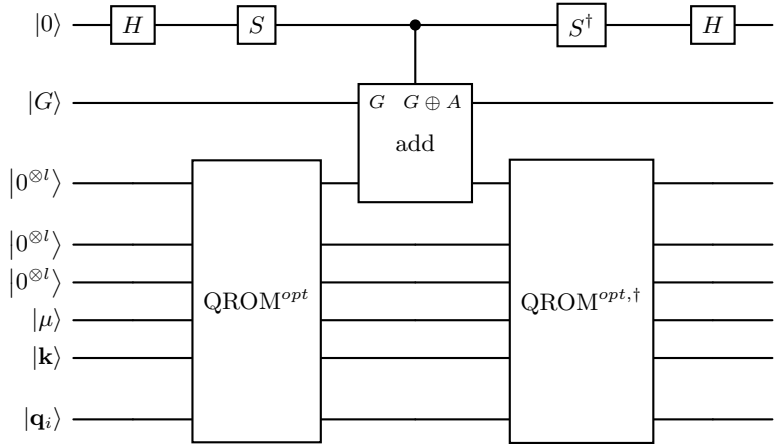
\begin{figure}[h]
\begin{quantikz}
\lstick{\ket{0}} & \gate{H} & \gate{S} & \ctrl{1} &\gate{S^\dagger} & \gate{H} & \\
\lstick{\ket{G}} & & & \gate[2][1.5cm] {\textrm{add}}\gateinput{$G$}\gateoutput{$G\oplus A$}& & & \\
\lstick{\ket{0^{\otimes l}}} & & \gate[6]{\textrm{QROM}^{opt}} & & \gate[6]{\textrm{QROM}^{opt,\dagger}} & & \\
\lstick{\ket{0^{\otimes l}}} & & &  &  & & \\
\lstick{\ket{0^{\otimes l}}} & & & & & &  \\
\lstick{\ket{\mu}} &  & & & & & \\
\lstick{\ket{\mathbf{k}}} & & & & & &  \\
\lstick{\ket{\mathbf{q}_i}} & & & & & & \\
\end{quantikz}
\caption{Quantum circuit for $O_A$ using optimized QROM routine, QROM$^{opt}$ (see Fig.~\ref{fig:qrom-opt}) and controlled phase kickback.}
\label{fig:oa}
\end{figure}

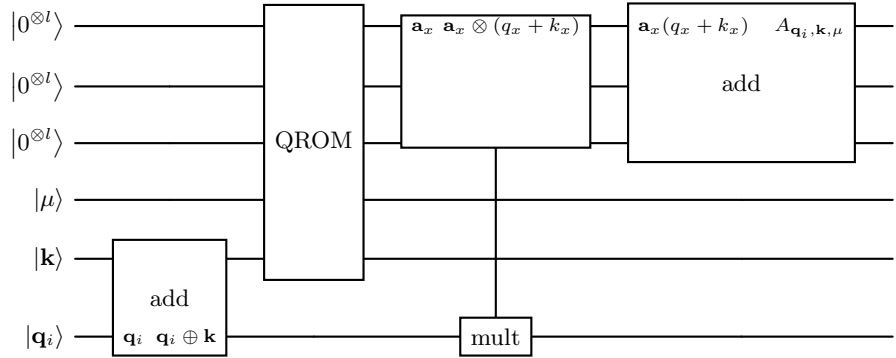
\begin{figure}[h]
\begin{quantikz}
\lstick{\ket{0^{\otimes l}}} & & \gate[5]{\textrm{QROM}} & \gate[3][2.5cm]{}\gateoutput{$\mathbf{a}_x\otimes(q_x+k_x)$}\gateinput{$\mathbf{a}_x$} & \gate[3][3cm]{\textrm{add}}[1.5cm]\gateinput{$\mathbf{a}_x(q_x+k_x)$}\gateoutput{$A_{\mathbf{q}_i,\mathbf{k},\mu}$} & \\
\lstick{\ket{0^{\otimes l}}} & & &  &  & \\
\lstick{\ket{0^{\otimes l}}} & & &\ctrl{3} & &\\
\lstick{\ket{\mu}} &  & & & & \\
\lstick{\ket{\mathbf{k}}} & \gate[2][1.5cm]{\textrm{add}} & & & & \\
\lstick{\ket{\mathbf{q}_i}} & \gateinput{$\mathbf{q}_i$}\gateoutput{$\mathbf{q}_i\oplus \mathbf{k}$} & &  \gate{\textrm{mult}} & & \\
\end{quantikz}
\caption{Optimized QROM$^{opt}$ circuit using QROM and some additional arithmetic. For the arithmetic to work properly we must have $l \geq \log_2 \ngridel/3$.}
\label{fig:qrom-opt}
\end{figure}

In our case, the circuit $\mathcal{A}$ is applied in two steps.
First, the optimized routine above is used to load the value $A_{\mathbf{q},\mathbf{k},\mu}$ into an auxiliary register.
Then a binary addition is carried between this loaded register value and the phase gradient register $|G\rangle$ controlled on an additional qubit $|0\rangle$.
This carries out the controlled $R_z$ rotation with phase $A_{\mathbf{q},\mathbf{k},\mu}$.
This is couched between a $HS$ fixup to map to a $R_y$ rotation, thereby giving us the desired action of $O_A$.
The circuit representation of this sequence of operations is shown in Fig.~\ref{fig:oa} with the optimized QROM protocol in Fig.~\ref{fig:qrom-opt}.

Our implementation of $O_A$ thereby requires the same set of auxiliary qubits $\mu,\mathbf{k}$ as $O_c$, with an additional $3l = \log_2\ngridel$ auxiliary qubits for the optimized QROM protocol and phase kickback.
The T cost is
\begin{equation}
T_{O_A} = 2T_\mathrm{QROM^{opt}} + T_\mathrm{add}(\log_2\ngridel).
\end{equation}
The optimized QROM requires 
\begin{equation}
T_\mathrm{QROM} = 3T_\mathrm{add}(\log_2\ngridel) + 3T_\mathrm{mult}(\log_2\ngridel) + (4(6\ngridph-4)),
\end{equation}
wherein we use the cost of QROM to be $4D - 4$ where $D = 6\ngridph$ is the number of binary strings we need to store~\cite{babbush2018}.
The total cost of $O_A$ is thereby
\begin{subequations}
\begin{align}
&T_{O_A} = 2\left(3(4\log_2\ngridel - 4) + 3(4(\log_2\ngridel)^2 + 16\log_2\ngridel+12) + 4(6\ngridph-4) \right) + 4\log_2\ngridel - 4 \\
& Q_{O_A} = 2\log_2(2\ngridph) + \log_2\ngridel + 1.
\end{align}
\label{eq:costs-oa}
\end{subequations}
Here we use the cost of a recent schoolbook multiplication circuit~\cite{litinski2024quantum}, which is $n^2+4n+3$ Toffoli gates or $4$ times as many T gates.

\paragraph{Full block-encoding cost}
The full block-encoding cost is:
\begin{subequations}
\begin{align}
& T_B(\epsilon) = T_\mathrm{sawtooth}(\log_2\nel) + \nel (T_{O_A}+T_{O_c}) + 2T_{\textrm{el-QFT}}(\nel \log_2\ngridel) \\
&Q_B(\epsilon) = 2\log_2\nel + 2\log_2(2\ngridph) + \log_2\ngridel + 1.
\end{align}
\end{subequations}
The sawtooth structure here is for the SELECT portion of the LCU over $\nel$ wherein we have $\nel$ controlled applications of $\mathcal{J}^{(i)}$ composed of one call to $O_A, O_c$ each.
Finally, we couch all of this between an electronic QFT and its inverse, since all of our block encodings take the momentum basis for the electrons, but our computational basis is the position basis for the electrons.
Substituting in the cost for the sawtooth logic and the QFT based on the required number of single qubit rotations, $
T_{\textrm{el-QFT}} = 4\log(1/\epsilon) \times (\nel \log_2\ngridel)^2$, 
we have the total cost of our novel block-encoding circuit
\begin{subequations}
\begin{align}
& T_B(\epsilon) = (4\log_2\nel - 4) + \nel (T_{O_A}+T_{O_c}) +  8\log(1/\epsilon)\nel^2 (\log_2\ngridel)^2,\\
&Q_B(\epsilon) = 2\log_2\nel + 2\log_2(2\ngridph) + \log_2\ngridel + 1,
\end{align}
\label{eq:costs-ob}
\end{subequations}
where the expressions for $T_{O_A},T_{O_c}$ are provided in Eqs.~\eqref{eq:costs-oa},~\eqref{eq:costs-oc}.
As demonstrated in App.~\ref{app:resources-int}, the T-gate cost is almost entirely dominated by the Fourier transformation, and so the the block encoding cost for the iron opacity instance reduces to:
\begin{equation}
T_B(\epsilon) \cong 2T_\mathrm{el-QFT}(\nel \log_2 \ngridel, \epsilon).
\end{equation}

\subsubsection{Summary of dynamical simulation costs}
The expressions in Eqs.~\eqref{eq:interaction-simulation-cost} and~\eqref{eq:interaction-simulation-cost-reduced} give the resource cost of the dynamical simulation using the interaction picture for iron opacity.
The values of the norms $\alpha_A$ and $\alpha_B$ are found in Eqs.~\eqref{eq:alphaA} and~\eqref{eq:alphaB}; the gate costs $T_A(t,\epsilon)$ are in Eqs.~\eqref{eq:ham-A-resources},~\eqref{eq:ham-ph-resources}, and~\eqref{eq:ham-el-resources}; and the gate costs $T_B(\epsilon)$ are in Eqs.~\eqref{eq:costs-ob},~\eqref{eq:costs-oc}, and ~\eqref{eq:costs-oa}.
All of these quantities are expressed in terms of the physical parameters of the iron opacity simulation which we detail in App.~\ref{app:parameters_for_solar_iron}.

\subsection{Observable estimation}
\label{subapp:observable_estimation}

First, we discuss the relationship between the opacity and the expectation value of the photon number operator at the end of the simulation.
The opacity $\opacity(\bk)$ describes the attenuation of light passing through a medium and determines the transmission probability $\trans_\mathrm{macro}(\bk)$ for a photon passing through a macroscopic sample of mass density $\rho$ and thickness $L$ through the following well-known relationship:
\begin{equation}
    \trans_\mathrm{macro}(\bk) = e^{-\opacity\rho L}.
\end{equation}
Equivalently, the opacity can be expressed in terms of the macroscopic transmission probability as
\begin{equation}
    \opacity(\bk) = - \frac{\ln(\trans_\mathrm{macro}(\bk))}{\rho L}.
    \label{eq:opacity_macro}
\end{equation}

Since we are approximating the response of a macroscopic material by the response of a single atom, we need an expression for $\opacity$ without any macroscopic length scales like $L$.
After traveling a distance $L$, the photon beam would have interacted with $L \rho_i^{1/3}$ atoms on average, where $\rho_i$ is the ion number density of the plasma.
Given our average-atom approximation, we assume that the interaction of the light beam with a given atom does not affect the interaction with a neighboring atom, meaning that 
\begin{equation}
    \trans_\mathrm{macro}(\bk) = (\trans_\mathrm{atom}(\bk))^{L \rho_i^{1/3}},
\end{equation}
where $\trans_\mathrm{atom}$ is the transmission probability for a photon passing through a single atom.
Substituting this relationship between $\trans_\mathrm{macro}$ and $\trans_\mathrm{atom}$ into Eq.~\eqref{eq:opacity_macro}, we find that
\begin{equation}
    \opacity(\bk) = - \rho_i^{1/3} \ln(\trans_\mathrm{atom}(\bk))/\rho.
    \label{eq:opacity_atom}
\end{equation}
The atomic transmission probability directly relates to the expected fraction of photons remaining at the end of the simulation
\begin{equation}
    \trans_\mathrm{atom}(\bk) = \frac{\bra{\psi_\mathrm{f}}\hat{n}_\bk\ket{\psi_\mathrm{f}}}{\bra{\psi_\mathrm{i}}\hat{n}_\bk\ket{\psi_\mathrm{i}}},
\end{equation}
where $\ket{\psi_\mathrm{f}}$ is the final state after time evolution of the initial state $\ket{\psi_\mathrm{i}}$ and 
\begin{equation}
    \hat{n}_\bk=\sum_{\mu=\pm 1} \acr{\bk}{\mu}\aan{\bk}{\mu}
\end{equation}
is the helicity-unresolved photonic number operator.
If we assume that scattering processes are negligible, the atomic absorption probability is given by
\begin{equation}
    \abs(\bk) = 1-\trans_\mathrm{atom}(\bk) = \frac{\bra{\psi_\mathrm{i}}\hat{n}_\bk\ket{\psi_\mathrm{i}} - \bra{\psi_\mathrm{f}}\hat{n}_\bk\ket{\psi_\mathrm{f}}}{\bra{\psi_\mathrm{i}}\hat{n}_\bk\ket{\psi_\mathrm{i}}}.
\end{equation}
Therefore, by computing the absorption or transmission probabilities, which we can compute given the initial and final photonic number distributions, we can then get the opacity $\opacity$ through Eq.~\eqref{eq:opacity_atom}.

In the binary encoding that we are working with, the expectation value of the occupation number of the modes with momentum $\bk$ is
\begin{equation}
    \langle \hat{n}_\bk \rangle = \sum \limits_{m=0}^{\log_2\maxocc -1} 2^m \sum \limits_{\mu=\pm 1} \frac{1+\langle \hat{Z}_{\bk,\mu,m}\rangle}{2},
\end{equation}
where $\hat{Z}_{\bk,\mu,m}$ is the Pauli $Z$ operator acting on the $m$th qubit in the photonic register associated with the mode with quantum numbers $\bk$ and $\mu$ and the expectation is evaluated with respect to whichever state is relevant ($\ket{\psi_\mathrm{i}}$ or $\ket{\psi_\mathrm{f}}$).
The standard error in $\langle \hat{n}_\bk \rangle$ after $\ns$ repetitions is 
\begin{equation}
    \epsilon_{\textrm{est},{\bk}}=\sqrt\frac{\text{Var}(\hat{n}_\bk)}{\ns}.
\end{equation}
Thus choosing $\ket{\psi_\mathrm{i}}$ such that each photonic mode is close to an occupation number eigenstate is advantageous.
The initial state in our protocol has 
\begin{equation}
    \text{Var}(\hat{n}_\bk) = \maxocc p_\mathrm{i}(\bk) (1-p_\mathrm{i}(\bk)) \lesssim 8\times10^{-4} \maxocc,
    \label{eq:var_nk_i}
\end{equation}
where $p_\mathrm{i}(\bk)$ is a Gaussian distribution (see Section~\ref{subapp:initial-state_prep}).
In contrast, the least judicious choice would have $\mathcal{O}(\maxocc^2)$ variance.
We expect similar variance in the final state as well, due to the perturbative nature of the light-matter interaction.

To estimate how large $\ns$ should be to resolve differences in the opacity relevant to the Bailey et al.~\cite{bailey2015higher} experiment on solar iron, we consider a simplified model wherein changes in the final photon numbers are due to single absorption events.
That is, we approximate the final state by 
assuming small absorption probabilities $\abs(\bk)\ll 1/\maxocc$.
Since experimentally measured iron opacities correspond to absorption probabilities on the order of $10^{-5}$ \cite{bailey2015higher}, this assumption holds for $\maxocc\lesssim 10^4$.
In this regime, the final state after passing through the iron ion will be
\begin{align}
    \ket{\psi_\mathrm{f}} \equiv 
     & \left(\sqrt{p_I}\, e^{-it\hel} \otimes e^{-it\hph} + \sum_{\mu=\pm 1}\sum_{\bk\in\mathbb{K}}\sqrt{p_\mathrm{abs}(\bk,\mu)}\, \hat{U}_{\bk,\mu}^\mathrm{el} \otimes \hat{U}_{\aan{\bk}{\mu}}^\mathrm{ph} +\mathcal{O}\left(\abs(\bk)\maxocc\right) \hat{U}_\mathrm{>1}\right) \ket{\psi_\mathrm{i}},
     \label{eq:psif_ansatz}
\end{align}
where the first term corresponds to no absorption events; $\hat{U}_{\bk,\mu}^\mathrm{el}$ and $\hat{U}_{\aan{\bk}{\mu}}^\mathrm{ph}$ represent the electronic and photonic evolution operators corresponding to absorption of a photon with momentum $\bk$ and helicity $\mu$, respectively; and $\hat{U}_\mathrm{>1}$ represents the composite evolution operator corresponding to higher-order processes.
To lowest order in $\abs(\bk)\maxocc$, the probabilities in Eq.~\eqref{eq:psif_ansatz} are
\begin{equation}
    p_\mathrm{abs}(\bk,\mu) = \abs(\bk)\bra{\psi_\mathrm{i}^\mathrm{ph}} \hat{n}_{\bk,\mu}\ket{\psi_\mathrm{i}^\mathrm{ph}} \leq \abs(\bk)\maxocc
\end{equation}
and
\begin{equation}
    p_I = 1-\sum_{\mu=\pm 1}\sum_{\bk\in\mathbb{K}} p_\mathrm{abs}(\bk,\mu).
\end{equation}
Tracing out the electronic subsystem gives the final photonic density matrix
\begin{equation}
    \rho_\mathrm{f}^\mathrm{ph}\equiv\mathrm{Tr}_\mathrm{el}(\ket{\psi_\mathrm{f}}\bra{\psi_\mathrm{f}})
    =p_I \ket{\psi_\mathrm{i}^\mathrm{ph}}\bra{\psi_\mathrm{i}^\mathrm{ph}} 
    + \sum_{\mu=\pm 1}\sum_{\bk\in\mathbb{K}} p_\mathrm{abs}(\bk,\mu) \, \hat{U}_{\aan{\bk}{\mu}}\ket{\psi_\mathrm{i}^\mathrm{ph}}\bra{\psi_\mathrm{i}^\mathrm{ph}}\hat{U}_{\acr{\bk}{\mu}}
    + \mathcal{O}\left((\abs(\bk)\maxocc)^2\right).
\end{equation}
The absorption probability can be inferred from the initial and final photonic states as
\begin{equation}
    \abs(\bk) = \frac{\bra{\psi_\mathrm{i}} \hat{n}_\bk \ket{\psi_\mathrm{i}} - \bra{\psi_\mathrm{f}} \hat{n}_\bk\ket{\psi_\mathrm{f}} }{ \bra{\psi_\mathrm{i}} \hat{n}_\bk \ket{\psi_\mathrm{i}} }
    = \frac{\langle \psi_\mathrm{i}^\mathrm{ph} |\hat{n}_\bk |\psi_\mathrm{i}^\mathrm{ph}\rangle - \mathrm{Tr}(\hat{n}_\bk\,\rho_\mathrm{f}^\mathrm{ph}) }{\langle \psi_\mathrm{i}^\mathrm{ph} |\hat{n}_\bk |\psi_\mathrm{i}^\mathrm{ph}\rangle },
    \label{eq:absorption}
\end{equation}
where
\begin{align}
    \bra{\psi_\mathrm{f}} \hat{n}_\bk\ket{\psi_\mathrm{f}} =& \mathrm{Tr}(\hat{n}_\bk\,\rho_\mathrm{f}^\mathrm{ph}) \nonumber\\
    =& p_I \bra{\psi_\mathrm{i}^\mathrm{ph}} \hat{n}_\bk\ket{\psi_\mathrm{i}^\mathrm{ph}}
    + \sum_{\mu=\pm 1}\sum_{\bk'\in\mathbb{K}} p_\mathrm{abs}(\bk',\mu)\,\bra{\psi_\mathrm{i}^\mathrm{ph}}\hat{U}_{\acr{\bk'}{\mu}} \hat{n}_\bk \, \hat{U}_{\aan{\bk'}{\mu}}\ket{\psi_\mathrm{i}^\mathrm{ph}} \nonumber\\
    =& (1-\abs(\bk))\bra{\psi_\mathrm{i}^\mathrm{ph}} \hat{n}_\bk\ket{\psi_\mathrm{i}^\mathrm{ph}}. \label{eq:absorption2}
\end{align}
Note that in Eq.~\eqref{eq:absorption}, both the numerator and denominator scale with $\maxocc$ and thus the dependence on $\maxocc$ cancels out.

Because the functional form of $\ket{\psi_\mathrm{i}}$ is known, $\bra{\psi_\mathrm{i}}\hat{n}_\bk\ket{\psi_\mathrm{i}}$, $\bra{\psi_\mathrm{i}}\hat{n}^2_\bk\ket{\psi_\mathrm{i}}$, and the relevant variances can be computed a priori, as done for the bound in Eq.~\eqref{eq:var_nk_i}.
The occupation number means and variances in the final state will have to be evaluated through $\ns$ repetitions of our protocol, and then $\abs(\bk)$ and its variance can be estimated when the samples from each photonic mode are combined with the analytically known statistics for the initial state using Eq.~\eqref{eq:absorption2}.

We can think of an absorption event as decrementing the occupation number of a particular mode, which involves flipping the least-significant bit in the binary encoding with probability $\abs(\bk)$.
For a mode with $\bra{\psi_\mathrm{i}}\hat{n}_\bk\ket{\psi_\mathrm{i}}=p_\mathrm{i}(\bk)\maxocc$ (where $0\leq p_\mathrm{i}(\bk)\leq 1$) we can consider a single sample from our protocol as $\lfloor p_\mathrm{i}(\bk) \maxocc\rceil$ independent Bernoulli trials with probability $\abs(\bk)$ of detecting a flip in each trial, where $\lfloor \rceil$ denotes the nearest integer.
Repeating the protocol $\npf$ times increases the number of Bernoulli trials by a multiplicative factor of $\npf$.
The expected value of the associated Bernoulli process is $\lfloor p_\mathrm{i}(\bk)\maxocc\rceil\npf \abs(\bk)$ and its variance is $\lfloor p_\mathrm{i}(\bk)\maxocc\rceil\npf \abs(\bk)(1-\abs(\bk))$.
The maximum likelihood estimator for $\abs(\bk)$ after $\lfloor p_\mathrm{i}(\bk)\maxocc\rceil\npf$ trials in which $x(\bk)$ successful trials are observed is
\begin{equation}
    \tilde{\abs}(\bk) = \frac{x(\bk)}{\lfloor p_\mathrm{i}(\bk)\maxocc\rceil\npf},
\end{equation}
with standard error
\begin{equation}
    \epsilon_{\textrm{std}}(\bk) = \sqrt{\frac{\abs(\bk)(1-\abs(\bk))}{\lfloor p_\mathrm{i}(\bk)\maxocc\rceil\npf}} \label{eq:std_error}.
\end{equation}
We note that the subscript on $\npf$ is intended to indicate that this is the number of samples per feature, and an entire spectrum will be comprised of $\ns$ shots across runs for many individual features that will have similar statistics.

Accounting for the error in the estimate of the opacity in Eq.~\ref{eq:opacity_atom}, we find that
\begin{equation}
    \opacity(\bk) = - \rho_i^{1/3} \ln(1-\abs(\bk) \pm \epsilon_{\textrm{std}}(\bk))/\rho=\rho_i^{1/3} \ln(1-\abs(\bk)(\bk))/\rho  \pm \epsilon_{\textrm{op}}(\bk),
\end{equation}
where the additive error in the opacity is
\begin{equation}
    \epsilon_{\textrm{op}}(\bk) \approx \rho_i^{1/3}/\rho \sqrt{\frac{\abs(\bk)}{\lfloor p_\mathrm{i}(\bk)\maxocc\rceil\npf}} + \mathcal{O}(\abs(\bk)),
\end{equation}
for $\epsilon_{\textrm{std}}(\bk)<1$ and $\abs(\bk) \ll 1$.
This suggests that $\lfloor p_\mathrm{i}(\bk)\maxocc\rceil\npf$ ought to be greater than $\abs(\bk)^{-1}$, which is consistent with the intuition that the free parameters need to be set such that there is a reasonable probability of observing at least one absorption event.
Increasing either $\maxocc$ or $\npf$ can reduce the statistical error, but we note that $p_\mathrm{i}(\bk)$ will vary from mode to mode and it might be that we cannot let $\maxocc$ get too large if we need to remain in the linear response regime for a particular set of conditions.
Thus $\npf$ ultimately sets the error floor.

The absorption probability in the contested region of iron's spectrum is around $\abs(\bk) = 4\times 10^{-5}$ and thus $\alpha = 1/2$ and $\maxocc = 10^4$ are reasonable values for the photonic subsystem parameters. Ensuring that $\epsilon_\textrm{std} \leq 0.1 \ \abs(\bk)$ would require $\npf \geq 500$.
However, this makes no use of other prior information, which we can use to reduce $\npf$.
For example, we might know where the center of a certain spectral feature is, and/or its line shape, but the magnitude and width might be uncertain.
In that setting, it might only be necessary for us to be able to estimate the peak height and peak width parameters, given a peak shape. 
This could be achieved with a smaller value of $\npf$, effectively leveraging the fact that observing an absorption event at a particular momentum can constrain the absorption probability at nearby momenta given a model of the line shape.

\begin{figure}
    \centering
    \includegraphics[width=1.0\textwidth]{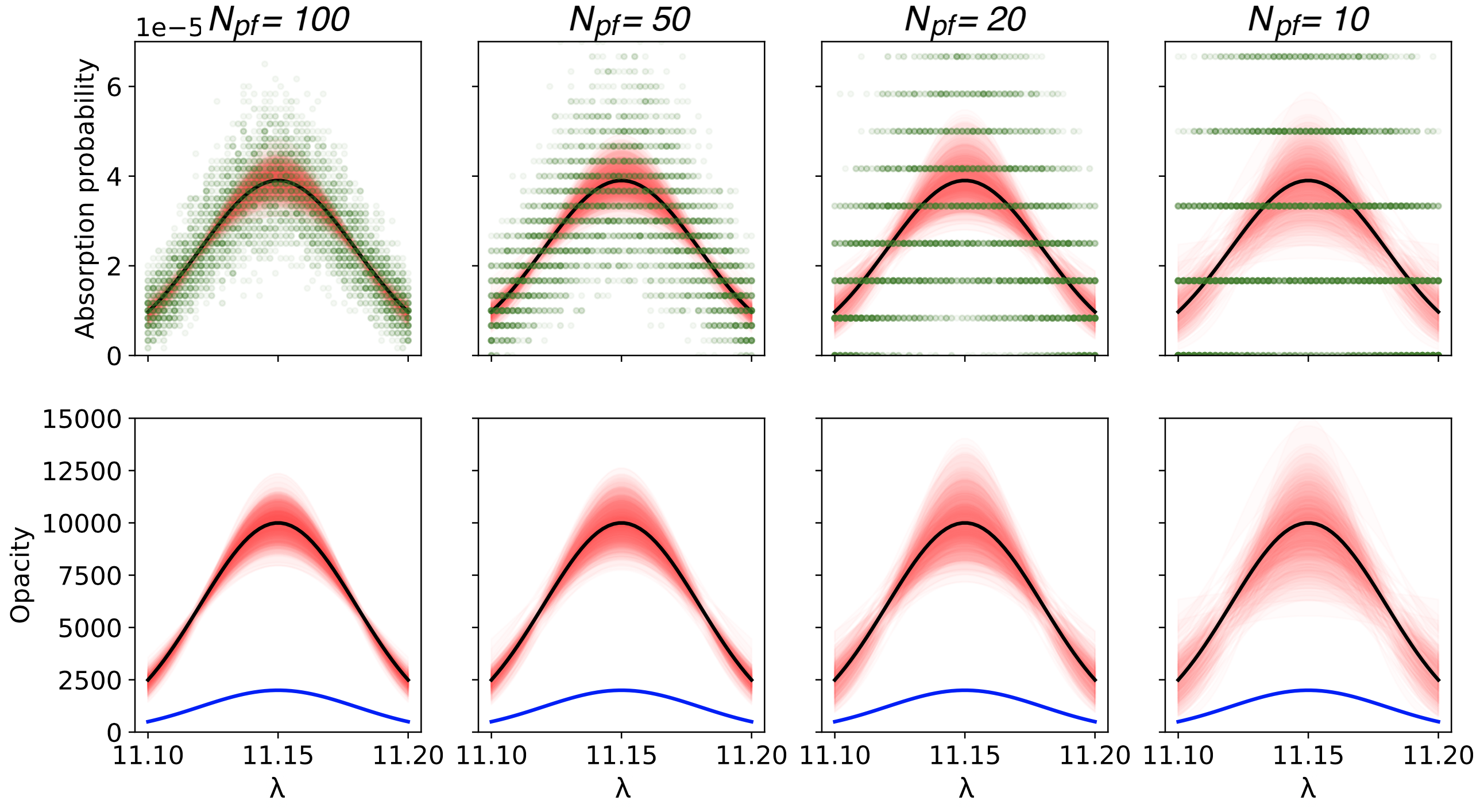}
    \caption{Bernoulli statistics involved in estimating $\npf$. The top row shows estimated absorption probabilities, where the green points show $100$ Bernoulli samples at each wavelength $\lambda$ assuming $\maxocc=10^4$ and $p_\mathrm{i}(\bk)=1/2$.
    The red curves show Gaussian fits to each set of Bernoulli samples, with translucent red shading indicating $\pm 2\sigma$ co-variances from the curve fits to account for statistical variations across the $\npf$ shots comprising each sample.
    The black curves indicate the targeted distribution, which is representative of experimental data at the lowest density from Bailey et al.~\cite{bailey2015higher}.
    The bottom row shows the corresponding opacity curves in units of cm$^2$/g. 
    The blue curves are representative of the state-of-the-art classical computation~\cite{hansen2007hybrid} for this spectral feature from Bailey et al.~\cite{bailey2015higher} 
    We note that the statistical variance at $\npf = 10$ is sufficiently large that it nearly overlaps the classical modeling results at the edges of the depicted wavelength range, indicating that $\npf \geq 10$ is required to improve accuracy beyond classical simulation.
}
\label{fig:observable_estimation}
\end{figure}

To estimate the reduction in $\npf$ achievable using this sort of prior information, we report the results of a numerical experiment in Fig.~\ref{fig:observable_estimation}.
Here, our objective is to fit a Gaussian curve to Bernoulli-sampled data, reproducing a peak similar to one in the experimental data in Ref.~\cite{bailey2015higher}.
We find that for $\npf \geq 10$ we achieve small enough statistical error to meaningfully resolve a feature relative to the reported disagreement between theory and experiment. 

\clearpage
\section{Simulation setup and parameters for solar iron opacity}
\label{app:parameters_for_solar_iron}

\renewcommand{\theequation}{D\arabic{equation}}
\renewcommand{\thefigure}{D\arabic{figure}}
\renewcommand{\thetable}{D\arabic{table}}
\setcounter{figure}{0}
\setcounter{table}{0}

\begin{wrapfigure}{r}{0.5\textwidth}
    \centering
    \vspace{-0.15in}
    \includegraphics[width=0.425\textwidth]{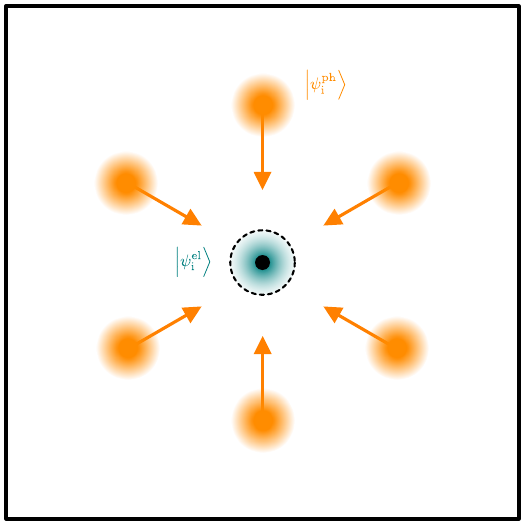}\\
    \includegraphics[width=0.425\textwidth]{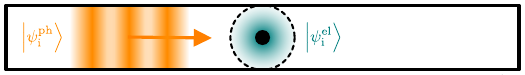}
    \vspace{-0.1in}
    \caption{Simulation setups considered in this work.
    In the top panel, $N_\mathrm{wp}$ fully localized 3D Gaussian wave packets (orange) are in a spherical arrangement and approach the central atom (blue) from different directions.
    In the bottom panel, $N_\mathrm{wp}$ planar wave packets are arranged collinearly and approach the central atom with the same average momentum $\bar{\bk}$.
    The spatial distribution of each wave packet is Gaussian along the direction of motion and uniform along transverse axes, with the simulation cell width given by the atom's diameter.
    }
    \vspace{-0.1in}
    \label{fig:bothsetups}
\end{wrapfigure}

We considered the two simulation setups illustrated in Fig.~\ref{fig:bothsetups}.
The first setup consists of a cubical simulation cell containing the iron ion at the center and $N_\mathrm{wp}$ 3D Gaussian photonic wave packets approaching the atom from different directions.
Including multiple wave packets within this cubic setup allows for more photons to interact with the atom within a fixed simulation time, reducing the number of shots required for statistical sampling of absorption events.
However, the number of occupied photonic modes $\ngridph$ scales with $N_\mathrm{wp}$.
Also, each wave packet must be sufficiently localized in spatial directions orthogonal to its average momentum to ensure that the photons pass through the atom, i.e., $\sigma_r^\perp = 1/(2\sigma_k^\perp)\lesssim\rws$.
This constraint places a lower bound on $\sigma_k^\perp$ and thus limits the number of wave packets that fit within the relevant momentum modes.

The second setup involves a smaller, elongated simulation cell where a train of $N_\mathrm{wp}$ planar photonic wave packets approaches the central atom from a single direction.
This setup dramatically reduces the required simulation cell volume $\Omega$ and electronic grid points $\ngridel$, since the width of the simulation cell is given by the diameter of the average atom.
The collinear motion of the photonic wave packets and their spatial delocalization along the orthogonal directions (i.e., $\sigma_k^\perp=0$) also dramatically reduces $\ngridph$, which becomes independent of $N_\mathrm{wp}$.
However, in this setup, the simulation time varies linearly with $N_\mathrm{wp}$ because wave packets that begin further from the atom require more time to propagate through and emerge on the other side.

The conditions and results of state-of-the-art iron opacity experiments \cite{bailey2015higher} determine several parameters informing quantum resource requirements for each simulation setup, particularly regarding the photonic degrees of freedom.
We discuss these experimental considerations and the resulting requirements for the momentum resolution $\Delta_k$, simulation cell volume $\Omega$, number of photonic modes $\ngridph$, evolution time $t$, number of wave packets $N_\mathrm{wp}$, and time-evolution accuracy $\epsilon$ in Sec.~\ref{subapp:parameters_experimental}.
Estimating the parameters describing the electronic subsystem\,---\,namely, the grid spacing $\Delta_r$, number of grid points $\ngridel$, and regularization constant $\ereg$\,---\,involved classical numerics as described in Sec.~\ref{subapp:parameters_numerical}.
Throughout, we compare the requirements of computing the full opacity spectrum (spanning wavelengths between 7 and \SI{13}{\angstrom}) to the requirements of a single spectral feature (strategically chosen at a wavelength around \SI{11}{\angstrom}).

Tables~\ref{tab:appendix_parameters_cubic} and \ref{tab:appendix_parameters_long} summarize the parameters required for both simulation setups and both spectral ranges based on the analyses in Sec.~\ref{subapp:parameters_experimental} and \ref{subapp:parameters_numerical}.
Throughout, we assume $\eta=26$ electrons corresponding to charge neutrality for the iron plasma, though the simulation protocol would also apply for other values of $\eta$ representing thermal fluctuations in the local charge.
To minimize resource requirements per shot, the main text discusses only the second, elongated setup (Table \ref{tab:appendix_parameters_long}) for the full spectrum with $N_\mathrm{wp}=1$.

\begin{table}[h]
    \centering
    \bgroup
    \def\arraystretch{1.5}
    \begin{tabular}{l|c|c|c|c|c|c|c|c|c|c}
        & $\nel$ & $\Omega$ & $\Delta_r$ & $\ngridel$ & $\ereg$ & $\Delta_k$ & $\ngridph/N_\mathrm{wp}$ & max $N_\mathrm{wp}$ & \tee & $\epsilon$ \\\hline
         full spectrum & 26 & $3.3\times 10^{13}$ & $10^{-2}$ & $3.3\times 10^{19}$ & $10^{-3}$ & $1.9\times 10^{-4}$ & $9.9\times 10^{10}$ & 13 & 0.30 & $2.5\times 10^{-7}$\\\hline
         single feature & 26 & $1.3\times 10^{13}$ & $10^{-2}$ & $2.2\times 10^{19}$ & $10^{-3}$ & $2.7\times 10^{-4}$ & $4.9\times 10^8$ & 9 & 15 & $2.5\times 10^{-7}$
    \end{tabular}
    \caption{Parameters determining the quantum resources required to predict iron opacity at solar conditions using the \emph{cubic} simulation cell.
    All non-dimensionless values are given in atomic units.
    }
    \label{tab:appendix_parameters_cubic}
    \egroup
\end{table}

\begin{table}[h]
    \centering
    \bgroup
    \def\arraystretch{1.5}
    \begin{tabular}{l|c|c|c|c|c|c|c|c|c|c}
        & $\nel$ & $\Omega$ & $\Delta_r$ & $\ngridel$ & $\ereg$ & $\Delta_k$ & $\ngridph$ & max $N_\mathrm{wp}$ & \tee & $\epsilon$ \\\hline
         full spectrum & 26 & $1.3\times 10^{7}$ & $10^{-2}$ & $1.3\times 10^{13}$ & $10^{-3}$ & $1.9\times 10^{-4}$ & 4500  & 2967 & $0.42+0.08 (N_\mathrm{wp}-1)$ & $2.5\times 10^{-7}$\\\hline
         single feature & 26 & $9.6\times 10^{6}$ & $10^{-2}$ & $1.2\times 10^{13}$ & $10^{-3}$ & $2.7\times 10^{-4}$ & 40 & 22 & $23+6 (N_\mathrm{wp}-1)$ & $2.5\times 10^{-7}$
    \end{tabular}
    \caption{Parameters determining the quantum resources required to predict iron opacity at solar conditions using the \emph{elongated} simulation cell.
    All non-dimensionless values are given in atomic units.
    }
    \label{tab:appendix_parameters_long}
    \egroup
\end{table}

\subsection{Experimental considerations}
\label{subapp:parameters_experimental}

The iron opacity experiments \cite{bailey2015higher} include wavelengths $\lambda$ in the 7\,--\,\SI{13}{\angstrom} range and the measured spectral lines are about \SI{0.1}{\angstrom} wide.
The absorption probabilities corresponding to the measured opacity are on the order of $10^{-5}$.
As detailed below, these properties constrain the momentum resolution $\Delta_k$, simulation cell volume $\Omega$, the number of photonic modes $\ngridph$, the evolution time $t$, the number of wave packet $N_\mathrm{wp}$, and the time-evolution precision $\epsilon$.

\subsubsection{Momentum resolution $\Delta_k$ and simulation cell volume $\Omega$}
Resolving opacity features of \SI{0.1}{\angstrom} spectral width requires a wavelength resolution of $\Delta_\lambda \lesssim \SI{0.01}{\angstrom}$.
The corresponding momentum resolution is 
\begin{equation}
    \Delta_k = \frac{2\pi}{\lambda_\mathrm{max}^2} \Delta_\lambda \lesssim 
    \left\{\begin{array}{ll}
    1.9\times 10^{-4}\,\SI{}{\au} & \quad\textrm{full spectrum}\\
    2.7\times 10^{-4}\,\SI{}{\au} & \quad\textrm{single feature}
    \end{array} ,\right.
    \label{eq:Delta_k}
\end{equation}
where the maximum wavelength of interest $\lambda_\mathrm{max}$ is \SI{13}{\angstrom} for the full opacity spectrum and \SI{11.2}{\angstrom} for the case of a single spectral feature.
In the latter case, a somewhat larger $\Delta_k$ of up to $7\times 10^{-4}$\,\SI{}{\au} would be possible by selecting a feature at small $\lambda$.
However, the low-wavelength portion of the experimentally measured opacity spectrum \cite{bailey2015higher} is largely smooth and featureless, limiting the practical utility of focusing on that regime.
Instead, for the single-feature case we choose to focus on the 11.1\,--\,\SI{11.2}{\angstrom} wavelength range, which exhibits a sharp feature with a significant discrepancy between experimental measurements and classical predictions depending on the plasma conditions~\cite{bailey2015higher}.

Achieving this momentum resolution within the plane-wave photonic basis in turn requires a cubic simulation cell volume of
\begin{equation}
    \Omega_\mathrm{cube} = \left(\frac{2\pi}{\Delta_k}\right)^3 \approx 
    \left\{\begin{array}{ll}
    3.3\times 10^{13}\,\SI{}{\au} & \quad \textrm{full spectrum}\\
    1.3\times 10^{13}\,\SI{}{\au} & \quad \textrm{single feature}
    \end{array}\right.
    \label{eq:omega_estimate_cube}
\end{equation}
or an elongated simulation cell volume of
\begin{equation}
    \Omega_\mathrm{long} = \frac{2\pi}{\Delta_k} (2\rws)^2 \approx 
    \left\{\begin{array}{ll}
    1.3\times 10^{7}\,\SI{}{\au} & \quad\textrm{full spectrum} \\
    9.6\times 10^{6}\,\SI{}{\au} & \quad\textrm{single feature}
    \end{array}\right. .
    \label{eq:omega_estimate_long}
\end{equation}
Notably, these volumes are much larger than would be needed to contain the average atom by itself: for the lowest electron density of $\rho_e=7.1\times 10^{21}\,\SI{}{\per\centi\meter\cubed}$ probed in Ref.~\onlinecite{bailey2015higher}, the volume per iron ion is only $26/\rho_e = 2.5\times 10^{4}\,\SI{}{\au}$.

\subsubsection{Number of photonic modes $\ngridph$}
\label{subapp:ngamma}
The number of modes needed to represent the photonic state depends on the spectral resolution from Eq.~\eqref{eq:Delta_k} and the spectral width of the photonic wave packets:
\begin{equation}
    \ngridph \propto \left\{\begin{array}{ll}
    N_\mathrm{wp}\sigma_k^\shortparallel (\sigma_k^\perp)^2 / \Delta_k^3 & \quad\textrm{cubic setup}\\
    \sigma_k^\shortparallel/\Delta_k & \quad\textrm{elongated setup}
    \end{array}\right. ,
    \label{eq:ngamma_naive}
\end{equation}
where $\sigma_k^\shortparallel$ and $\sigma_k^\perp$ denote the momentum-space standard deviations of each Gaussian wave packet along directions parallel and perpendicular to its average momentum, respectively (see Eq.~\eqref{eq:initial_photonic_state_general}). 
Notably, enough momentum modes must be included to adequately capture the Gaussian tails of the initial photonic state, enable sufficient spatial localization of the photons, and mitigate the overhead in the number of shots required to resolve the opacity at the edges of the targeted spectral range. 
In this section, we determine appropriate choices for $\sigma_k^\shortparallel$, $\sigma_k^\perp$, and the proportionality constants in Eq.~\eqref{eq:ngamma_naive} to arrive at the required values for $\ngridph$.

First, we determine $\sigma_k^\shortparallel$ by constraining overheads associated with the spectral edges.
Let $k_\mathrm{min}^\mathrm{targ}=2\pi/\lambda_\mathrm{max}^\mathrm{targ}$ and $k_\mathrm{max}^\mathrm{targ}=2\pi/\lambda_\mathrm{min}^\mathrm{targ}$ denote the minimum and maximum wave numbers of interest for the opacity calculation, with $[k_\mathrm{min}^\mathrm{targ},k_\mathrm{max}^\mathrm{targ}] = [0.26, 0.47]$\,\SI{}{\au} for the full spectrum and $[0.297, 0.300]$\,\SI{}{\au} for the single feature.
Each photonic wave packet will be centered at wave number $\bar{k}\equiv|\bar{\bk}_J|=(k_\mathrm{min}^\mathrm{targ}+k_\mathrm{max}^\mathrm{targ})/2$.
Since momentum modes further from $\bar{\bk}_J$ will have lower average initial occupations, resolving opacity near the edges of the momentum range of interest will require more shots and higher simulation precision than at the center of the wave packet.
We choose to limit these overheads to a factor of $2$ by setting
\begin{equation}
    \sigma_k^\shortparallel = \frac{k_\mathrm{max}^\mathrm{targ}-k_\mathrm{min}^\mathrm{targ}}{(8 \ln(2))^{1/2}}
    \label{eq:sigmakpar}
\end{equation}
so that the average initial occupation at the spectral edges is
\begin{equation}
     \langle \hat{n}_{\bk} \rangle  \Big|_{|\bk|=k_\mathrm{min}^\mathrm{targ}}
    =
    \langle \hat{n}_\bk \rangle \Big|_{|\bk|=k_\mathrm{max}^\mathrm{targ}}
    = \frac{N_\mathrm{wp}\maxocc}{\mathcal{N}} e^{-(k_\mathrm{max}^\mathrm{targ}-\bar{k})^2/2(\sigma_k^\shortparallel)^2} 
    = \frac{1}{2} \langle \hat{n}_{\bk} \rangle  \Big|_{|\bk|=\bar{k}}.
\end{equation}
Table~\ref{tab:photon_sigmas} lists the resulting numerical values of $\sigma_k^\shortparallel$ for both the full spectrum and single feature cases.

Next, we determine $\sigma_k^\perp$ for the cubic simulation setup by considering the spatial localization of the 3D Gaussian wave packets.
We use
\begin{equation}
    \sigma_k^\perp = \left\{\begin{array}{ll}
         1/\rws & \textrm{cubic setup} \\
         0 & \textrm{elongated setup}
    \end{array}\right.
    \label{eq:sigmakperp}
\end{equation}
so that the transverse spatial width of each 3D wave packet in the cubic case is $\sigma_r^\perp = 1/(2\sigma_k^\perp)=\rws/2$ and most of the photons actually pass through the central atom.
Notably, this $\sigma_k^\perp$ value is close to the $\sigma_k^\shortparallel$ value for the full spectrum (see Table~\ref{tab:photon_sigmas}); in this case, the photonic wave packets are nearly isotropic.
However, for the case of a single feature, the photonic wave packets are asymmetric, with $\sigma_k^\perp \gg \sigma_k^\shortparallel$ and $\sigma_r^\perp \ll \sigma_r^\shortparallel$.

While our $\sigma_k^\shortparallel$ and $\sigma_k^\perp$ values represent reasonable choices for this problem instance, further optimization of the underlying space-time tradeoffs may be possible.
Although smaller $\sigma_k^\shortparallel$ and/or $\sigma_k^\perp$ could reduce $\ngridph$, maintaining adequate photon statistics would then require increasing $\maxocc$ or $\ns$.
Reducing $\sigma_k^\shortparallel$ would also increase $\sigma_r^\shortparallel$ and require a longer simulation time (see Sec.~\ref{subapp:time}).
These considerations would ultimately produce only minor adjustments to our final resource estimates.

\begin{table}[]
    \centering
    \begin{tabular}{c|c|c}
         & $\sigma_k^\shortparallel$ & $\sigma_r^\shortparallel$ \\\hline
         full spectrum & 0.093 & 5.37 \\
         single feature & 0.0011 & 440. 
    \end{tabular}
    \qquad\qquad
    \begin{tabular}{c|c|c}
         &  $\sigma_k^\perp$ & $\sigma_r^\perp$ \\\hline
         cubic setup & 0.099 & 5.08 \\
         elongated setup & 0 & $\infty$
    \end{tabular}
    \caption{Standard deviations of the Gaussian wave packets comprising the initial photonic state given by Eq.~\eqref{eq:initial_photonic_state_general}.
    In momentum space, $\sigma_k^\shortparallel$ and $\sigma_k^\perp$ denote spreads parallel and perpendicular to the wave packet's direction of motion, respectively.
    The corresponding real-space values are given by $\sigma_r^\shortparallel = 1/(2\sigma_k^\shortparallel)$ and $\sigma_r^\perp = 1/(2\sigma_k^\perp)$.
    While $\sigma_k^\shortparallel$ and $\sigma_r^\shortparallel$ depend on the spectral range of interest (see Eq.~\eqref{eq:sigmakpar}), $\sigma_k^\perp$ and $\sigma_r^\perp$ depend on the simulation setup (see Fig.~\ref{fig:bothsetups}) and the plasma density (see Eq.~\eqref{eq:sigmakperp}).
    All values are given in atomic units.
    }
    \label{tab:photon_sigmas}
\end{table}

\begin{figure}
    \centering
    \includegraphics[width=0.45\textwidth]{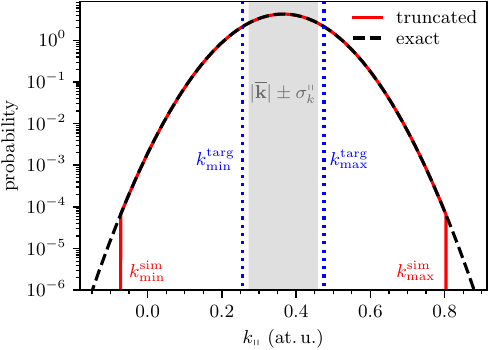}\hfill
    \includegraphics[width=0.45\textwidth]{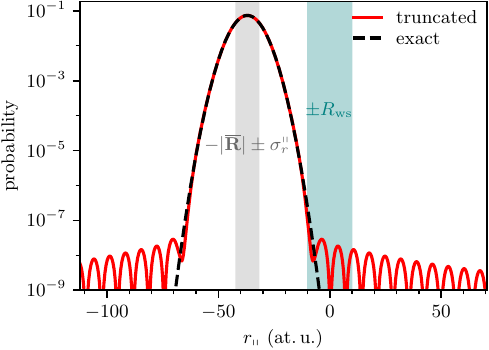}
    \caption{Initial photonic wave packet in momentum space (left) and real space (right) along its direction of motion in the full spectrum case at $k_\perp=0$.
    Solid red curves show distributions for a truncated Gaussian defined over the finite momentum range $k_\shortparallel\in[k_\mathrm{min}^\mathrm{sim}, k_\mathrm{max}^\mathrm{sim}]$ included in the simulation, while dashed black curves correspond to a full Gaussian in the limit of an infinite momentum range.
    Gray shading indicates momenta and positions within one standard deviation of the mean.
    Blue dotted lines indicate the minimum ($k_\mathrm{min}^\mathrm{targ}$) and maximum ($k_\mathrm{max}^\mathrm{targ}$) wave numbers corresponding to the spectral range of interest.
    Teal shading indicates positions inside the central atom, $|r_\shortparallel|<\rws$.
    }
    \label{fig:photon_truncation}
\end{figure}

Finally, spatial localization of the initial photonic state also determines the proportionality constants in Eq.~\eqref{eq:ngamma_naive}.
To accommodate the tails of the Gaussian photonic wave packets, the photon momentum range included in the quantum simulation must be wider than the relevant standard deviations.
As shown in Fig.~\ref{fig:photon_truncation}, truncating the Gaussian tails leads to delocalization and spurious satellites in the corresponding spatial distribution of the photons.
These artifacts affect the opacity calculation in two ways.
First, many of the photons away from the main peak of the spatial distribution will not pass through the iron atom: the spurious weight starting on the wrong side of the atom ($r_\shortparallel>\rws$) will move away from it over time, while the spurious weight starting further from the atom than the main peak ($r_\shortparallel \lesssim -|\bar{\bR}|-\sigma_r^\shortparallel$) will not reach the atom by the end of the simulation.
The computed opacity can easily be adjusted to account for the small fraction of photons that do not meaningfully participate in the light-matter interaction.

\begin{figure}
    \centering
    \includegraphics{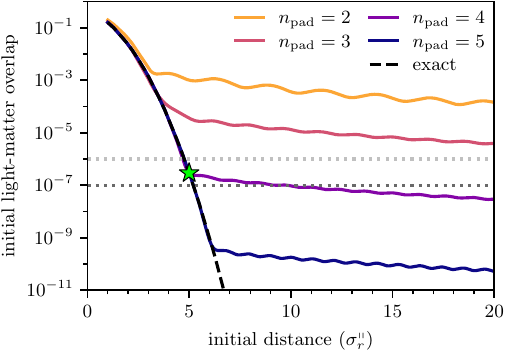}
    \caption{Overlap between the closest initial photonic wave packet and the iron atom as a function of the wave packet's initial distance from the atom sphere, $|\bar{\bR}_J|-\rws$.
    The distance is given in units of the wave packet's width, $\sigma_r^\shortparallel=1/(2\sigma_k^\shortparallel)$.
    Solid curves show results for truncated wave packets with a momentum range padded according to Eq.~\eqref{eq:kpad}, while the dashed black curve shows results for the full Gaussian in the limit of an infinite momentum range.
    Light (dark) gray dotted lines show thresholds corresponding to 10\% (1\%) of typical absorption probabilities.
    The green star indicates the optimal parameters chosen for the quantum simulation protocol.
    }
    \label{fig:photon_overlap}
\end{figure}

More importantly, the initial spatial distribution of the photons inevitably overlaps with the iron atom, potentially biasing the opacity result relative to the physically-relevant limit of all photons approaching the atom from infinitely far away.
This effect would occur even without truncating the tails of the Gaussian wave packets, and it can be systematically reduced by initializing the wave packets further from the atom (see Fig.~\ref{fig:photon_overlap}).
While the impact of this overlap on the computed opacity is difficult to predict, the worst-case scenario is that all photons beginning inside the atom get absorbed artificially.
To ensure the accuracy of the opacity calculation, we select photonic parameters that keep the fraction of photons starting inside the atom small compared to typical $\sim 10^{-5}$ absorption probabilities measured in opacity experiments.
This conservative strategy requires an initial fractional light-matter overlap below $10^{-6}$ for a 10\% precision target for computed absorption rates.

To achieve this accuracy threshold, we pad the momentum range according to
\begin{equation}
    k_\mathrm{max}^\mathrm{sim}-k_\mathrm{min}^\mathrm{sim} = n_\mathrm{pad} (k_\mathrm{max}^\mathrm{targ}-k_\mathrm{min}^\mathrm{targ}) = (8\ln(2))^{1/2} n_\mathrm{pad} \sigma_k^\shortparallel,
    \label{eq:kpad}
\end{equation}
where $k_\mathrm{min}^\mathrm{sim}$, $k_\mathrm{max}^\mathrm{sim}$ denote the minimum and maximum values of $k_\shortparallel$ included in the simulation while $n_\mathrm{pad}$ is a constant to be optimized.
The initial light-matter overlap also depends on the initial distance between the photonic wave packet and the atom sphere, $|\bar{\bR}_J|-\rws$.
As shown in Fig.~\ref{fig:photon_overlap}, there is a space-time tradeoff between this initial distance and $n_\mathrm{pad}$. 
Decreasing the momentum range padding $n_\mathrm{pad}$ reduces $\ngridph$ and the number of photonic qubits.
However, decreasing $n_\mathrm{pad}$ also requires the photonic wave packets to begin further from the atom to maintain a small initial light-matter overlap and thus increases the simulation time and attendant gate counts.
We conclude that $n_\mathrm{pad}=4$ and a minimum initial distance of
\begin{equation}
    \min_J|\bar{\bR}_J| - \rws = 5\sigma_r^\shortparallel
    \label{eq:initial_wp_dist}
\end{equation}
achieves a favorable balance between these considerations.

Applying the same momentum range padding of $8(2\ln(2))^{1/2}$ to each of the transverse directions in the cubic setup, we obtain
\begin{equation}
    \ngridph = \left\{\begin{array}{ll}
    512\ln(2) N_\mathrm{wp}(k_\mathrm{max}^\mathrm{targ}-k_\mathrm{min}^\mathrm{targ}) (\sigma_k^\perp)^2 / \Delta_k^3 & \quad\textrm{cubic setup}\\
    4(k_\mathrm{max}^\mathrm{targ}-k_\mathrm{min}^\mathrm{targ})/\Delta_k & \quad\textrm{elongated setup}
    \end{array}\right. .
    \label{eq:ngamma_final}
\end{equation}
The cubic setup requires $\ngridph/N_\mathrm{wp}>10^8$ even for the case of a single spectral feature, while the elongated setup provides the full opacity spectrum with only $\ngridph\approx 4500$.
Since the number of qubits in the second-quantized encoding of the photonic subsystem scales linearly with $\ngridph$, we proceed with the more practical elongated setup for our quantum resource estimates.

\subsubsection{Evolution time $t$}
\label{subapp:time}
The coupled electron-photon dynamics must be simulated long enough for all of the photonic wave packets to propagate toward the atom, pass through it, and emerge far enough away on the other side that further light-matter interactions can be considered negligible.
The peak of the furthest wave packet will reach the origin after time $\max_J |\overline{\bR}_J| /c$ and cross the atom sphere boundary after additional time $\rws/c$.
Then, the simulation must continue for time $\propto \sigma_r^\shortparallel/c$ to allow the tails of the Gaussian wave packet to pass through the atom.

At the end of the time evolution, the final spatial distribution of the photons will inevitably overlap with the iron atom, similar to the initial overlap considered in Sec.~\ref{subapp:ngamma} and Fig.~\ref{fig:photon_overlap}.
Any photons remaining inside the atom will have only partially interacted with it, again potentially biasing the opacity result.
However, now the worst case scenario is that none of these photons get absorbed, a much less severe error.
Evolving the system long enough for the peak of each photonic wave packet to pass through the atom and emerge $2\sigma_r^\shortparallel$ away from it already constrains the resulting bias in the absorption rate to less than 3\%.

Thus, we set the total simulation time to
\begin{equation}
    t = \frac{1}{c}\left(\max_J |\overline{\bR}_J| + \rws + 2\sigma_r^\shortparallel \right).
\end{equation}
In the cubic setup, all $|\overline{\bR}_J|$ are equal and the simulation time does not depend on the number of wave packets.
However, in the elongated set up, the wave packets begin at different distances and we choose to maintain a uniform spatial separation of $2\sigma_r^\shortparallel$ between adjacent Gaussian peaks.
Thus, using Eq.~\eqref{eq:initial_wp_dist}, the initial distance of each wave packet from the origin is
\begin{equation}
    |\overline{\bR}_J| = \left\{\begin{array}{ll}
    \rws + 5\sigma_r^\shortparallel &\quad\textrm{cubic setup} \\
    \rws + (3+2J) \sigma_r^\shortparallel &\quad\textrm{elongated setup}
    \end{array}\right. .
\end{equation}
Therefore, the total simulation time is
\begin{equation}
    t = \left\{\begin{array}{ll}
         (2\rws + 7\sigma_r^\shortparallel)/c &\quad\textrm{cubic setup} \\
         (2\rws + (5 + 2N_\mathrm{wp}) \sigma_r^\shortparallel)/c &\quad\textrm{elongated setup}
    \end{array}\right. .
\end{equation}

\subsubsection{Number of wave packets $N_\mathrm{wp}$}
\label{subapp:paramters_Nwp}

The widths of the Guassian wave packets given in Table~\ref{tab:photon_sigmas} limit the number of wave packets that fit within the simulation.
In the cubic setup, momentum-space overlap between wave packets can produce artifacts associated with the occupancy cutoff $\maxocc$: the overlapping Gaussian tails should in principle contain support over larger occupations, but these are only represented modulo $\maxocc$.
However, these incorrectly treated high-occupancy states would have very low probability such that some modest overlap among photonic wave packets is tolerable.
To estimate the maximum number of 3D wave packets in the cubic setup, we consider the number of spheres of radius $2\sigma_k^\perp$ that can be packed onto the spherical surface of radius $\bar{k}$:
\begin{equation}
    N_\mathrm{wp}^\mathrm{cube} \lesssim \frac{4\pi\bar{k}^2}{\pi(2\sigma_k^\perp)^2} = \left(\frac{\bar{k}}{\sigma_k^\perp}\right)^2 .
\end{equation}
We find that $N_\mathrm{wp}^\mathrm{cube}\sim 10$ for both the full spectrum and single feature simulations.

In the elongated setup, all the wave packets occupy the same momentum modes and $\maxocc$ must scale with $N_\mathrm{wp}$.
The number of wave packets is instead constrained by the real-space degrees of freedom.
The maximum number of planar wave packets separated by $2\sigma_r^\shortparallel$ that can be packed along the supercell length $2\pi/\Delta_k$ while maintaining a distance of at least $5\sigma_r^\shortparallel$ from the atom sphere is
\begin{equation}
    N_\mathrm{wp}^\mathrm{long} \leq \frac{2\pi/\Delta_k - 2\rws - 10\sigma_r^\shortparallel}{2\sigma_r^\shortparallel} +1
    = \frac{\pi/\Delta_k - \rws}{\sigma_r^\shortparallel} - 4.
\end{equation}
We find that the elongated setup can support about 3000 and 20 wave packets in the full spectrum and single feature cases, respectively.

Ultimately, we minimize the cost per shot by selecting the elongated setup with a single wave packet.

\subsubsection{Time-evolution accuracy $\epsilon$}
\label{subapp:parameters_epsilon}

In our simulation, we will not obtain the exact final state $\ket{\psi_\mathrm{f}}$, but instead an approximate final state $\ket{\psi_{f'}}$ containing additive error $\ket{\epsilon}$:
\begin{equation}
    |\psi_{f'}\rangle \equiv |\psi_\mathrm{f}\rangle + |\epsilon\rangle, \quad \braket{\epsilon}{\epsilon} \leq \epsilon^2.
\end{equation}
Given the approximate final state, we can carry out the same inference as in Eq.~\eqref{eq:absorption} for the approximate absorption probability $\abs'(\bk)$:
\begin{equation}
\abs'(\bk) \equiv \frac{\langle \psi_\mathrm{i} |\hat{n}_\bk |\psi_\mathrm{i}\rangle - \langle \psi_{f'} |\hat{n}_\bk |\psi_{f'}\rangle }{\langle \psi_\mathrm{i} |\hat{n}_\bk |\psi_\mathrm{i}\rangle } 
= \abs(\bk) - \frac{2 \Re[\bra{\epsilon}\hat{n}_\bk\ket{\psi_\mathrm{f}}]}{\bra{\psi_\mathrm{i}} \hat{n}_\bk \ket{\psi_\mathrm{i}}} + \mathcal{O}(\epsilon^2).
\end{equation}
To lowest order in simulation error $\epsilon$, the error in the absorption probability is
\begin{equation}
    |\abs'(\bk) - \abs(\bk)| 
    = 2\frac{\Re[\bra{\epsilon}\hat{n}_\bk\ket{\psi_\mathrm{f}}]}{\bra{\psi_\mathrm{i}} \hat{n}_\bk \ket{\psi_\mathrm{i}}} 
    \leq 2\epsilon \frac{\bra{\psi_\mathrm{f}}\hat{n}_\bk^2\ket{\psi_\mathrm{f}}^{1/2}}{\bra{\psi_\mathrm{i}} \hat{n}_\bk \ket{\psi_\mathrm{i}}} 
    \leq 2\epsilon \frac{\bra{\psi_\mathrm{i}}\hat{n}_\bk^2\ket{\psi_\mathrm{i}}^{1/2}}{\bra{\psi_\mathrm{i}} \hat{n}_\bk \ket{\psi_\mathrm{i}}}.
\end{equation}
The first inequality is understood by noting that $|\epsilon\rangle$ is bounded in norm by $\epsilon$, and the largest 
possible overlap between $\ket{\epsilon}$ and $\hat{n}_\bk\ket{\psi_\mathrm{f}}$ occurs when $\ket{\epsilon}\propto\hat{n}_\bk\ket{\psi_\mathrm{f}}$, i.e., when $\ket{\epsilon}=\epsilon\hat{n}_\bk\ket{\psi_\mathrm{f}}/|\hat{n}_\bk\ket{\psi_\mathrm{f}}|$.
Given the Gaussian initial photonic state of Eq.~\eqref{eq:initial_photonic_state_planar} and the parameters determined earlier in this appendix, within the spectral range of interest we have $\bra{\psi_\mathrm{i}} \hat{n}_\bk \ket{\psi_\mathrm{i}}\geq \maxocc/2$ and $\bra{\psi_\mathrm{i}}\hat{n}_\bk^2\ket{\psi_\mathrm{i}}\leq\maxocc^2$ so that
\begin{equation}
    |\abs'(\bk) - \abs(\bk)| \leq 4\epsilon.
\end{equation}
Note that the final expression for the error is independent of $\maxocc$.

Given that the typical absorption probability we are interested in is $\mathcal{O}(10^{-5}$), it suffices then to take $\epsilon = 2.5\times10^{-7}$ for the Hamiltonian simulation to get accurate results for the absorption probability, and thereby opacity.
In \mbox{App. \ref{subapp:Hamiltonian_simulation_via_interaction}}, we found that most of the costs involved in the time evolution ($T_B(\epsilon)$, $K$, and $\log_2 M$) scale logarithmically in $\epsilon$.
The only non-logarithmic scaling in $\epsilon$ occurs in the Trotterized electronic simulation cost $T_A(t,\epsilon)$, which scales as $\epsilon^{1/\tilde{k}_{\textrm{el}}}$ for $\tilde{k}_{\textrm{el}}$-th order Trotter.
The choice of $\maxocc$ is primarily relevant to observable estimation, as the expectation values required to compute $\abs(\bk)$ are found by sampling multiple simulations with successful absorption processes.

\subsection{Classical numerics for electronic parameters}
\label{subapp:parameters_numerical}

To estimate parameters involving the electronic subsystem (and evaluate state-dependent Trotter bounds in App.~\ref{app:state-dep-trotter}), we performed classical computations based on a spin-degenerate mean-field average-atom implementation similar to Ref.~\onlinecite{callow2022atomec}.
Just like the atomic model used in our quantum simulation protocol, this classical average-atom model considers the electronic degrees of freedom for a single atom.
The nucleus lies at the origin, and the electrons are confined to a radius $\rws = (3/(4\pi \rho_i))^{1/3}$ corresponding to the plasma ion density $\rho_i$.
To make the calculations classically tractable, the electronic state is described by Mermin-Kohn-Sham density functional theory \cite{mermin1965thermal}, where single-particle orbitals $\ket{\phi_\alpha}$ have temperature-dependent Fermi occupations $f_\alpha$ and the electron-electron interaction is approximated by a mean-field effective potential.
Since these classical calculations focus on properties of the electronic subsystem alone, there are no symmetry-breaking electron-photon interactions.
Therefore, we further exploit spherical symmetry to reduce dimensionality.

Specifically, the mean-field average-atom orbitals satisfy Schr\"odinger-like equations
\begin{equation}
\hat{H}_\mathrm{AA}[\eden] \ket{\phi_\alpha} = \epsilon_\alpha \ket{\phi_\alpha},
\end{equation}
where $\epsilon_\alpha$ are single-particle eigenenergies.
The average-atom Hamiltonian
\begin{equation}
	\hat{H}_\mathrm{AA}[\eden]({\bf r}) = -\frac{1}{2} \nabla^2 - \frac{Z}{|{\bf r}|} + V_\mathrm{hxc}[\eden](\bf{r}),
\end{equation}
is a functional of the electron density
\begin{equation}
	\eden({\bf r}) = \sum_\alpha f_\alpha |\phi_\alpha({\bf r})|^2.
\end{equation}
Here, the effective potential $V_\mathrm{hxc}[\eden]({\bf r})$ contains the Hartree potential capturing the classical Coulombic electron-electron interaction and the exchange-correlation potential capturing approximate quantum-mechanical corrections.
We used the local density approximation (LDA) to exchange and correlation \cite{perdew1992accurate} as implemented in LIBXC \cite{lehtola2018recent}.

Under spherical symmetry, the single-particle orbitals factorize as
\begin{equation}
	\phi_\alpha({\bf r}) = R(r)_{n_\alpha}^{\ell_\alpha} \, Y_{\ell_\alpha}^{m_\alpha}(\theta,\varphi),
    \label{eq:aa_dof_separation}
\end{equation}
where $Y_{\ell_\alpha}^{m_\alpha}(\theta,\varphi)$ is a spherical harmonic and $R(r)_{n_\alpha}^{\ell_\alpha}$ is a radial wavefunction.
The indices $n_\alpha$, $\ell_\alpha$, $m_\alpha \in \mathbb{Z}$ are quantum numbers satisfying $0\leq \ell_\alpha < n_\alpha$ and $-\ell_\alpha \leq m_\alpha \leq \ell_\alpha$.
As commonly done in classical computations based on the average-atom model, we assume a spherically symmetric electron density $\eden({\bf r})=\eden(r)$ and only solve for the radial wavefunctions $R_{n_\alpha}^{\ell_\alpha}(r)$, which satisfy the Schr\"odinger-like equation
\begin{equation}
    \hat{H}^{\ell_\alpha}_\mathrm{AA}[\eden] \, R_{n_\alpha}^{\ell_\alpha}(r) = \epsilon_\alpha R_{n_\alpha}^{\ell_\alpha}(r),
\end{equation}
\begin{equation}
    \hat{H}^{\ell_\alpha}_\mathrm{AA}[\eden] = -\frac{1}{2}\frac{d^2}{dr^2} - \frac{Z}{r} + V_\mathrm{hxc}[\eden] + \frac{\ell_\alpha(\ell_\alpha+1)}{r^2}.
\end{equation}
Without loss of generality, we take $R_{n_\alpha}^{\ell_\alpha}$ to be real.
For compatibility with the quantum protocol, where the atom is contained within a much larger photonic cell, we impose a Dirichlet boundary condition at the Wigner-Seitz radius,
\begin{equation}
    R_{n_\alpha}^{\ell_\alpha}(\rws) = 0.
\end{equation}

\subsubsection{Spatial discretization $\Delta_r$ and number of electronic grid points $\ngridel$}
\label{subapp:parameters_Deltar_NG}
\begin{figure}
    \includegraphics{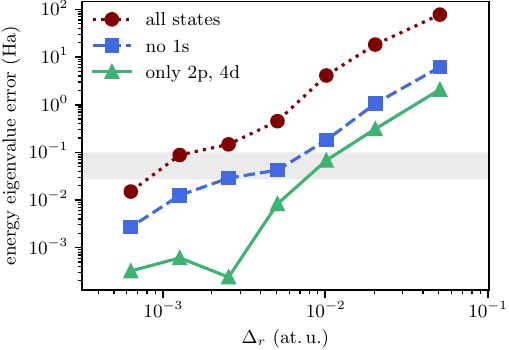}
    \caption{\label{fig:delta}
    Maximum single-particle eigenvalue errors in classical average-atom calculations using different grid spacings $\Delta_r$.
    Gray shading indicates the target precision of 0.03\,--\,0.1 Ha needed to obtain accurate spectral line positions within $\Delta_\lambda=0.01$~\AA.
    When considering all states (red circles), the largest error occurs for the 1s orbital which does not participate in processes within the spectral range of interest.
    When ignoring the 1s eigenvalue (blue squares), the largest error usually occurs for the 2s orbital which does participate.
    However, the dominant opacity features arise from transitions between 2p and 4d orbitals, and the corresponding eigenvalue errors (green triangles) achieve target precision with a coarser grid spacing.
}
\end{figure}

To estimate the spatial discretization $\Delta_r$ required for the electronic subsystem, we evaluated the convergence of the single-particle eigenvalues $\epsilon_\alpha$ with respect to $\Delta_r$ as shown in Fig.~\ref{fig:delta}.
Obtaining spectral lines at accurate wavelengths within precision $\Delta_\lambda$ requires an eigenvalue precision of 
\begin{equation}
    \Delta_\epsilon = \frac{1}{2}\left(\frac{hc}{\lambda-\Delta_\lambda} - \frac{hc}{\lambda+\Delta_\lambda}\right) .
\end{equation}
Given the spectral range and resolution discussed in Section~\ref{subapp:parameters_experimental},
we have $\Delta_\epsilon\approx$ 0.03\,--\,\SI{0.1}{\hartree}.
Achieving this target precision for every single-particle eigenvalue requires $\Delta_r\approx 10^{-3}\SI{}{\au}$.

The 1s orbital dominates the eigenvalue errors because it is highly localized near the nucleus.
However, transitions involving the 1s orbital occur at wavelengths of \SI{2}{\angstrom} or less, far outside the spectral range of interest.
Ignoring the inconsequential errors in the 1s eigenvalue allows coarser discretization with $\Delta_r\approx 5 \times 10^{-3}\SI{}{\au}$.
Further restricting our attention to eigenvalue errors in the 2p and 4d orbitals, which dominate opacity features within the spectral range of interest, allows $\Delta_r\approx 10^{-2}\,\SI{}{\au}$.

Together with the $\Omega$ values from Eq.~\eqref{eq:omega_estimate_cube} and \eqref{eq:omega_estimate_long}, the $\Delta_r$ estimate determines the number of electronic grid points, 
\begin{equation}
    \ngridel = \frac{\Omega}{\Delta_r^3}.
\end{equation}
Since $\ngridel$ enters the qubit count logarithmically, the qubit overhead from extending the electronic grid over the much larger photonic volume is only about a factor of 2.

\subsubsection{Regularization constant $\ereg$}
We determine a physical upper bound on $\ereg$ by assessing its influence on average-atom orbitals and eigenenergies.
Specifically, we evaluate errors incurred by regularizing the electron-ion and Hartree potentials:
\begin{equation}
    V_\mathrm{ion}(r,\ereg) = -\frac{Z}{(r^2+\ereg^2)^{1/2}},
\end{equation}
\begin{align}
    V_\mathrm{h}[\eden](r,\ereg) &= \int d\vec{r}'^3 \frac{\eden(r')}{(|\vec{r}-\vec{r}'|^2+\ereg^2)^{1/2}} \\
    &= \frac{2\pi}{r} \int dr' \eden(r') r' \left[ ((r+r')^2 + \ereg^2)^{1/2} - ((r-r')^2 + \ereg^2)^{1/2}\right],
\end{align}
where $\ereg\rightarrow 0$ approaches the exact Coulomb interaction.
The orbital infidelity is taken as
\begin{equation}
    \epsilon_\mathrm{state}(\ereg) = \max_\alpha \left(1- |\langle\phi_\alpha^{\ereg} | \phi_\alpha)\rangle^2 \right)
    \label{eq:app:reg-state-error}
\end{equation}
and the eigenenergy error is taken as
\begin{equation}
    \epsilon_\mathrm{energy}(\ereg) 
    = \max_\alpha \left| \langle\phi_\alpha^{\ereg} | \hat{H}^{\ell_\alpha}_\mathrm{AA}(\ereg) | \phi_\alpha^{\ereg}\rangle 
    - \langle\phi_\alpha | \hat{H}^{\ell_\alpha}_\mathrm{AA} | \phi_\alpha\rangle \right| .
    \label{eq:app:reg-energy-error}
\end{equation}

\begin{figure}
    \centering
    \includegraphics{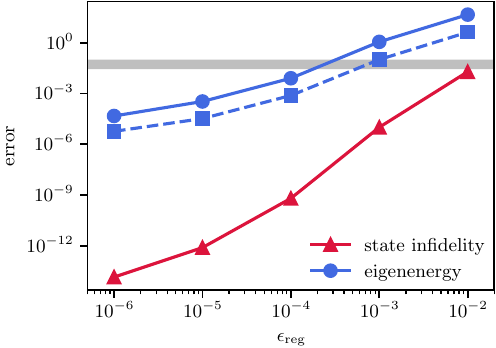}
    \caption{
    Orbital infidelity $\epsilon_\mathrm{state}$ and eigenenergy error $\epsilon_\mathrm{energy}$ (see Eqs.~\eqref{eq:app:reg-state-error} and \eqref{eq:app:reg-energy-error}) due to regularizing the Coulomb interaction as estimated by average-atom calculations.
    The eigenenergy error is given in Hartree, and the gray bar indicates the energy precision required for an iron opacity simulation (see Eq.~\eqref{eq:app:reg-precision}).
    The dashed blue squares assume the $1s$ shell is pseudized and exclude it from the eigenenergy error.
    }
    \label{fig:app:reg-error}
\end{figure}

Given that we are targeting opacities within the $\lambda=$7\,--\,13\AA\ wavelength range, including features on the order of $\Delta\lambda\approx 0.1$\AA\ wide \cite{bailey2015higher}, the eigenenergy error should be within
\begin{equation}
    \Delta E \approx \frac{hc}{\lambda-\Delta\lambda} - \frac{hc}{\lambda} \approx 0.03\text{\,--\,}\SI{0.1}{\hartree} .
    \label{eq:app:reg-precision}
\end{equation}
As shown in Fig. \ref{fig:app:reg-error}, $\ereg\lesssim 10^{-4}$ leaves the average-atom orbitals nearly unchanged and preserves sufficiently accurate eigenenergies relative to the bare, $\ereg=0$ case.
Notably, the $1s$ shell dominates the eigenenergy errors but is not expected to contribute to the opacity within the wavelength range of interest.
Ignoring the $1s$ eigenenergy error allows a more lenient $\ereg\lesssim 10^{-3}$.
Thus, we proceed to use $\ereg= 10^{-3}$ in our resource estimates.

\clearpage
\section{Resource estimates for interaction picture simulation}
\label{app:resources-int}

\renewcommand{\theequation}{E\arabic{equation}}
\renewcommand{\thefigure}{E\arabic{figure}}
\renewcommand{\thetable}{E\arabic{table}}
\setcounter{figure}{0}
\setcounter{table}{0}

At this point we can combine the simulation protocol and associated analytical costs described in App.~\ref{app:quantum_alg_for_opacity} with the parameters determined in App.~\ref{app:parameters_for_solar_iron} to compute the total logical resource cost of computing iron opacity at solar conditions.
We include the state preparation costs from App.~\ref{subapp:initial-state_prep}, the dynamical simulation costs from App.~\ref{subapp:Hamiltonian_simulation_via_interaction}, and the observable estimation overheads from App.~\ref{subapp:observable_estimation}.
As specified in App.~\ref{app:parameters_for_solar_iron}, the cubic simulation cell is prohibitively expensive because of the extremely large number of photonic modes required in that simulation setup.
Thus, we will consider resource estimates only for the elongated simulation cell.
We begin by discussing the total T-gate count and qubit count for the elongated simulation cell, and then go into more granular detail about contributions to the total costs.

There are three independent variables that we can adjust to optimize the resource estimates in our simulation.
The first is $\ngridph$, the number of photonic modes: increasing $\ngridph$ to simulate a wider spectral range increases the interaction picture norm $\alpha_B$ but dramatically decreases the total simulation time because the photons become more spatially localized.
Here we consider two extremes, where $\ngridph=4500$ corresponds to the full spectral range measured in opacity experiments and $\ngridph=40$ corresponds to a single spectral feature (see App.~\ref{subapp:ngamma}).
The second is $\maxocc$, the maximum photonic occupancy: increasing $\maxocc$ increases the total number of photons in the simulation and hence the probability of observing an absorption event, thus reducing the number of shots required for observable estimation, at the cost of increasing $\alpha_B$ with $\maxocc^{1/2}$ scaling.

The third independent variable is $N_\mathrm{wp}$, the number of wave packets. 
Increasing $N_\mathrm{wp}$ while keeping $\maxocc$ fixed strictly increases costs, since it does not increase the total number of photons in the elongated simulation setup but only spreads them out in space and thereby increases the simulation time.
However, increasing $N_\mathrm{wp}$ under fixed $\maxocc$ may carry a physical advantage by decreasing the power incident on the atom and enabling control of any nonlinear optical effects.
Instead, we consider the cost-reducing strategy of increasing $N_\mathrm{wp}$ while keeping $\maxocc/N_\mathrm{wp}$ fixed, which increases the number of photons and decreases the number of shots required at fixed incident power.
Notably, $N_\mathrm{wp}$ is limited by the spatial width of the wave packets, and the single feature case ($\ngridph=40$) only accommodates at most 22 spatially delocalized wave packets (see App.~\ref{subapp:paramters_Nwp}).

\begin{figure}[b]
    \centering
    \includegraphics[width=0.8\linewidth]{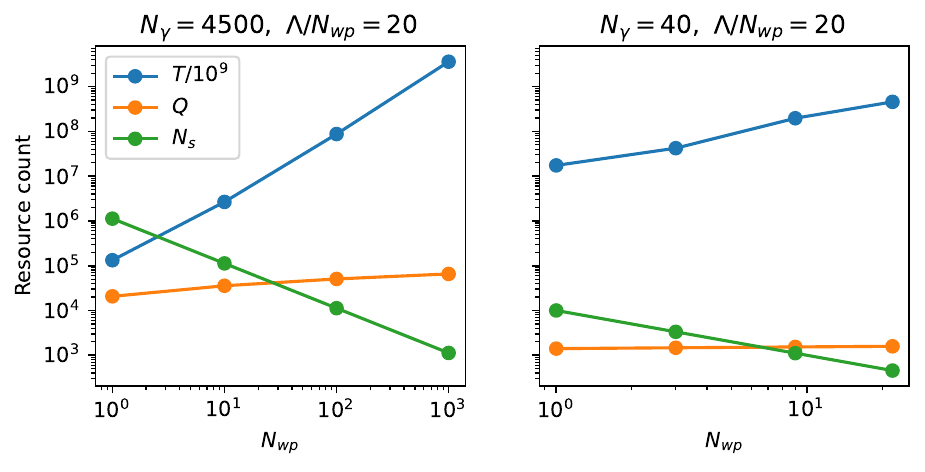}
    \caption{Total logical resource costs of the all-electron simulation of iron opacity for the full spectrum (left panel) and one spectral feature (right panel).
    Shown are the T-gate counts in billions ($T$) including state preparation and interaction picture dynamics for a single shot, the logical qubit counts ($Q$), and the number of shots required for observable estimation ($\ns$).
    The independent variable is the number of wave packets in the simulation $N_\textrm{wp}$, while the number of photonic modes $\ngridph$ and the maximum photonic occupancy per wave packet $\maxocc/N_\mathrm{wp}$ are held fixed as labeled.} 
    \label{fig:fig1-app}
\end{figure}

Fig.~\ref{fig:fig1-app} shows the T-gate counts, logical qubit counts, and total shot counts for the simulation with a fixed $\maxocc/N_\textrm{wp} = 20$ while varying $N_\textrm{wp}$.
Both cases share some qualitative behavior, namely that the simulation time increases and hence the T gate cost increases with $N_\textrm{wp}$, and that the number of shots required decreases.
Contrasting the two cases, we see a clear tradeoff between resources.
The single feature simulation with $\ngridph = 40$ requires fewer qubits and fewer shots in total, but its T-gate cost is about two orders of magnitude greater compared to the full spectrum $\ngridph=4500$ case.
The lower qubit counts are primarily due to a smaller photonic register, the dominant memory cost, whereas the increased T counts are due to the longer simulation time required by the larger spatial width of the wave packets with constrained momentum.
Meanwhile, the reduction in the number of shots is simply proportional to the reduction in the number of spectral features, where we assume that classical pre-processing can adequately identify electronic configurations contributing to the selected opacity feature.

\begin{figure}
    \centering
    \includegraphics[width=0.8\linewidth]{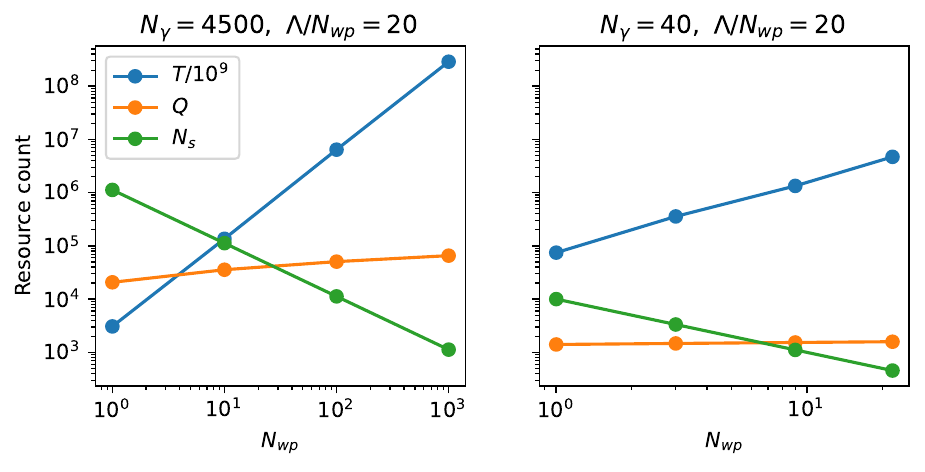}
    \caption{Total logical resource costs of the iron opacity simulation with pseudized 1s electrons for the full spectrum (left panel) and one spectral feature (right panel).
    Shown are the T-gate counts in billions ($T$) including state preparation and interaction picture dynamics for a single shot, the logical qubit counts ($Q$), and the number of shots required for observable estimation ($\ns$).
    The independent variable is the number of wave packets in the simulation $N_\textrm{wp}$, while the number of photonic modes $\ngridph$ and the maximum photonic occupancy per wave packet $\maxocc/N_\mathrm{wp}$ are held fixed as labeled.} 
    \label{fig:fig2-app}
\end{figure}

In Fig.~\ref{fig:fig2-app}, we consider the same parameters as Fig.~\ref{fig:fig1-app}, but with the K-shell electrons pseudized according to the approach described in App.~\ref{app:pseudization}.
The K-shell does not participate in the processes determining iron opacity at the thermodynamic conditions of interest, but it increases simulation costs by requiring higher spatial and temporal resolutions for the electronic subsystem.
The primary benefit of pseudization is an increase in the grid spacing, as described in App.~\ref{subapp:parameters_Deltar_NG}, from $\Delta_r = 10^{-3}$ for the all-electron case to $\Delta_r = 10^{-2}$ when the $1s$ electrons are pseudized.
The larger $\Delta_r$ reduces the total number of T gates by 2\,--\,3 orders of magnitude without appreciably changing the number of qubits or the number of shots.

\begin{figure}
    \centering
    \includegraphics[width=0.5\columnwidth]{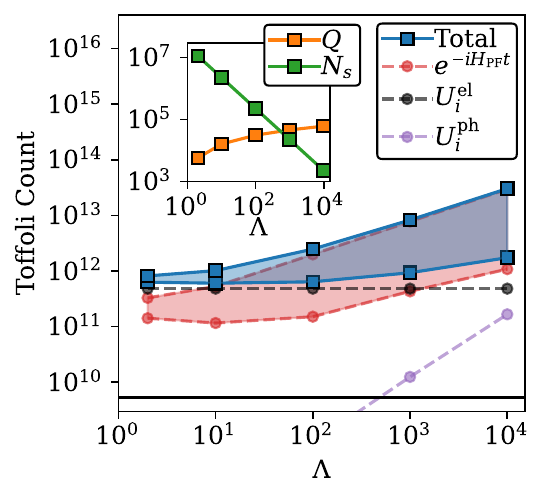}
    \caption{\textit{Reproduced from main text, Fig.~\ref{fig:full_cost}}: Logical resource estimates for the opacity simulation of solar iron using the interaction picture algorithm with pseudized K-shell electrons. 
    Total Toffoli counts are shown for a single shot of the full simulation as a function of the maximum photonic occupancy ($\maxocc$). 
    The total simulation cost is decomposed into contributions from the simulation cost of the Pauli-Fierz Hamiltonian ($e^{-i\hpf t}$) and the state preparation costs for the electronic ($U_\mathrm{i}^\mathrm{el}$) and photonic ($U_\mathrm{i}^\mathrm{ph}$) subsystems. 
    Shading indicates ranges between pessimistic and optimistic bounds involved in the Hamiltonian simulation cost.
    The horizontal solid black line shows the simulation cost of only the electronic part of the Hamiltonian ($e^{-i\hel t}$) for reference. 
    The inset shows total qubit counts $(Q)$ and the number of shots required $(\ns)$, including observable estimation and thermal sampling overheads.
    \label{fig:fig3-app}}
\end{figure}

Moving forward, we will examine the $\ngridph = 4500$ and $N_\textrm{wp} =1$ case, which minimizes circuit depth per shot, shown in Fig.~\ref{fig:fig3-app}, a reproduction of Fig.~\ref{fig:full_cost} in the main text.
The number of qubits stays relatively constant as $\maxocc$ increases due to the logarithmic scaling.
Meanwhile, the number of shots rapidly decreases for higher $\maxocc$, with a three order of magnitude reduction over the range of $\maxocc$ shown.
The shaded range of T counts corresponds to different choices for the minimum relevant momentum, $\min|\bk|$. 
The higher curve corresponds to the strict upper bound of $\min|\bk| = \Delta_k$ given by the spectral resolution, while the lower curve corresponds to a more optimistic bound of $\min|\bk| = k_\textrm{min}^\textrm{targ}$ given by the targeted spectral range.
Although the Gaussian wave packet shown in Fig.~\ref{fig:photon_truncation} has support over momentum modes with $|\bk|=\Delta_k$, the weight at this strict minimum is about $10^3$ times smaller than at $k_\textrm{min}^\textrm{targ}$.
Thus, choosing $\min|\bk| = k_\textrm{min}^\textrm{targ}$ effectively offers a state-dependent estimate, and we expect that the true simulation cost is closer to the lower curve.

The apparent divergence between the two bounds for the total T count as $\maxocc$ increases can be understood through the component-wise decomposition of the total T gate cost shown in the right panel of Fig.~\ref{fig:fig3-app}.
The value of $\min|\bk|$ influences the T gate cost of Hamiltonian simulation, but not state preparation.
For $\maxocc \leq 10^3$, the lower estimate for the total cost is dominated by the state preparation cost, which remains nearly constant in this regime because the dominant electronic contribution does not depend on $\maxocc$.
Meanwhile the upper estimate is dominated by the cost of simulating dynamics, which scales as $\maxocc^{1/2}$.
For $\maxocc > 10^3$, the lower estimate becomes dominated by the dynamics as well. 
This behavior explains the divergence in the total T gate cost between the two values of $\min|\bk|$: the upper estimate is dominated by the polynomial scaling dynamics cost, whereas the lower estimate is dominated by the nearly constant state preparation cost.

\clearpage
\section{Pseudization of 1s electrons}
\label{app:pseudization}

\renewcommand{\theequation}{F\arabic{equation}}
\renewcommand{\thefigure}{F\arabic{figure}}
\renewcommand{\thetable}{F\arabic{table}}
\setcounter{figure}{0}
\setcounter{table}{0}

The general form for a non-local pseudopotential is
\begin{equation}
    \hat{V}_\mathrm{pp} = \hat{V}_\mathrm{loc}(|\br|) + \sum_{\ell m} \hat{P}_{\ell m}  \hat{V}_\mathrm{non}^{\ell m}(|\br|), \qquad\hat{P}_{\ell m} \equiv |Y_\ell^m\rangle \langle Y_\ell^m|. \label{eq:nonlocal_defn}
\end{equation}
Here $\hat{V}_\mathrm{loc}, \hat{V}_\mathrm{non}^{\ell m}$ are radial operators for the local and non-local part of the pseudopotential and $\hat{P}_{\ell m}$ are projectors onto the $Y_{\ell}^{m}$ angular momentum channels.
Our objective is to implement the exponential of this operator.
In doing so, we first note that
\begin{equation}
\left[ \hat{V}_\mathrm{loc},\;\hat{P}_{\ell m}  \hat{V}_\mathrm{non}^{\ell m}\right] = \left[\hat{V}_\mathrm{loc}, \hat{V}_\mathrm{non}^{\ell m} \right] \hat{P}_{\ell m}  = 0,
\end{equation}
and that the commutator between different $\ell m$ channels trivially vanishes.
As such, $\exp(i\hat{V}_\mathrm{pp})$ (suppressing time for simplicity) is simply a product of exponentials of each term in Eq.~\ref{eq:nonlocal_defn}.
The exponential of the non-local terms $\hat{U}_\mathrm{non}$ can be expressed as
\begin{equation}
\hat{U}_\mathrm{non} =e^{i\sum_{\ell m}\hat{P}_{\ell m}  \hat{V}_\mathrm{non}^{\ell m}(r) } = \prod_{\ell m} \left(\id + \hat{P}_{\ell m}  \sum_{k=1}^{\infty}\frac{\left(i\hat{V}^{\ell m}_\mathrm{non}\right)^k}{k!}\right) = \prod_{\ell m} \left((\id - \hat{P}_{\ell m}) + \hat{P}_{\ell m} e^{i\hat{V}^{\ell m}_\mathrm{non}}\right) = \prod_{\ell m} \hat{U}_\textrm{non}^{\ell m}.
\end{equation}

Each $\hat{U}_\textrm{non}^{\ell m}$ can be implemented using two primitives: the exponential unitary $\hat{E}^{\ell m} = e^{i\hat{V}^{\ell m}_\mathrm{non}}$ and a block encoding of the projector $\hat{W}^{\ell m}$ such that $\langle 0^a|\hat{W}^{\ell m}|0^a\rangle = \hat{P}_{\ell m}$.
The unitary $\hat{U}_\textrm{non}^{\ell m}$ is then achieved by sandwiching an $a$-qubit controlled $\hat{E}^{\ell m}$ by a compute and uncompute pair of $(\hat{W}^{\ell m})^\dagger$ and $\hat{W}^{\ell m}$.
We will assume that $\hat{E}^{\ell m}$ is easy to implement, as it is a radial 1-body function,  utilizing a similar routine to the 1-body Coulomb potential computed for the interaction picture simulation, and so we will focus on constructing efficient block-encodings $\hat{W}^{\ell m}$.

Since we are working with a Cartesian real-space grid, whereas the projector operators are over spherical harmonics, we will begin by defining shells of width $\Delta_b$ and a shell index for a given Cartesian grid point
\begin{equation}
b(\br) \equiv \lfloor|\br|/\Delta_b \rfloor.
\end{equation}
Each shell will then constitute an approximation to a radial grid at radius $r = b \Delta_b$ containing $n_{b(\br)}$ points. 
This approximation will improve as the grid spacing $\Delta_r$ and the shell spacing $\Delta_b$ trend to zero with $\Delta_r/\Delta_b \gg 1$.
The projection operator $\hat{P}^{\ell m}$ can be approximated over this shell construction, and has matrix element values
\begin{equation}
\langle \br|\hat{P}^{\ell m}|\br'\rangle \approx \frac{\delta_{b(\br), b(\br')}}{n_{b(\br)}}\times 4\pi Y_\ell^m(\theta,\varphi)\bar{Y}_\ell^m(\theta', \varphi').
\end{equation}
The only non-zero matrix elements are when $\br, \br'$ are within the same shell, and for generic $\ell,m$ there is variation in the precise matrix element value depending on the angular components of $\br, \br'$.
The matrix elements significantly simplify in the case of interest in this work, $\ell = m = 0$, wherein the angular contributions vanish
\begin{equation}
\langle \br|\hat{P}^{00}|\br'\rangle \approx \frac{\delta_{b(\br), b(\br')}}{n_{b(\br)}}.
\end{equation}
Importantly, we see here that the operator only requires knowing whether $\br, \br'$ are in the \textit{same shell} without any further need for angular data, which will greatly simplify the logic needed to implement the projector.

Due to the suppression of the angular components, we can construct an algorithm for the block-encoding of $\hat{P}^{00}$ within a first-quantized approach efficiently.
We first note that the projector can be written as a linear combination of shell-projectors
\begin{equation}
\hat{P}^{00} = \sum_{b=0}^{B-1}|\psi_b\rangle \langle \psi_b|, \qquad |\psi_b \rangle \equiv \frac{1}{\sqrt{n_b}}\sum_{\br \in b}|\br\rangle.
\end{equation}
This can be written in LCU format as:
\begin{equation}
\hat{P}^{00} = \sum_{b=0}^{B-1} \frac{1}{2}\id  + \frac{1}{2}\hat{R}_b, \qquad \hat{R}_b \equiv 2|\psi_b\rangle \langle \psi_b| - \id,
\end{equation}
where $\hat{R}_b$ are reflectors on each shell.

We can then compute this using the usual PREPARE and SELECT circuits and $1 + \log_2B$  ancillae qubits.
The PREPARE circuit acts as
\begin{equation}
\textrm{PREPARE}|0\rangle\otimes |0^{\log_2B}\rangle = \frac{1}{\sqrt{2}}(|0\rangle + |1\rangle) \otimes \frac{1}{\sqrt{B}} \sum_{b=0}^{B-1}|b\rangle,
\end{equation}
which is implemented using $1 + \log_2 B$ Hadamard gates.
The SELECT circuit is written as
\begin{equation}
\textrm{SELECT} = |0\rangle \langle 0| \otimes I_b \otimes I + |1\rangle \langle 1| \otimes \sum_{b=0}^{B-1} \left(|b\rangle \langle b|\otimes \hat{R}_b \right).
\end{equation}
This can be implemented via the circuit $\hat{R}$ defined below controlled on the single remaining ancilla qubit
\begin{equation}
\hat{R} \equiv \sum_{b=0}^{B-1} |b\rangle \langle b|\otimes \hat{R}_b.
\end{equation}

The operator $\hat{R}$ consists of reflectors $\hat{R}_b$ controlled on a target shell $b$ specified by the ancilla register $|b\rangle$, and can be implemented as such.
Starting with the system register $|\br\rangle$ we compute the square distance $\br^2$ in an ancilla.
Comparison logic is then carried out to determine whether $(b\Delta_b)^2 \leq |\br|^2 \leq ((b+1)\Delta b)^2$, which flips an ancilla flag qubit if True.
As such, if we initialize the ancilla flag qubit in the $|-\rangle$ state, the operations above would flip the phase of the state if True, and keep it fixed if False, implementing the reflector $\hat{R}_b$.
This sequence of operations is dominated by the cost of squaring over the $\log_2 \ngridel$ qubit electronic register of a single electron, which takes $(\log_2 \ngridel)^2$ Toffolis to compute, and the ancillas utilized in squaring can be borrowed from those required in the electronic structure simulation.

A single controlled-$\hat{R}$ operation is required per application of $\hat{P}^{00}$, two of which are required to implement $\hat{U}^{00}_\textrm{non}$ for a single electron.
This process needs to be carried out for each electron, and hence the cost of implementing the non-local pseudopotential with $\ell =m = 0$ is dominated by $2\nel$ calls to a controlled-$\hat{R}$ gate.
In our problem $\ngridel = 1.3 \times 10^{13}$ and so each $\hat{R}$ call requires $\sim 1900$ Toffoli gates, giving a total of $\sim 10^5$ Toffolis for each call of $U_\textrm{non}^{00}$.
When compared to the dominant cost of the electronic simulation, the 2-body Coulomb term, in Eq.~\eqref{eq:ham-el-resources}, the pseudopotential calculation is quadratically better in $\nel$ and carries smaller prefactors, such that the pseudopotential calculation is about 10 times cheaper than the 2-body Coulomb implementation.
In this way, we find that the pseudopotential calculation is about the same cost as implementing the bare 1-body Coulomb term, both of which result in a second digit correction to the total resource estimates for the Hamiltonian simulation of $\hel$.

The simplicity of the construction for $\ell = 0$ relies on the fact that $|\psi_b\rangle $ is a uniform superposition over the states in the shell.
As such, reflection about this state is mathematically equivalent to checking whether a basis state is in the shell.
When we take an arbitrary $\ell \neq 0$, the same formulation above can be utilized except that $\hat{R}_b$ now reflects about the state
\begin{equation}
|\psi_b^{\ell m}\rangle \equiv \frac{1}{\sqrt{n_b}} \sum_{\br \in b}Y_\ell^m(\theta, \varphi)|\br\rangle.
\end{equation}
It is thereby necessary in this case to be able to prepare such a state efficiently, which generally requires sophisticated techniques to retain poly-logarithmic scaling in $\ngridel$~\cite{Berry_2024}.

\clearpage
\section{Resource estimates for fully Trotterized simulation}
\label{app:trotterization}

\renewcommand{\theequation}{G\arabic{equation}}
\renewcommand{\thefigure}{G\arabic{figure}}
\renewcommand{\thetable}{G\arabic{table}}
\setcounter{figure}{0}
\setcounter{table}{0}

Although we do not utilize these in our main results, we have carried out extensive analysis of fully Trotterized simulations, wherein instead of using the interaction picture algorithm to incorporate the interaction term, we instead use Trotter products.
We also carry out state-dependent analysis.
We first give a brief overview of Trotter-based methods, then we provide details on the Trotterization of $\hat{H}_\textrm{PF}^\textrm{AA}$, and conclude with logical T-gate and qubit counts for the Hamiltonian simulation algorithm.

\subsection{Review of product formulas}
Trotter formulas approximate the dynamics from a given Hamiltonian $\hat{H}$ via a $\tilde{k}$-th order product formula.
Specifically, let $\hat{H} = \sum_{\gamma=1}^{\Gamma} \hat{H}_\gamma$.
One can implement a $\tilde{k}$th-order product formula to approximate the dynamics of $\hat{H}$ which satisfies an asymptotic error scaling:
\begin{equation}
\mathcal{S}_{\tilde{k}}(t) = e^{-it\hat{H}} + O(t^{p+1}).
\end{equation}
Explicit forms of the product formulae can be written recursively \cite{childs2021}:
\begin{equation}
\begin{split}
&\mathcal{S}_1(t) = e^{-it\hat{H}_\Gamma} \ldots e^{-it\hat{H}_2}e^{-it\hat{H}_1} \\
&\mathcal{S}_2(t) = e^{-i(t/2)\hat{H}_1} \ldots e^{-i(t/2)\hat{H}_\Gamma}e^{-i(t/2)\hat{H}_\Gamma}...e^{-i(t/2)\hat{H}_1} \\
&\mathcal{S}_{2\tilde{k}}(t) = \mathcal{S}^2_{2\tilde{k}-2}(u_{\tilde{k}} t)S_{2\tilde{k}-2}((1 - 4u_{\tilde{k}}) t)S^2_{2\tilde{k}-2}(u_{\tilde{k}} t),
\label{eq:trotter_formulas}
\end{split}
\end{equation}
where $u_{\tilde{k}} = 1/(4 - 4^{1/(2\tilde{k}-1)}).$
In most cases the simulation time is not small, and so the asymptotic error may be large.
To ameliorate this, dynamics are simulated by breaking the simulation into $\tilde{r}$ steps of time $t/\tilde{r} \ll 1$, such that:
\begin{equation}
\left|\left|e^{-it\hat{H}} - \Big(\mathcal{S}_{\tilde{k}}(t/\tilde{r})\Big)^{\tilde{r}} \right|\right| = O(t^{\tilde{k}+1}/\tilde{r}^{\tilde{k}}).
\end{equation}

To compute resource estimates, we will need to replace the asymptotic scaling on the right hand side by an explicit upper bound, which is accomplished by the introduction of a Trotter bound $\mathcal{T}_{\tilde{k}}$ satisfying:
\begin{equation}
   \left|\left|e^{-it\hat{H}} - \Big(\mathcal{S}_{\tilde{k}}(t/\tilde{r})\Big)^{\tilde{r}} \right|\right| \leq \mathcal{T}_{\tilde{k}} (t^{\tilde{k}+1}/\tilde{r}^{\tilde{k}}).
   \label{eq:trotter_formula}
\end{equation}
A total simulation error $\epsilon$ requires a number of steps
\begin{equation}
\tilde{r} \geq \left(\frac{\mathcal{T}_{\tilde{k}}}{\epsilon}\right)^{1/{\tilde{k}}} t^{\tilde{k} + 1/{\tilde{k}}}. 
\label{eq:trotter_steps}
\end{equation}
Computing a tight bound $\mathcal{T}_{\tilde{k}}$ is thereby a primary task in efficient Trotterized simulation, with established theoretical results based on operator norms and nested commutators \cite{childs2021} as well as through explicit numerical simulations \cite{rubin2024quantum}; Trotter bounds can even be state dependent \cite{burgarth2023state}.

In this work we will only consider first- and second-order Trotter products with two non-commuting terms.
Within this restricted set of Hamiltonians there are known tight Trotter bounds with small prefactors \cite{childs2021} which we make use of.
For first-order Trotterization, we have the expression for a generic Hamiltonian:
\begin{equation}
\left|\left|\mathcal{S}_1(t) - e^{-it\hat{H}}\right|\right| \leq \frac{t^2}{2} \sum_{\gamma_1 = 1}^{\Gamma} \left|\left| \sum_{\gamma_2 = \gamma_1 + 1}^{\Gamma} [\hat{H}_{\gamma_2}, \hat{H}_{\gamma_1}] \right|\right|.
\end{equation}
This yields a Trotter bound when $\Gamma = 2$:
\begin{equation}
\mathcal{T}_1 \leq \frac{1}{2} \Big| \Big| [\hat{H}_1, \hat{H}_2] \Big| \Big|.
\label{eq:first-order-trotter}
\end{equation}
For second-order Trotterization, we have the expression:
\begin{align}
\left|\left|\mathcal{S}_2(t) - e^{-it\hat{H}}\right|\right| \leq \frac{t^3}{12} & \sum_{\gamma_1 = 1}^{\Gamma} \left|\left| \sum_{\gamma_3 = \gamma_1 + 1}^\Gamma \left[ \hat{H}_{\gamma_3}, \sum_{\gamma_2 = \gamma_1 + 1}^\Gamma [\hat{H}_{\gamma_2}, \hat{H}_{\gamma_1}] \right]\right|\right| + \nonumber \\
\frac{t^3}{24} & \sum_{\gamma_1 = 1}^{\Gamma} \left|\left| \left[ \hat{H}_{\gamma_1}, \left[ \hat{H}_{\gamma_1}, \sum_{\gamma_2 = \gamma_1 + 1}^\Gamma \hat{H}_{\gamma_2} \right] \right]\right|\right|.
\end{align}
This yields a Trotter bound for $\Gamma = 2$:
\begin{equation}
\mathcal{T}_2 \leq \frac{1}{24} \Big|\Big| [\hat{H}_1 + 2\hat{H}_2, [\hat{H}_1, \hat{H}_2]] \Big| \Big|.
\label{eq:second-order-trotter}
\end{equation}

\subsection{Trotter formulae for the average-atom Pauli-Fierz Hamiltonian}
We will now construct explicit formulae for first- and second-order Trotter bounds for the average-atom Pauli-Fierz Hamiltonian.
Our first objective is to decompose $\hat{H}_\textrm{PF}^\textrm{AA}$ into the primitive terms $\hat{H}_\gamma$.
We choose the following decomposition:
\begin{equation}
\begin{split}
&\hat{H}_\textrm{PF}^\textrm{AA} = \hat{H}_1 + \hat{H}_2 = \left(\hel + \hph\right) + \helph \\
& \hel = \sum_{j=1}^{\nel} \left(\frac{\bpo{j}^2}{2} - \frac{Z}{(|\bro{j}|^2 + \ereg^2)^{1/2}}\right) +  \sum_{j = 1}^{\nel}\sum_{k \neq j}^{\nel} \frac{1}{2}\frac{1}{(|\bro{j} - \bro{k}|^2 + \ereg^2)^{1/2}} \\
& \hph = \frac{1}{2} \int d^3 \vec{r} [\hat{\mathbf{E}}(\vec{r})^2 + \Bfld(\vec{r})^2] \\
& \helph = -\frac{1}{2c} \sum_{j=1}^{\nel}\Big( \bpo{j} \cdot \Apot(\bro{j})  + \Apot(\bro{j}) \cdot \bpo{j} + \spin{j} \cdot \Bfld(\bro{j}) - \frac{2}{c} \Apot(\bro{j})^2\Big).
\label{eq:trotter_split}
\end{split}
\end{equation}
Since $\hel, \hph$ commute, they can be included into a single term $\hat{H}_1$, with the only non-commuting term being $\hat{H}_2 = \helph.$
Note that the electronic term $\hel$ has been altered relative to its bare form following a form proposed in prior literature \cite{Childs2022quantumsimulationof}.
This regularization removes divergences which occur when electrons of opposite spin occupy the same grid point in the real-space discretization utilized in our simulation.
In App.~\ref{app:state-dep-trotter} we show that $\ereg=10^{-3}$ is sufficiently small as to avoid significantly distorting realistic single particle orbitals or associated energies.

In App.~\ref{app:trotterbounds} we carry out a lengthy derivation to get the dominant contributions to the commutators in the Trotter bounds for the Pauli-Fierz Hamiltonian.
We find:
\begin{align}
& [\hat{H}_1, \hat{H}_2] \cong \frac{\nel}{c}\sum_{l=1}^{\nel}  \vec{\mathcal{E}}(\bro{l}) \cdot \Apot(\bro{l}) + \frac{1}{c} \sum_{l=1}^{\nel}\sum_{j > l}   \vec{\mathcal{E}}(\bro{l}-\bro{j}) \cdot \left(\Apot(\bro{l}) - \Apot(\bro{j})\right)~\text{and} \\
&[\hat{H}_1 + 2\hat{H}_2, [\hat{H}_1, \hat{H}_2]] \approx  \frac{\nel}{c}\sum_{l=1}^{\nel} \vec{\mathcal{D}}(\bpo{l}, \bro{l}) \cdot \Apot(\bro{l}) + \frac{1}{c} \sum_{l=1}^{\nel}\sum_{j > l}  \vec{\mathcal{D}}(\bpo{l}, \bro{l}-\bro{j}) \cdot \left(\Apot(\bro{l}) - \Apot(\bro{j})\right),
\end{align}
where 
\begin{align}
\vec{\mathcal{E}}(\vec{r}) \equiv \frac{\vec{r}}{(|\vec{r}|^2 + \ereg^2)^{3/2}}~\text{and} \ \vec{\mathcal{D}}(\vec{p}, \vec{r}) \equiv \frac{1}{2}\left( (\vec{p} - 3(\vec{p}\cdot\hat{r})\hat{r}) \, \frac{1}{(|\vec{r}|^2 + \epsilon^{2}_\textrm{reg})^{3/2}} + \frac{1}{(|\vec{r}|^2 + \epsilon^{2}_\textrm{reg})^{3/2}} \, (\vec{p} - 3\hat{r}(\hat{r}\cdot\vec{p})) \right).
\label{eq:e-d-defn}
\end{align}
Note the conceptual similarity of $\vec{\mathcal{E}}, \vec{\mathcal{D}}$ to electric monopole and dipole moment operators.
As such, the first- and second-order Trotter bounds in Eqs.~\eqref{eq:first-order-trotter}, ~\eqref{eq:second-order-trotter} for the Pauli-Fierz Hamiltonian can be written as:
\begin{align}
&\mathcal{T}_1 \leq \frac{||\Apot||}{2}(e_1 + e_2), \ e_1 \equiv \frac{\nel}{c} \left|\left|\sum_{l=1}^{\nel} \vec{\mathcal{E}}(\bro{l})\right|\right|, \  e_2 \equiv \frac{1}{c}\sum_{l=1}^{\nel}\sum_{j>l}\left|\left|\vec{\mathcal{E}}(\bro{l}-\bro{j}) \right|\right|~\text{and}\\
&\mathcal{T}_2 \leq \frac{||\Apot||}{24}(d_1 + d_2), \ d_1
\equiv  \frac{\nel}{c}\sum_{l=1}^{\nel} \Big| \Big| \vec{\mathcal{D}}(\bpo{l}, \bro{l})\Big|\Big|, \ d_2 \equiv \frac{1}{c} \sum_{l=1}^{\nel}\sum_{j>l} \Big| \Big| \vec{\mathcal{D}}(\bpo{l},\bro{l}-\bro{j}) \Big|\Big|.
\label{eq:trotter-formulae}
\end{align}

We also find in App.~\ref{app:trotterbounds} that expressions for state-dependent Trotter bounds are a simple generalization of the state-independent formulae specifically due to the low absorption probability of photons in the opacity simulation:
\begin{equation}
\mathcal{T}_{1,\psi_\mathrm{i}} \leq \frac{1}{2} \Big| \Big| [\hat{H}_1, \hat{H}_2]|\psi_\mathrm{i}\rangle \Big| \Big|,\ \mathcal{T}_{2,\psi_\mathrm{i}} \leq \frac{1}{24} \Big|\Big| [\hat{H}_1 + 2\hat{H}_2, [\hat{H}_1, \hat{H}_2]] |\psi_\mathrm{i}\rangle \Big| \Big|,
\end{equation}
where $|\psi_\mathrm{i}\rangle$ is the initial state the simulation begins in.
As such, the equations~\eqref{eq:trotter-formulae} generalize simply where $e_1, e_2, d_1, d_2$ now are not just operator norms, but norms of a new state with the operator acting on $|\psi_\mathrm{i}\rangle$:
\begin{align}
&\mathcal{T}_{1,\psi_\mathrm{i}} \leq \frac{||\Apot||}{2}(e_{1,\psi_\mathrm{i}} + e_{2,\psi_\mathrm{i}}), \ e_{1,\psi_\mathrm{i}} \equiv \frac{\nel}{c} \left|\sum_{l=1}^{\nel} \vec{\mathcal{E}}(\bro{l})|\psi_\mathrm{i}\rangle \right|, \  e_{2,\psi_\mathrm{i}} \equiv \frac{1}{c}\sum_{l=1}^{\nel}\sum_{j>l}\left|\vec{\mathcal{E}}(\bro{l}-\bro{j}) |\psi_\mathrm{i}\rangle \right|, \\
&\mathcal{T}_{2,\psi_\mathrm{i}} \leq \frac{||\Apot||}{24}(d_{1,\psi_\mathrm{i}} + d_{2,\psi_\mathrm{i}}), \ d_{1,\psi_\mathrm{i}}
\equiv  \frac{\nel}{c}\sum_{l=1}^{\nel} \Big| \vec{\mathcal{D}}(\bpo{l}, \bro{l})|\psi_\mathrm{i}\rangle \Big|, \ d_{2,\psi_\mathrm{i}} \equiv \frac{1}{c} \sum_{l=1}^{\nel}\sum_{j>l} \Big| \vec{\mathcal{D}}(\bpo{l},\bro{l}-\bro{j})|\psi_\mathrm{i}\rangle \Big|.
\label{eq:trotter-formulae-statedep}
\end{align}
We will typically supress the index $\psi_\mathrm{i}$ when it is clear from context we are using the state-dependent norms.

\subsection{First- and second-order Trotter formulae for $\hat{H}_\textrm{PF}^{\textrm{AA}}$ \label{app:trotterbounds}}
\subsubsection{Analytic first-order formulae}
\label{app:trotterbounds-firstorder}
\paragraph{$[\hel, \helph]$}
\label{sec:commutator_hel_helph}
For the electronic term, we start with the contributions of the potential terms in $\hel.$
The potential energy commutes with the spin-dependent part of $\helph$, leaving only the momentum-dependent part.
The full commutator then takes the form:
\begin{equation}
[\hat{V}_{el}, \helph] = \left[-\sum_{j = 1}^{\nel} \frac{Z}{(|\bro{j}|^2 + \ereg^2)^{1/2}} + \sum_{j = 1}^{\nel} \sum_{k \neq j}^{\nel} \frac{1}{2}\frac{1}{(|\bro{j} - \bro{k}|^2 + \ereg^2)^{1/2}}, -\frac{1}{2c}\sum_{l = 1}^{\nel} \left(\bpo{l}\cdot \Apot(\bro{l}) + \Apot(\bro{l})\cdot \bpo{l}\right)\right],
\end{equation}
which we can write more neatly as:
\begin{equation}
[\hat{V}_{el}, \helph] = -\frac{1}{2c}\sum_{l = 1}^{\nel} (\vec{C}_l\cdot \Apot(\bro{l}) + \Apot(\bro{l})\cdot \vec{C}_l).
\end{equation}
where
\begin{equation}
\vec{C}_l\equiv \left[-\sum_{j = 1}^{\nel} \frac{Z}{(|\bro{j}|^2 + \ereg^2)^{1/2}} + \sum_{j = 1}^{\nel} \sum_{k \neq j}^{\nel} \frac{1}{2}\frac{1}{(|\bro{j} - \bro{k}|^2 + \ereg^2)^{1/2}}, \bpo{l}\right].
\end{equation}
Inserting the identity $\bpo{l} = -i\nabla_l$, we can reduce $\vec{C}_l$ as
\begin{equation}
\vec{C}_l = i\left(Z\frac{\bro{l}}{(|\bro{l}|^2 + \ereg^2)^{3/2}} - \sum_{j\neq \ell} \frac{\bro{j} - \bro{l}}{(|\bro{j} - \bro{l}|^2 + \ereg^2)^{3/2}}\right).
\end{equation}
Inserting this into our above expression for the commutator, we find
\begin{equation}
[\hat{V}_{el}, \helph] = -\frac{i}{c}\sum_{l = 1}^{\nel} \left(Z\frac{\bro{l}}{(|\bro{j}|^2 + \ereg^2)^{3/2}} - \sum_{j\neq l} \frac{\bro{j} - \bro{l}}{(|\bro{j} - \bro{l}|^2 + \ereg^2)^{3/2}}\right)\cdot \Apot(\bro{l}).
\label{eq:commutator_el_elph}
\end{equation}
Note that we do not need to keep track of the non-commutation of $\vec{C}_l, \Apot(\bro{l})$ since both are functions of position only.

As for the kinetic energy term, we have four terms in the commutator:
\begin{equation}
[\hat{T}_{el},\helph] = -\frac{1}{4c} \sum_{l=1}^{\nel} \left[\bpo{l}^2, \bpo{l} \cdot \Apot(\bro{l}) + \Apot(\bro{l}) \cdot \bpo{l} + \spin{j}\cdot \Bfld(\bro{l}) - \frac{2}{c}\Apot(\bro{j})^2 \right].
\end{equation}
We can drop the last term as we know it is an order of magnitude smaller than the rest due to the $c^{-1}$ term, and the remaining terms can be computed as:
\begin{equation}
[\hat{T}_{el},\helph] = \frac{1}{4c}\sum_{l=1}^{\nel} \left[i\vec{\nabla}_l \cdot \vec{\nabla}^2\Apot(\bro{l}) + 2i\vec{\nabla}_l^\mu \vec{\nabla}_l^\nu \vec{\nabla}_{l,\mu}\Apot_\nu(\bro{l}) + \spin{l} \cdot \vec{\nabla}^2\Bfld(\bro{l}) \right] + \textrm{transpose}.
\label{eq:commutator_el_elph2}
\end{equation}
To avoid ambiguity of dot products in the middle term, we use Einstein summation notation.
The notation ``transpose" here refers to transposing the order of the electronic and EM degrees of freedom.

\paragraph{$[\hph, \helph]$}
\label{sec:commutator_hph_helph}
As for the photonic terms, the commmutator is:
\begin{align}
[\hph, \helph] = \frac{-1}{2c}[\hph, \sum_{j=1}^{\nel} \bpo{j} \cdot \Apot(\bro{j}) + \Apot(\bro{j}) \cdot \bpo{j} + \spin{j}\cdot \Bfld(\bro{j}) - \frac{2}{c}\Apot(\bro{j})^2 ].
\end{align}
Firstly, we drop the $\Apot^2$ term, since it is much smaller than the other terms.
This is due to the extra $1/c \sim 1/137$ factor of this term, as well as the fact that, when compared to $\bpo{} \cdot \Apot$, it is smaller by a factor of $||\Apot||/||\bpo{}|| \sim 10^{-3}$ as shown in Eqs.~\eqref{eq:fieldnorms}.
Then, we can re-write everything in terms of two field commutators
\begin{equation}
[\hph, \helph] = \frac{-1}{2c} \sum_{j=1}^{\nel}( \bpo{j} \cdot \vec{F}_j + \vec{F}_j \cdot \bpo{j} + \spin{j} \cdot \vec{G}_j),
\end{equation}
where 
\begin{align}
& \vec{F}_j = [\hph, \Apot(\bro{j})] \\
& \vec{G}_j = [\hph, \Bfld(\bro{j})].
\end{align}

Starting with the definition of $\Apot$ in Eq.~\eqref{eq:aebfields}, we can compute $\vec{F}_j$ as:
\begin{equation}
\vec{F}_j = \Omega_\gamma^{-1/2} \sum_{\nu \bk} \frac{\helvec{\bk}{\mu}}{\sqrt{2\omega(\bk)}} [\hph, \acr{\bk}{\nu} e^{-i\bk \cdot \bro{j}} + \aan{\bk}{\nu}e^{i\bk \cdot \bro{j}}].
\end{equation}
Recalling the standard commutation rules between the photonic Hamiltonian and the creation and annihilation operators, we find:
\begin{equation}
\vec{F}_j = \Omega_\gamma^{-1/2} \sum_{\nu \bk}\sqrt{\frac{\omega(\bk)}{2}} \helvec{\bk}{\mu}  [\acr{\bk}{\nu}e^{-i\bk \cdot \bro{j}} - \aan{\bk}{\nu}e^{i\bk \cdot \bro{j}}] = -i\hat{\mathbf{E}}(\bro{j}),
\end{equation}
as defined in Eq.~\eqref{eq:aebfields}.
This is perhaps an unsurprising result, since $\Apot, \hat{\mathbf{E}}$ are conjugate variables in canonical quantization of the EM field.
To compute $\vec{G}_j$, we make use of the observation:
\begin{equation}
\vec{G}_j = [\hph, \nabla_j \times \Apot(\bro{j})] = \nabla_j \times [\hph, \Apot(\bro{j})] = -i\nabla_j \times \hat{\mathbf{E}}(\bro{j}).
\end{equation}
Note that we can take the curl outside of the commutator because it only acts on the electronic degree of freedom $\bro{j}$, which is not present in $\hph$ at all.
Combining these expressions, we get:
\begin{equation}
[\hph, \helph] = \frac{i}{2c} \sum_{j=1}^{\nel} \left(\bpo{j} \cdot \hat{\mathbf{E}}(\bro{j}) + \hat{\mathbf{E}}(\bro{j}) \cdot \bpo{j} + \spin{j} \cdot (\nabla_j \times \hat{\mathbf{E}}(\bro{j})) \right).
\label{eq:commutator_ph_elph}
\end{equation}

\paragraph{Reduction via relative norms}
We reduce the complexity of the commutator by identifying terms which are dominant and removing small sub-dominant contributions.
Taking the norm of the sum of the expressions for the commutators in Eqs.~\eqref{eq:commutator_el_elph},~\eqref{eq:commutator_el_elph2},~\eqref{eq:commutator_ph_elph}, we find that the total commutator norm is:
\begin{align}
||[\hat{H}_1,\hat{H}_2]|| \leq & ||(\hat{\mathbf{E}}_1 + \hat{\mathbf{E}}_2)\cdot \Apot(\vec{r})||+ \nonumber \\
& \frac{\nel}{2c}\left|\left|\vec{p} \cdot \vec{\nabla}^2\Apot(\vec{r}) + \vec{\sigma} \cdot \vec{\nabla}^2\Bfld(\vec{r}) + 2\vec{p}^\mu \vec{p}^\nu \vec{\nabla}_\mu \Apot_\nu(\vec{r}) \right|\right|+ \nonumber \\
&\frac{\nel}{c} \left|\left|\vec{p} \cdot \hat{\mathbf{E}}(\vec{r})\right|\right| + \frac{\nel}{2c} \left|\left|\spin{j} \cdot (\nabla \times \hat{\mathbf{E}}(\vec{r}))\right|\right|.
\end{align}
Here we have defined:
\begin{equation}
   \hat{\mathbf{E}}_1
  = \frac{\nel}{c}\sum_{l=1}^{\nel}   \frac{\bro{l}}{(|\bro{l}|^2+\ereg)^{3/2}}~\text{and} \ \ 
   \hat{\mathbf{E}}_2 
  = \frac{1}{c} \sum_{l=1}^{\nel}\sum_{j\neq l}    \frac{\bro{l} - \bro{j}}{(|\bro{l} - \bro{j}|^2+\ereg^2)^{3/2}}.
\end{equation}

Taking our discretized parameters, we can get actual values for these expressions.
Starting with the electronic terms, any power of $|\vec{p}|^n$ or $|\vec{r}|^{-n}$ should have norm $(\Delta_r^2 + \ereg^2)^{-n/2}$.
The electromagnetic field terms are more sophisticated, but we can bound them by bounding each of the $\ngridph$ creation and annihilation operator terms, with norm $\maxocc$, with the appropriate sums and prefactors:
\begin{align}
& ||\Apot || \leq 4 \Omega_\gamma^{-1/2} \ngridph \maxocc (ck_\mathrm{max} - ck_\mathrm{min})^{-1/2} \\ 
& ||\hat{\mathbf{E}}|| \leq 4 \Omega_\gamma^{-1/2} \ngridph \maxocc (ck_\mathrm{max} - ck_\mathrm{min})^{1/2} \\ 
& ||\Bfld|| \sim ||\hat{\mathbf{E}}||/c.
\label{eq:fieldnorms}
\end{align}
Spatial derivatives of any field multiply the norm by corresponding powers of $k_\mathrm{max}$ since all spatial dependence in the fields is $\exp(i\vec{k}\cdot \vec{r}).$
Combining all of these factors, we find a final expression:
\begin{align}
||[\hat{H}_1, \hat{H}_2]|| \leq &\frac{2\nel^2}{c} \frac{||\Apot(\vec{r})||}{\Delta_r^2 + \ereg^2} + \nonumber \\
& \frac{\nel}{2c} \left( \frac{k_\mathrm{max}^2||\Apot(\vec{r})||}{(\Delta_r^2 + \ereg^2)^{1/2}} +  \frac{k_\mathrm{max}^2||\hat{\mathbf{E}}(\vec{r})||}{c} + \frac{2k_\mathrm{max}||\Apot(\vec{r})||}{(\Delta_r^2 + \ereg^2)} + \frac{2||\hat{\mathbf{E}}(\vec{r})||}{(\Delta_r^2 + \ereg^2)^{1/2}} + k_\mathrm{max}||\hat{\mathbf{E}}(\vec{r})|| \right).
\label{eq:commutatornorm_first_order}
\end{align}
Taking the values of $\Delta_r = 0.02, \ereg = 10^{-3}, \nel = 26$ and $k_\mathrm{max} \leq 0.5$, we find that $||\hat{\mathbf{E}}(\vec{r})||/||\Apot(\vec{r})|| \sim 10^2$, and that the ratio of the first term in Eq.~\eqref{eq:commutatornorm_first_order} to the 2nd through 5th terms are on the order of: $10^4, 10^4, 10^2, 50, 10^3$.
This means that the first term due to the potential energy is dominant by almost two orders of magnitude to all other terms and hence we have to a good approximation:
\begin{equation}
[\hat{H}_1, \hat{H}_2] \cong \frac{\nel}{c}\sum_{l=1}^{\nel}  \vec{\mathcal{E}}(\bro{l}) \cdot \Apot(\bro{l}) + \frac{1}{c} \sum_{l=1}^{\nel}\sum_{j > l}   \vec{\mathcal{E}}(\bro{l}-\bro{j}) \cdot \left(\Apot(\bro{l}) - \Apot(\bro{j})\right),
\end{equation}
with
\begin{equation}
\vec{\mathcal{E}}(\vec{r}) \equiv \frac{\vec{r}}{(|\vec{r}|^2 + \ereg^2)^{3/2}}.
\end{equation}
Note that conceptually this expression looks like the matter monopole electric field $\vec{\mathcal{E}}$ coupling to the photon vector potential $\Apot$ with regularization.

\subsubsection{Analytic second-order formulae}
\label{app:trotterbounds-secondorder}
For the second-order formula, we need to compute the following nested commutators:
\begin{equation}
[\hel, [\hat{H}_1, \hat{H}_2]],\  [\hph, [\hat{H}_1, \hat{H}_2]], \ [\helph, [\hat{H}_1, \hat{H}_2]].
\end{equation}
Instead of computing all of the relevant terms, we will first do an analysis similar to the end of the last section, computing the order of magnitude differences in the terms and thereby reducing the problem to only the dominant contributions to the Trotter bounds.
To do so, we first need to study the terms in $[\hat{H}_1, \hat{H}_2]$. 
Based on the norm expression in Eq.~\eqref{eq:commutatornorm_first_order}, there are four major determining factors in the norm of a term.
\begin{enumerate}
    \item The power of $\Delta_r^{-1} = 0.02$, coming from gradients, curls, momenta $\vec{p}$ and position $\vec{r}$ operators, 
    \item The power of $\nel = 26$ coming from Coulombic potentials, 
    \item The power of $c = 137$, coming from nested commutators, 
    \item The ratio of $||\hat{\mathbf{E}}||/||\Apot|| = ck_\mathrm{max} \sim 50$.
\end{enumerate}
By just tabulating these four elements, we can easily determine that the $\Apot$ dependent terms in Eq.~\eqref{eq:commutatornorm_first_order} should be roughly 25 times larger than the second term.

We can try doing the same thing for $[\hel, [\hat{H}_1, \hat{H}_2]]$, and then for the nested commutators for $\hph, \helph$.
For $\hel$, the commutator $[\hel, [\hel, \helph]]$ applies a momentum squared operator to the $\Apot$-dependent term in the first order commutator, increasing the power of $\Delta_r^{-1}$ by 2.
Since the operations commute except when the electronic indices are identical, no powers of $\nel$ are increased, and since no factors of $c$ are present and no photonic operators, the latter two factors do not affect anything either.
As for the $[\hel, [\hel, \helph]]$ term, we also see that the largest contribution is a $\vec{p}^2$ action, but now only on the $\hat{\mathbf{E}}$-dependent terms in the first-order commutator, increasing the power of $\Delta_r^{-1}$ by 2, with nothing else increasing.
Thereby, we expect that the term $[\helph, [\hel, \helph]]$ would be a subleading contribution to $[\hel, [\hat{H}_1, \hat{H}_2]]$, which can be verified by our large calculation in the previous section.
At this stage, the leading factor contains two additional powers of $\Delta_r^{-1}$ relative to the first-order expression, giving a total expression along the lines of $O(1) \times \nel^2 ||\Apot||/(c \Delta_r^4).$

Moving now towards $[\hph, [\hat{H}_1, \hat{H}_2]]$, this term amounts to taking all $\Apot \rightarrow \hat{\mathbf{E}}, \hat{\mathbf{E}} \rightarrow \nabla \times \hat{\mathbf{E}}$, as in the previous section.
At most, then, it yields an increasing in one power of $ck_\mathrm{max}$ or one power of $\Delta_r$, which would still be 50 times smaller than the two factors of $\Delta_r$ from the $[\hel, [\hel, \hph]]$ term.
As such, we can ignore this contribution to the commutator as well.
Finally, we have $[\helph, [\hat{H}_1, \hat{H}_2]].$
As for the first part, $[\helph, [\hel, \hph]]$ this term yields a single increase in $\Delta_r^{-1}$ coming from the non-commutation of the momentum operator in $\helph$.
Due to the canonical commutation relations between $\hat{\mathbf{E}}, \Apot$, the field-dependent parts are not affected by this term.
Further, one gets an additional $1/c$ factor from $\helph$, making this contribution 500 times smaller than the primary contribution from the $[\hel, [\hel, \hph]]$ term.
Evidently, the entire commutator $[\hat{H}_1 + 2\hat{H}_2, [\hat{H}_1, \hat{H}_2]]$ can be well approximated as:
\begin{equation}
[\hat{H}_1 + 2\hat{H}_2, [\hat{H}_1, \hat{H}_2]] \approx [\hel, [\hat{H}_1, \hat{H}_2]] \approx [\hel, [\hel, \helph]].
\end{equation}
Furthermore, following the previous section, the dominant cost will be the potential term, i.e. $[\hel, [\hat{V}_{el}, \helph]]$ which we will compute in the following section.

\paragraph{$[\hel, [\hat{H}_1, \hat{H}_2]]$}

We can break this into two parts itself, firstly:
\begin{align}
[\hel, [\hat{V}_{el}, \helph]] &= \left[\frac{1}{2}\sum_{k = 1}^{\nel} \bpo{k}^2, -\frac{i}{c}\sum_{l = 1}^{\nel} \left(Z\frac{\bro{l}}{(|\bro{l}|^3+\ereg^2)^{3/2}} - \sum_{j\neq l} \frac{\bro{l} - \bro{j}}{(|\bro{l} - \bro{j}|^2+\ereg^2)^{3/2}}\right)\cdot \Apot(\bro{l})\right] \nonumber \\
&= -\frac{i}{2c}\sum_{k = 1}^{\nel} (\bpo{k}\cdot \vec{\mathcal{C}}_k + \vec{\mathcal{C}}_k\cdot \bpo{k})
\label{eq:electronic_1}
\end{align}
where
\begin{align}
\vec{\mathcal{C}}_k &\equiv \left[\bpo{k}, \sum_{l = 1}^{\nel} \left(Z\frac{\bro{l}}{(|\bro{l}|^2+\ereg^2)^{3/2}} - \sum_{j\neq l} \frac{\bro{l} - \bro{j}}{(|\bro{l} - \bro{j}|^2+\ereg^2)^{3/2}}\right)\cdot \Apot(\bro{l})\right] \nonumber \\
&= -iZ\nabla_k\left(\frac{\bro{k}\cdot \Apot(\bro{k})}{(|\bro{k}|^2+\ereg^2)^{3/2}}\right) + i\nabla_k\sum_{l = 1}^{\nel}\sum_{j\neq l} \frac{(\bro{l} - \bro{j})\cdot \Apot(\bro{l})}{(|\bro{l} - \bro{j}|^2+\ereg^2)^{3/2}} \nonumber \\
&= -iZ\nabla_k\left(\frac{\bro{k}\cdot \Apot(\bro{k})}{(|\bro{k}|^2+\ereg^2)^{3/2}}\right) + i\nabla_k\left(\sum_{j\neq k} \frac{(\bro{k} - \bro{j})\cdot \Apot(\bro{k})}{(|\bro{k} - \bro{j}|^2+\ereg^2)^{3/2}} + \sum_{l\neq k} \frac{(\bro{l} - \bro{k})\cdot \Apot(\bro{l})}{(|\bro{l} - \bro{k}|^2+\ereg^2)^{3/2}}\right) \nonumber \\
&= -iZ\nabla_k\left(\frac{\bro{k}\cdot \Apot(\bro{k})}{(|\bro{k}|^2+\ereg^2)^{3/2}}\right) + i\nabla_k\sum_{j\neq k} \frac{(\bro{k} - \bro{j})\cdot (\Apot(\bro{k}) - \Apot(\bro{j}))}{(|\bro{k} - \bro{j}|^2+\ereg^2)^{3/2}} \nonumber \\
&= -iZ\left(\frac{\Apot(\bro{k}) + \bro{k}\cdot \nabla\Apot(\bro{k})}{(|\bro{k}|^2+\ereg^2)^{3/2}} - \frac{3\bro{k}(\bro{k}\cdot \Apot(\bro{k}))}{(|\bro{k}|^2+\ereg^2)^{5/2}}\right) \nonumber \\
&+ i\sum_{j\neq k} \left(\frac{\Apot(\bro{k}) - \Apot(\bro{j}) + (\bro{k} - \bro{j})\cdot \nabla\Apot(\bro{k})}{(|\bro{k} - \bro{j}|^2+\ereg^2)^{3/2}} - \frac{3(\bro{k} - \bro{j})((\bro{k} - \bro{j})\cdot (\Apot(\bro{k}) - \Apot(\bro{j})))}{(|\bro{k} - \bro{j}|^2+\ereg^2)^{5/2}}\right).
\label{eq:dominant-second-order}
\end{align}
Note that the final expression involves the Jacobian $\nabla \Apot$.

Next, we have
\begin{align}
[\hel, [\hph, \helph]] &= \left[\sum_{j = 1}^{\nel}\left(\frac{1}{2}\bpo{j}^2 - \frac{Z}{(|\bro{j}|^2+\ereg^2)^{1/2}}\right) + \sum_{j = 1}^{\nel} \sum_{k \neq j}^{\nel} \frac{1}{2}\frac{1}{(|\bro{j} - \bro{k}|^2+\ereg^2)^{1/2}}, \frac{i}{2c} \sum_{l = 1}^{\nel} \bpo{l} \cdot \hat{\mathbf{E}}(\bro{l}) + \hat{\mathbf{E}}(\bro{l}) \cdot \bpo{l} \right] \nonumber \\
&+ \left[\sum_{j = 1}^{\nel} \frac{1}{2}\bpo{j}^2, \frac{i}{2c} \sum_{l = 1}^{\nel} \spin{l} \cdot \nabla_l \times \hat{\mathbf{E}}(\bro{l})\right].
\end{align}
We compute the various terms:
\begin{align}
\left[\sum_{j = 1}^{\nel} \frac{1}{2}\bpo{j}^2, \frac{i}{2c} \sum_{l = 1}^{\nel} \bpo{l} \cdot \hat{\mathbf{E}}(\bro{l})\right] &= \frac{i}{4c}\sum_{j = 1}^{\nel} \bpo{j} \cdot \left[\bpo{j}^2, \hat{\mathbf{E}}(\bro{j})\right] \nonumber \\
&= \frac{i}{4c}\sum_{j = 1}^{\nel} \bpo{j} \cdot (\bpo{j}\cdot [\bpo{j}, \hat{\mathbf{E}}(\bro{j})] + [\bpo{j}, \hat{\mathbf{E}}(\bro{j})]\cdot \bpo{j}) \nonumber \\
&= \frac{1}{4c}\sum_{j = 1}^{\nel} \bpo{j} \cdot (\bpo{j}\cdot \nabla_j\hat{\mathbf{E}}(\bro{j}) + \nabla_j\hat{\mathbf{E}}(\bro{j})\cdot \bpo{j})
\end{align}
and
\begin{align}
\left[-\sum_{j = 1}^{\nel} \frac{Z}{(|\bro{j}|^2+\ereg^2)^{1/2}} \right. & \left. + \sum_{j = 1}^{\nel} \sum_{k \neq j}^{\nel} \frac{1}{2}\frac{1}{(|\bro{j} - \bro{k}|^2+\ereg^2)^{1/2}}, \frac{i}{2c} \sum_{l = 1}^{\nel} \bpo{l} \cdot \hat{\mathbf{E}}(\bro{l})\right] \nonumber \\
&= -\frac{i}{2c}\sum_{l = 1}^{\nel} \left[\bpo{l}, -\frac{Z}{(|\bro{l}|^2+\ereg^2)^{1/2}} + \sum_{j\neq l} \frac{1}{(|\bro{l} - \bro{j}|^2+\ereg^2)^{1/2}}\right]\cdot \hat{\mathbf{E}}(\bro{l}) \nonumber \\
&= -\frac{1}{2c}\sum_{l = 1}^{\nel} \nabla_l\left(-\frac{Z}{(|\bro{l}|^2+\ereg^2)^{1/2}} + \sum_{j\neq l} \frac{1}{(|\bro{l} - \bro{j}|^2+\ereg^2)^{1/2}}\right)\cdot \hat{\mathbf{E}}(\bro{l}) \nonumber \\
&= -\frac{1}{2c}\sum_{l = 1}^{\nel} \left(Z\frac{\bro{l}}{(|\bro{l}|^2+\ereg^2)^{3/2}} - \sum_{j\neq l} \frac{\bro{l} - \bro{j}}{(|\bro{l} - \bro{j}|^2+\ereg^2)^{3/2}}\right)\cdot \hat{\mathbf{E}}(\bro{l})
\end{align}
and
\begin{align}
\left[\sum_{j = 1}^{\nel} \frac{1}{2}\bpo{j}^2, \frac{-i}{4c} \sum_{l = 1}^{\nel} \spin{l} \cdot \nabla_l \times \hat{\mathbf{E}}(\bro{l}) \right] &= \frac{-i}{8c}\sum_{j = 1}^{\nel} \left[\bpo{j}^2, \spin{j} \cdot \nabla_j \times \hat{\mathbf{E}}(\bro{j}) \right] \nonumber \\
&= -\frac{1}{8c}\sum_{j = 1}^{\nel} (\bpo{j}\cdot \nabla_j(\spin{j} \cdot \nabla_j \times \hat{\mathbf{E}}(\bro{j})) + \nabla_j(\spin{j} \cdot \nabla_j \times \hat{\mathbf{E}}(\bro{j}))\cdot \bpo{j}).
\end{align}
Hence
\begin{align}
[\hel, [\hph, \helph]] &= \frac{1}{4c}\sum_{j = 1}^{\nel} \left(\bpo{j} \cdot (\bpo{j}\cdot \nabla\hat{\mathbf{E}}(\bro{j}) + \nabla\hat{\mathbf{E}}(\bro{j})\cdot \bpo{j}) + (\bpo{j}\cdot \nabla\hat{\mathbf{E}}(\bro{j}) + \nabla\hat{\mathbf{E}}(\bro{j})\cdot \bpo{j}) \cdot \bpo{j} \right)  \nonumber \\
&-\frac{1}{2c}\sum_{l = 1}^{\nel} \left(Z\frac{\bro{l}}{(|\bro{l}|^2+\ereg^2)^{3/2}} - \sum_{j\neq l} \frac{\bro{l} - \bro{j}}{(|\bro{l} - \bro{j}|^2+\ereg^2)^{3/2}}\right)\cdot \hat{\mathbf{E}}(\bro{l}) \nonumber \\
&- \frac{1}{8c}\sum_{j = 1}^{\nel} (\bpo{j}\cdot \nabla_j(\spin{j} \cdot \nabla_j \times \hat{\mathbf{E}}(\bro{j}) + \nabla_j(\spin{j} \cdot \nabla_j \times \hat{\mathbf{E}}(\bro{j}))\cdot \bpo{j}).
\label{eq:electronic_2}
\end{align}

\paragraph{Reduction via relative norms}
The full commutator for $[\hel, [\hat{H}_1,\hat{H}_2]]$ can be constructed via the sum of the two terms in Eq.~\eqref{eq:dominant-second-order},~\eqref{eq:electronic_2}.
The prior dominates the latter via a similar argument used in the first-order Trotter section, and within Eq.~\eqref{eq:dominant-second-order} the two terms with Jacobian of $\Apot$ are sub-leading.
\begin{align}
[\hat{H}_1 + 2\hat{H}_2, [\hat{H}_1, \hat{H}_2]] \approx  \frac{\nel}{c}\sum_{l=1}^{\nel} \vec{\mathcal{D}}(\bro{l}) \cdot \Apot(\bro{l}) + \frac{1}{c} \sum_{l=1}^{\nel}\sum_{j > l}  \vec{\mathcal{D}}(\bro{l}-\bro{j}) \cdot \left(\Apot(\bro{l}) - \Apot(\bro{j})\right),
\end{align}
with 
\begin{equation}
\vec{\mathcal{D}}(\vec{r}) \equiv \frac{1}{2}\left( (\vec{p} - 3(\vec{p}\cdot\hat{r})\hat{r}) \, \frac{1}{(|\vec{r}|^2 + \epsilon^{2}_\textrm{reg})^{3/2}} + \frac{1}{(|\vec{r}|^2 + \epsilon^{2}_\textrm{reg})^{3/2}} \, (\vec{p} - 3\hat{r}(\hat{r}\cdot\vec{p})) \right).
\end{equation}
Note that conceptually this expression looks like the matter dipole electric field $\vec{\mathcal{D}}$ coupling to the photon vector potential $\Apot$.

\subsection{State-dependent formulae}
Consider starting with an initial state $|\psi_\mathrm{i}\rangle$.
After time evolution, we end up with a state:
\begin{equation}
    |\psi_\mathrm{f}^\prime \rangle = \mathcal{S}_{\tilde{k}}(\tee/\tilde{r})^{\tilde{r}}|\psi_\mathrm{i}\rangle = |\psi_\mathrm{f} \rangle + |\epsilon\rangle.
\end{equation}
Note that $|\psi_\mathrm{f}\rangle$ can approximately be written as:
\begin{equation}
    |\psi_\mathrm{f}\rangle \cong \sqrt{1-\abs^2}|\psi_\mathrm{i}\rangle + \sqrt{\abs}|\psi_{ex}\rangle + O(\abs^2),
    \label{eq:final_state}
\end{equation}
where $A = \sum_k A(k)$ is the total absorption probability.
The approximation above is valid as long as $\ngridph$ is small enough, namely since $A(k) \sim 10^{-5}$, $A \sim 10^{-5} \ngridph^r$, so $\ngridph^r \ll 10^5$ is fine.
To lowest order in $A$, one has only single-photon absorption events, meaning one can go from $|\psi_\mathrm{i}\rangle$ to a state with fewer photons, which is contained in $|\psi_{ex}\rangle$.
It is possible to move back into the same number of photon sector as $|\psi_\mathrm{i}\rangle$ with absorption events, however this is a second order correction, and hence can be left out.

Next, we can look at the structure of $|\epsilon\rangle$.
To get $|\epsilon\rangle$, we can look at the definition of the additive operator $\mathcal{A}(t)$ above.
Starting with first order Trotterization, we find that:
\begin{equation}
    \mathcal{S}_1(t) = e^{-itH} - \int_{0}^{t} d\tau e^{-i(t - \tau)H} [e^{-i\tau \hat{H}_2}, \hat{H}_1]e^{-i\tau \hat{H}_1},
\end{equation}
as described in Eq.~21 of ~\cite{childs2021}, where $H = \hat{H}_1 + \hat{H}_2$ contains two terms.
In our case, we approximate via $\left(\mathcal{S}_1(t/\tilde{r})\right)^{\tilde{r}}$, where $t/\tilde{r}$ is a small quantity, and $\hat{H}_1 = \hel + \hph, \hat{H}_2 = \helph$, reducing each step to:
\begin{equation}
    \mathcal{S}_1(t/\tilde{r}) - e^{-i(t/\tilde{r})H} \cong -\int_{0}^{t/\tilde{r}} d\tau \left(1 -i(t/\tilde{r} - \tau)H\right)[1 - i\tau \hat{H}_2, \hat{H}_1]\left(1 - i\tau \hat{H}_1\right) \cong i\frac{t^2}{2\tilde{r}^2}[\hat{H}_2, \hat{H}_1].
\end{equation}
The accumulated error then looks like:
\begin{equation}
    \mathcal{A}(t) = \left(\mathcal{S}_1(t/\tilde{r})\right)^{\tilde{r}} - e^{-itH} \cong \left( e^{-i(t/\tilde{r})H} + i\frac{t^2}{2r^2}[\hat{H}_2,\hat{H}_1] \right)^{\tilde{r}} - e^{-itH}, \ \ |\epsilon\rangle \equiv \mathcal{A}(t)|\psi_\mathrm{i}\rangle.
    \label{eq:epsilon_state}
\end{equation}

Our goal is then to evaluate Eq.~\eqref{eq:epsilon_state} such that we can find the support of $|\epsilon\rangle$, to lowest non-vanishing order, on $|\psi_{ex}\rangle$ and $|\psi_\mathrm{i}\rangle$, thereby making it easy to compute the overlap from Eq.~\eqref{eq:final_state}.
Starting with $\tilde{r} = 1$, we find:
\begin{equation}
     |\epsilon^{\tilde{r}=1}\rangle \equiv 
    \left(\mathcal{S}_1(t/\tilde{r})^1 - e^{-itH}\right)|\psi_\mathrm{i}\rangle = i \frac{t^2}{2\tilde{r}^2}[\hat{H}_2,\hat{H}_1]|\psi_\mathrm{i}\rangle.
\end{equation}
If we were to truncate our analysis here, we would find that $|\epsilon^{\tilde{r} = 1}\rangle$ would only contain terms proportional to $[\hat{H}_2,\hat{H}_1]|\psi_\mathrm{i}\rangle$.
If we look at the commutator $[\hat{H}_2, \hat{H}_1] = -[\hel + \hph, \helph]$, as computed in Eqs.~\eqref{eq:commutator_el_elph} and ~\eqref{eq:commutator_ph_elph}, we will find that $[\hat{H}_2,\hat{H}_1]$ only contains field operators of the type $\Apot, \hat{\mathbf{E}}, \nabla \times \hat{\mathbf{E}}$, all of which necessarily change the photonic numbers by $\pm$ 1.
Therefore, we would find that $\langle \psi_\mathrm{f} | \epsilon^{\tilde{r}=1}\rangle \leq \sqrt{\abs} \epsilon$, since $|\epsilon^{\tilde{r}=1}\rangle$ only has support on $|\psi_{ex}\rangle$.
However things are more complicated for $\tilde{r} = 2$ where we find:
\begin{equation}
    |\epsilon^{\tilde{r}=2}\rangle \equiv \left(\mathcal{S}_1(t/\tilde{r})^2 - e^{-itH}\right)|\psi_\mathrm{i}\rangle = i e^{-i(t/\tilde{r})H}\frac{t^2}{2\tilde{r}^2}[\hat{H}_2,\hat{H}_1]|\psi_\mathrm{i}\rangle + i \frac{t^2}{2\tilde{r}^2}[\hat{H}_2,\hat{H}_1]e^{-i(t/\tilde{r})H}|\psi_\mathrm{i}\rangle + O((t/\tilde{r})^3).
\end{equation}
We see here that to the lowest order in additive error, we have two terms, where in principle the number of photons may change in both terms, meaning that it is possible that $|\epsilon^{\tilde{r}=2}\rangle$ has non-zero weight on $|\psi_\mathrm{i}\rangle$.
However, there are some reductions when looking at lowest order in error.
For example (see paragraph after for explanation):
\begin{equation}
    i\frac{t^2}{2\tilde{r}^2}[\hat{H}_2,\hat{H}_1]\left[e^{-i(t/\tilde{r})H}|\psi_\mathrm{i}\rangle \right]\  ``\leq" \ i\frac{t^2}{2\tilde{r}^2}[\hat{H}_2,\hat{H}_1]\left[\sqrt{1-\abs^2}|\psi_\mathrm{i}\rangle + \sqrt{\abs} |\psi_{ex} \rangle\right] \cong i\frac{t^2}{2\tilde{r}^2}[\hat{H}_2,\hat{H}_1]|\psi_\mathrm{i}\rangle.
\end{equation}
The first inequality denoted $``\leq"$ is indicating that $\exp(-i(t/\tilde{r})H)|\psi_\mathrm{i}\rangle$ results in dynamics with fewer absorption events than $\exp(-itH)|\psi_\mathrm{i}\rangle$, with the latter being given in Eq.~\eqref{eq:final_state}.
We then see that the result is two terms, one with norm $\propto ||[\hat{H}_2, \hat{H}_1]||\sqrt{1-\abs^2}$ and the other with norm $\propto ||[\hat{H}_2, \hat{H}_1]||\sqrt{\abs}$.
Given that $\abs \sim 10^{-5}$, it is clear that the prior is the dominant term, and the latter can be discarded to lowest order.
A similar argument can be made for the term with the opposite order of operations.
As such, we may instead write 
\begin{equation}
    |\epsilon^{\tilde{r}=2}\rangle \equiv \left(\mathcal{S}_1(t/\tilde{r})^2 - e^{-itH}\right)|\psi_\mathrm{i}\rangle \cong 2 \times i\frac{t^2}{2\tilde{r}^2}[\hat{H}_2,\hat{H}_1]|\psi_\mathrm{i}\rangle +  O((t/\tilde{r})^3, 
    \sqrt{\abs}(t/\tilde{r})^2).
\end{equation}
We can continue this process, and we would find that:
\begin{equation}
    |\epsilon \rangle \cong \tilde{r} \times i \frac{t^2}{2\tilde{r}^2}[\hat{H}_2, \hat{H}_1]|\psi_\mathrm{i}\rangle + O((t/\tilde{r})^3, \sqrt{\abs}(t/\tilde{r})^2).
    \label{eq:first-order-state-dependent}
\end{equation}

A similar argument can be made for second order Trotterization.
In this case, the additive error is:
\begin{equation}
\mathcal{S}_2(t) - e^{-itH} = \int_{0}^{t} d\tau e^{-i(t-\tau)H}e^{-i\tau \hat{H}_1/2}\mathcal{T}_2(\tau) e^{-i\tau \hat{H}_2}e^{-i\tau \hat{H}_1/2},
\end{equation}
where 
\begin{equation}
\mathcal{T}_2(\tau) = e^{-i\tau \hat{H}_2}\left(-i\frac{\hat{H}_1}{2}\right)e^{i\tau \hat{H}_2} + i\frac{\hat{H}_1}{2} + e^{i\tau \hat{H}_1/2}(i\hat{H}_2)e^{-i\tau \hat{H}_1/2} - i\hat{H}_2,
\end{equation}
as described in Eqs.~L1, L2 of \cite{childs2021}.
We can do the same short time-trick, we will find that 
\begin{equation}
\mathcal{S}_2(t/\tilde{r}) - e^{-i(t/\tilde{r})H}\cong i \frac{t^3}{24\tilde{r}^3}[\hat{H}_1 + 2\hat{H}_2, [\hat{H}_1, \hat{H}_2]]. 
\end{equation}
Now effectively the same argument as above can be made.
By ignoring the small contributions from the $e^{-i(t/\tilde{r})H}$ terms in the total error, we find that state error looks like
\begin{equation}
|\epsilon \rangle \cong \tilde{r} \times i \frac{t^3}{24 \tilde{r}^3}[\hat{H}_1 + 2\hat{H}_2, [\hat{H}_1, \hat{H}_2]]|\psi_\mathrm{i}\rangle + O((t/\tilde{r})^4, \sqrt{\abs}(t/\tilde{r})^3).
\label{eq:second-order-state-dependent}
\end{equation}
From Eqs.~\eqref{eq:first-order-state-dependent},~\eqref{eq:second-order-state-dependent}, we find that in the regime we are working in where absorption is very rare $\abs \sim 10^{-5}$, the Trotter formulae for state-dependent bonds are constructed by simply applying the analytic formulae to the initial state.

\subsection{Numerical estimation of Trotter bounds for $\hat{H}_\textrm{PF}^{\textrm{AA}}$}
\label{app:state-dep-trotter}
To compute the resource estimates for the opacity simulation, we need to numerically estimate Trotter bounds.
For the first-order Trotter bound, we have one-body and two-body contributions $e_1$, $e_2$:
\begin{equation}
   e_1
  = \frac{\nel}{c}\sum_{l=1}^{\nel} \Big| \Big| \vec{\mathcal{E}}(\bro{l}) |\psi_\mathrm{i} \rangle \Big|\Big|
  = \frac{\nel}{c}\sum_{l=1}^{\nel}  \langle\psi_\mathrm{i} | \frac{|\bro{l}|^2}{(|\bro{l}|^2 + \ereg^2)^3} | \psi_\mathrm{i} \rangle ^{1/2}
  ,
  \label{eq:app:troter-e1}
\end{equation}
\begin{equation}
   e_2 
  = \frac{1}{c} \sum_{l=1}^{\nel}\sum_{j>l} \Big| \Big| \vec{\mathcal{E}}(\bro{l}-\bro{j}) |\psi_\mathrm{i} \rangle \Big|\Big|
  = \frac{1}{c} \sum_{l=1}^{\nel}\sum_{j\neq l}  \langle\psi_\mathrm{i} | \frac{|\bro{l} - \bro{j}|^2}{(|\bro{l} - \bro{j}|^2 + \ereg^2)^3}|\psi_\mathrm{i}\rangle ^{1/2}
  .
  \label{eq:app:trotter-e2}
\end{equation}
Similarly, for the second-order Trotter bound, we have one-body and two-body contributions $d_1$, $d_2$:
\begin{equation}
   d_1
  = \frac{\nel}{c}\sum_{l=1}^{\nel} \Big| \Big| \vec{\mathcal{D}}(\bpo{l}, \bro{l}) |\psi_\mathrm{i} \rangle \Big|\Big|
  = \frac{\nel}{c}\sum_{l=1}^{\nel}  \langle\psi_\mathrm{i} | \vec{\mathcal{D}}^\dagger(\bpo{l}, \bro{l}) \cdot \vec{\mathcal{D}}(\bpo{l}, \bro{l}) | \psi_\mathrm{i} \rangle ^{1/2}
  ,
  \label{eq:app:troter-d1}
\end{equation}
\begin{equation}
   d_2 
  = \frac{1}{c} \sum_{l=1}^{\nel}\sum_{j>l} \Big| \Big| \vec{\mathcal{D}}(\bpo{l},\bro{l}-\bro{j}) |\psi_\mathrm{i} \rangle \Big|\Big|
  = \frac{1}{c} \sum_{l=1}^{\nel}\sum_{j\neq l}  \langle\psi_\mathrm{i} | 
 \vec{\mathcal{D}}^\dagger(\bpo{l},\bro{l}-\bro{j}) \cdot \vec{\mathcal{D}}(\bpo{l},\bro{l}-\bro{j})
  |\psi_\mathrm{i}\rangle ^{1/2}
  ,
  \label{eq:app:trotter-d2}
\end{equation}
where
\begin{equation}
\vec{\mathcal{D}}(\vec{p},\vec{r}) \equiv \frac{1}{2}\left( \vec{p}\frac{1}{(|\vec{r}|^2 + \epsilon^{2}_\textrm{reg})^{3/2}} - 3(\vec{p}\cdot\vec{r})\vec{r} \frac{1}{(|\vec{r}|^2 + \epsilon^{2}_\textrm{reg})^{5/2}} + \frac{1}{(|\vec{r}|^2 + \epsilon^{2}_\textrm{reg})^{3/2}} \vec{p} - \frac{1}{(|\vec{r}|^2 + \epsilon^{2}_\textrm{reg})^{5/2}} 3\vec{r}(\vec{r}\cdot\vec{p})) \right) .
\end{equation}
The expressions above are state-dependent bounds, but the state vector can be excluded so the norms are operator norms, giving the expressions for state-independent bounds.
We will discuss how to estimate these parameters in this Appendix.

\subsubsection{State-independent Trotter bounds}
Starting with first order Trotter bounds, we evaluate the largest possible values of $e_1, e_2$ by looking at the operator $\vec{\mathcal{E}}$.
Since $\vec{\mathcal{E}}$ depend only on position, we can just specify the position of the electrons on our spatial grid.
If we look at the function:
\begin{equation}
|\vec{\mathcal{E}}|^2 = \frac{|\vec{r}|^2}{(|\vec{r}|^2 + \ereg^2)^3} \xrightarrow{\mathrm{max}} |\vec{r}| = \frac{\ereg}{\sqrt{2}}.
\end{equation}
As the grid spacing $\Delta_r \gg \ereg$, it means that we will generally not hit the maximum value here. 
The largest possible value is thereby either at $|\vec{r}| = 0$ or $|\vec{r}| = \Delta_r$.
It is easy to see it is not the prior since $\vec{\mathcal{E}}$ vanishes there, and so the worst case is when $|\vec{r}| = \Delta_r$.
In general it may not be possible to place all electrons (or pairs in the case of $e_2$) with this spacing relative to each other, but since the function monotonically decreases past the maximum we can still use this as an upper bound.
Therefore, we find that:
\begin{align}
& \textrm{max}(|\vec{\mathcal{E}}|^2) \leq \frac{\Delta_r^2}{(\Delta_r^2 + \ereg^2)^3} \\
& e_1, \ e_2 \leq \frac{\nel^2}{c} \frac{\Delta_r}{(\Delta_r^2 + \ereg^2)^{3/2}} = 1.2 \times 10^4.
\end{align}

The situation for the second order Trotter bound is different, since $\vec{\mathcal{D}}$ does not only depend on position.
However, we can note a few similarities to $e_1, e_2$, namely that the second and third terms necessarily vanish at $|\vec{r}| = 0$ and hence only have $\Delta_r^{-1}$ contributions, i.e. not scaling with $\ereg^{-1}$ to lowest order, so we expect them to be subdominant contributions.
The first term can also be made entirely position dependent by commuting the first order derivative through, and this yields a similar structure to $\vec{\mathcal{E}}$ but with 5/2 power in the denominator: this also vanishes at $|\vec{r}| = 0$ and hence does not scale inversely with $\ereg^{-1}$.
What remains is the last term which will be dominant in the worst case as it does not vanish at $\vec{r} = 0$.
Therefore we would have:
\begin{equation}
\textrm{max}(|\vec{\mathcal{D}}|) \leq \epsilon^{-3}_\textrm{reg}\Delta_r^{-1},
\end{equation}
here I have taken the maximum momentum norm to be $1/\Delta_r$.
Note again that this upper bound is likely not physical feasible.
For the one-body term $d_1$, this would require packing all electrons at zero distance, which is obviously impossible, and for the two-body term this would require every electron to be on the same site, also impossible.
Therefore, to be more precise, the worst case scenario here is going to be one where all the electrons are paired up and as close to each other as possible, which would be $\Delta_r$ distance away, with just one pair at the origin.
In this case, the contributions become:
\begin{align}
& d_1 \leq \frac{4}{c}\frac{1}{\ereg^{3}\Delta_r} + \frac{(\nel-2)^2}{c}\frac{1}{(\Delta_r^2 + \ereg^2)^{3/2}\Delta_r} = 1.4 \times 10^{9} \\
& d_2 \leq \frac{\nel}{c}\frac{1}{\ereg^{3}\Delta_r} + \frac{\nel^2- \nel}{c}\frac{1}{(\Delta_r^2 + \ereg^2)^{3/2}\Delta_r} = 9.5 \times 10^{9}.
\end{align}

\subsubsection{State-dependent Trotter bounds}
To estimate the Trotter error incurred for physically relevant initial states $|\psi_\mathrm{i}\rangle$, we evaluate state-dependent  bounds over a representative ensemble of classically computed initial states.
Consistent with the approximate thermal state preparation algorithm introduced by Ref.~\onlinecite{babbush2023quantum} and used in earlier resource estimates for another plasma simulation problem \cite{rubin2024quantum}, we use a Slater determinant ansatz
\begin{equation}
    |\psi_\mathrm{i} \rangle = |\chi_1, \chi_2, \cdots, \chi_{\nel}\rangle,
\end{equation}
where $|\chi_\alpha\rangle = |\phi_\alpha\rangle |\sigma_\alpha\rangle$ are single-particle spin-orbitals with $|\phi_\alpha\rangle$ and $|\sigma_\alpha\rangle$ denoting spatial orbitals and corresponding spin states.
The $|\phi_\alpha\rangle$ are obtained from a classical mean-field calculation and then sampled according to Boltzmann statistics as described in Section \ref{subapp:parameters_numerical}.
For each spatial orbital configuration sampled, the corresponding spin configuration is set randomly subject to Pauli exclusion constraints.

For this type of initial state, the expectation values within $e_1$ and $e_2$ can be expressed as
\begin{equation}
    \langle\psi_\mathrm{i} | \frac{|\bro{l}|^2}{(|\bro{l}|^2+\ereg^2)^3} | \psi_\mathrm{i} \rangle = \frac{(\nel-1)!}{\nel!} \sum_{\alpha=1}^{\nel}  \langle\phi_\alpha | \frac{|\vec{r}|^2}{(|\vec{r}|^2+\ereg^2)^3} | \phi_\alpha \rangle~\text{and}
\end{equation}
\begin{align}
    \langle\psi_\mathrm{i} | \frac{|\bro{l} - \bro{j}|^2}{(|\bro{l} - \bro{j}|^2+\ereg^2)^3} | \psi_\mathrm{i} \rangle 
    = \frac{(\nel-2)!}{\nel!} \sum_{\alpha=1}^{\nel} \sum_{\beta\neq \alpha} \Big[ &\langle\phi_\alpha \phi_\beta | \frac{|\vec{r}_1 - \vec{r}_2|^2}{(|\vec{r}_1-\vec{r}_2|^2+\ereg^2)^3} | \phi_\alpha \phi_\beta \rangle \nonumber \\
    -& \langle\phi_\alpha \phi_\beta | \frac{|\bro{l} - \bro{j}|^2}{(|\vec{r}_1-\vec{r}_2|^2+\ereg^2)^3} | \phi_\beta \phi_\alpha \rangle \delta_{\sigma_\alpha,\sigma_\beta} \Big],
\end{align}
where the $1/\nel!$ factors come from the Slater determinant normalization and the $(\nel-1)!$, $(\nel-2)!$ factors come from the number of ways to permute the other single-particle orbitals.
Each of the $\nel$ and $\nel(\nel-1)$ terms in the sums over $l$ and $l,j$ in Eqs.~\eqref{eq:app:troter-e1} and \eqref{eq:app:trotter-e2}, respectively, are identical.
Thus, the error terms become
\begin{equation}
    e_1
    = \frac{\nel^{3/2}}{c} \left( \sum_{\alpha=1}^{\nel}  \langle\phi_\alpha | \frac{|\vec{r}|^2}{(|\vec{r}|^2+\ereg^2)^3} | \phi_\alpha \rangle \right)^{1/2}
  ~\text{and}
\end{equation}
\begin{equation}
   e_2 
  = \frac{\nel^{1/2}(\nel-1)^{1/2}}{c} \left( \sum_{\alpha=1}^{\nel}\sum_{\beta\neq \alpha}  \langle\phi_\alpha \phi_\beta | \frac{|\vec{r}_1 - \vec{r}_2|^2}{(|\vec{r}_1-\vec{r}_2|^2+\ereg^2)^3} | \phi_\alpha \phi_\beta \rangle 
  - \langle\phi_\alpha \phi_\beta | \frac{|\vec{r}_1 - \vec{r}_2|^2}{(|\vec{r}_1-\vec{r}_2|^2+\ereg^2)^3} | \phi_\beta \phi_\alpha \rangle \delta_{\sigma_\alpha,\sigma_\beta} \right) ^{1/2}
  .
\end{equation}

Similarly, the second-order error terms can be expressed as
\begin{equation}
    d_1
    = \frac{\nel^{3/2}}{c} \left( \sum_{\alpha=1}^{\nel}  \langle\phi_\alpha | \vec{\mathcal{D}}^\dagger(\vec{p}, \vec{r}) \cdot \vec{\mathcal{D}}(\vec{p}, \vec{r}) | \phi_\alpha \rangle \right)^{1/2}
  ~\text{and}
\end{equation}
\begin{align}
   d_2 
   = \frac{\nel^{1/2}(\nel-1)^{1/2}}{c} \Big( \sum_{\alpha=1}^{\nel}\sum_{\beta\neq \alpha} &  \langle\phi_\alpha \phi_\beta | \vec{\mathcal{D}}^\dagger(\vec{p}_1, \vec{r}_1-\vec{r}_2) \cdot \vec{\mathcal{D}}(\vec{p}_1, \vec{r}_1-\vec{r}_2) | \phi_\alpha \phi_\beta \rangle \nonumber \\
  -& \langle\phi_\alpha \phi_\beta | \vec{\mathcal{D}}^\dagger(\vec{p}_1, \vec{r}_1-\vec{r}_2) \cdot \vec{\mathcal{D}}(\vec{p}_1, \vec{r}_1-\vec{r}_2) | \phi_\beta \phi_\alpha \rangle \delta_{\sigma_\alpha,\sigma_\beta} \Big) ^{1/2}
  .
\end{align}

We discuss numerical evaluation of the matrix elements involved in our estimates of $e_1$, $e_2$, $d_1$, and $d_2$ in Section \ref{app:aa-integrals}.
Finally, we present sampled thermal distributions for the one-body and two-body contributions as well as the total state-dependent error in Section \ref{app:state-dep-trotter-results}.

\paragraph{Evaluating average-atom matrix elements}
\label{app:aa-integrals}
Given the form in Eq.~\eqref{eq:aa_dof_separation} and orthonormality of spherical harmonics, the matrix elements involved in $e_1$ can be readily evaluated as
\begin{equation}
    e_\alpha \equiv \langle\phi_\alpha | \frac{|\vec{r}|^2}{(|\vec{r}|^2+\ereg^2)^3} | \phi_\alpha \rangle 
    = \int dr \; \frac{r^4}{(r^2+\ereg^2)^3} (R_{n_\alpha}^{\ell_\alpha}(r))^2.
\end{equation}

Meanwhile, the matrix elements involved in $e_2$ consist of 6-dimensional integrals that require further simplification for efficient evaluation.
We reduce them to radial integrals through generalized Laplace expansions.
The only angular dependence in $\mathcal{E}^2$ involves $\cos\theta_{12} = \hat{r}_1\cdot\hat{r}_2$, so we can expand the two-body operator in a Legendre polynomial basis that then expands into spherical harmonics:
\begin{align}
    \mathcal{E}^2(\vec{r}_1 - \vec{r}_2) &= \sum_{\ell=0}^\infty \mathcal{E}_\ell(r_1,r_2) P_\ell(\cos\theta_{12}) ~\nonumber \\
    &= \sum_{\ell=0}^\infty \sum_{m=-\ell}^\ell \frac{4\pi}{2\ell+1} \mathcal{E}_\ell(r_1,r_2) \bar{Y}_\ell^m(\theta_1,\varphi_1) Y_\ell^m(\theta_2,\varphi_2).
    \label{eq:e2_laplace}
\end{align}
The expansion function $\mathcal{E}_\ell$ is given by
\begin{align}
    \mathcal{E}_\ell(r_1,r_2) &= \frac{2\ell+1}{2} \int_0^\pi \sin\theta_{12} \, d\theta_{12} \, \mathcal{E}^2(\vec{r}_1 - \vec{r}_2) P_\ell(\cos\theta_{12}) \nonumber \\
    &= \frac{2\ell+1}{2} \left(
\left.\left[ \frac{1}{b(a-bu)} - \frac{\ereg^2}{2b(a-bu)^2}\right]P_\ell(u) + \frac{\epsilon^2}{2b^2(a-bu)} P_\ell'(u) + Q_{\ell}(u) + R_{\ell}\ln(a-bu)
\right)\right|_{u=-1}^{u=1},
    \label{eq:e2_expansion_integral}
\end{align}
where $u=\cos\theta_{12}$, $a=r_1^2 + r_2^2 + \ereg^2$, and $b=2r_1r_2$. 
The expression in Eq.~\eqref{eq:e2_expansion_integral} was obtained through repeated integration by parts.
$Q_{\ell}(u)$ and $R_{\ell}$ relate to the quotient and remainder in a polynomial division problem satisfying
\begin{align}
    -b R_{\ell} + Q_\ell'(u) (a-bu) = -\frac{P_\ell'(u)}{b} - \frac{\ereg^2 P_\ell''(u)}{2b^2}.
\end{align}

The expansion given by Eq.~\eqref{eq:e2_laplace} allows separation of variables
\begin{equation}
    \langle\phi_\alpha \phi_\beta | \mathcal{E}^2(\vec{r}_1 - \vec{r}_2) | \phi_\mu \phi_\nu \rangle 
    = \sum_{\ell=0}^\infty \sum_{m=-\ell}^{\ell} \frac{4\pi}{2\ell+1} I_{r_1,r_2}^{\alpha,\beta,\mu,\nu,\ell} \, I_{\theta_1,\varphi_1}^{\alpha,\mu,\ell,m} \, I_{\theta_2,\varphi_2}^{\beta,\nu,\ell,m}
    \label{eq:2body_matrix_element}
\end{equation}
with
\begin{equation}
    I_{r_1,r_2}^{\alpha,\beta,\mu,\nu,\ell}
    = \int_0^{\rws} r_1^2 dr_1 \int_0^{\rws} r_2^2 dr_2 \; \mathcal{E}_\ell(r_1,r_2) R_{n_\alpha}^{\ell_\alpha}(r_1) R_{n_\beta}^{\ell_\beta}(r_2) R_{n_\mu}^{\ell_\mu}(r_1) R_{n_\nu}^{\ell_\nu}(r_2),
    \label{eq:Irho}
\end{equation}
\begin{align}
    I_{\theta_1,\varphi_1}^{\alpha,\mu,\ell,m}
    =& \int_0^{2\pi} d\varphi_1 \int_0^\pi \sin\theta_1 d\theta_1 \bar{Y}_{\ell_\alpha}^{m_\alpha}(\theta_1,\varphi_1) \bar{Y}_{\ell}^{m}(\theta_1,\varphi_1) Y_{\ell_\mu}^{m_\mu}(\theta_1,\varphi_1) \nonumber \\
    =& (-1)^{m_\alpha+m} \left( \frac{(2\ell_\alpha+1)(2\ell+1)(2\ell_\mu+1)}{4\pi} \right)^{1/2} \wigner{\ell_\alpha}{\ell}{\ell_\mu}{0}{0}{0} \wigner{\ell_\alpha}{\ell}{\ell_\mu}{-m_\alpha}{-m}{m_\mu}
    \label{eq:Iangle1},~\text{and}
\end{align}
\begin{align}
    I_{\theta_2,\varphi_2}^{\beta,\nu,\ell,m}
    =& \int_0^{2\pi} d\varphi_2 \int_0^\pi \sin\theta_2 d\theta_2 \bar{Y}_{\ell_\beta}^{m_\beta}(\theta_2,\varphi_2) Y_{\ell}^{m}(\theta_2,\varphi_2) Y_{\ell_\nu}^{m_\nu}(\theta_2,\varphi_2) \nonumber \\
    =& (-1)^{m_\beta} \left( \frac{(2\ell_\beta+1)(2\ell+1)(2\ell_\nu+1)}{4\pi} \right)^{1/2} \wigner{\ell_\beta}{\ell}{\ell_\nu}{0}{0}{0} \wigner{\ell_\beta}{\ell}{\ell_\nu}{-m_\beta}{m}{m_\nu}
    \label{eq:Iangle2},
\end{align}
where the integrals of spherical harmonic products are taken from Ref.~\onlinecite{varshalovich1988quantum}.
Selection rules for the Wigner-3j symbols in Eqs.~\eqref{eq:Iangle1}, \eqref{eq:Iangle2} require that
\begin{equation}
    m=m_\mu-m_\alpha = m_\beta - m_\nu,
\end{equation}
\begin{equation}
    |\ell_\alpha-\ell_\mu| \leq \ell \leq \ell_\alpha+\ell_\mu,
\end{equation}
\begin{equation}
    |\ell_\beta-\ell_\nu| \leq \ell \leq \ell_\beta+\ell_\nu,~\text{and}
\end{equation}
\begin{equation}
    \ell \equiv \ell_\alpha + \ell_\mu \equiv \ell_\beta + \ell_\nu \mod 2.
\end{equation}
For the Coulomb term with $(\mu,\nu)=(\alpha,\beta)$, these selection rules reduce to $m=0$, $\ell\leq 2\min(\ell_\alpha,\ell_\beta)$, and $\ell\equiv 0\mod 2$.
Meanwhile, for the exchange term with $(\mu,\nu)=(\beta,\alpha)$, they reduce to $m=m_\beta-m_\alpha$, $|\ell_\alpha-\ell_\beta| \leq \ell\leq \ell_\alpha+\ell_\beta$, and $\ell\equiv \ell_\alpha+\ell_\beta\mod 2$.
In either case, the sum over $m$ in Eq.~\eqref{eq:2body_matrix_element} collapses into at most a single term while the sum over $\ell$ covers a finite range.


\paragraph{Thermal distributions of state-dependent Trotter error terms}
\label{app:state-dep-trotter-results}

For all-electron simulations in the $\Delta_r \rightarrow 0$ limit, the one-body contribution to the first-order Trotter error ($e_1$) is greater than the two-body contribution ($e_2$) by about an order of magnitude.
The tightly-bound 1s states strongly dominate $e_1$, and thus pseudization significantly reduces the error (see Table \ref{tab:first-order-trotter-errors}).
In contrast, for all-electron simulations using a realistic $\Delta_r=0.02$\AA\ grid spacing, $e_1$ and $e_2$ are comparable.
In this case, pseudization decreases the error estimate less dramatically.
Generally, the error is minimized for configurations where the low-lying s shells besides 1s are vacant.

\begin{table}[h]
\begin{tabular}{c|c|c|c|c|c|c}
$\Delta_r$ & $e_1$ AE & $e_1$ PP & $e_2$ AE & $e_2$ PP & $e_1+e_2$ AE & $e_1+e_2$ PP \\\hline
0.02\AA & 1490\,--1590 & 148\,--\,562 & 1680\,--\,1980 & 576\,--\,1110 & 3170\,--\,3560 & 730\,--\,1670 \\
$\rightarrow 0$ & 7420\,--\,7870 & 158\,--\,2630 & 985\,--\,1140 & 253\,--\,547 & 8400\,--\,8970 & 450\,--\,3050
\end{tabular}
\caption{
Range of one-body ($e_1$) and two-body ($e_2$) contributions to the first-order Trotter error when all electrons are included (AE) and 1s electrons are pseudized (PP).
Results are compared for the $\Delta_r=0.02$\AA\ grid spacing deemed adequate for converged eigenenergies and the $\Delta_r\rightarrow 0$ limit.
\label{tab:first-order-trotter-errors}
}
\end{table}

Unlike the first-order case, the one-body contribution $d_1$ to second-order Trotter errors appears negligible compared to the two-body contribution $d_2$ (see Table \ref{tab:second-order-trotter-errors}).
Pseudization again helps lower the total errors.

\begin{table}[h]
\begin{tabular}{c|c|c|c|c|c|c}
$\Delta_r$ & $d_1$ AE & $d_1$ PP & $d_2$ AE & $d_2$ PP & $d_1+d_2$ AE & $d_1+d_2$ PP \\\hline
0.02\AA & (2.67\,--\,2.95)$\times 10^6$ & (0.411\,--\,1.31)$\times 10^6$ & (1.62\,--\,2.03)$\times 10^9$ & (0.0759\,--\,1.04)$\times 10^9$ & (1.62\,--\,2.03)$\times 10^9$ & (0.0771\,--\,1.04)$\times 10^9$ \\
$\rightarrow 0$ & (4.95\,--\,5.25)$\times 10^9$ & (0.0278\,--\,1.76)$\times 10^9$ &  &  &  & 
\end{tabular}
\caption{
One-body ($d_1$) and two-body ($d_2$) contributions to the second-order Trotter error when all electrons are included (AE) and 1s electrons are pseudized (PP).
Results are compared for the $\Delta_r=0.02$\AA\ grid spacing deemed adequate for converged eigenenergies and the $\Delta_r\rightarrow 0$ limit.
\label{tab:second-order-trotter-errors}
}
\end{table}

\begin{figure}
    \centering
	\includegraphics{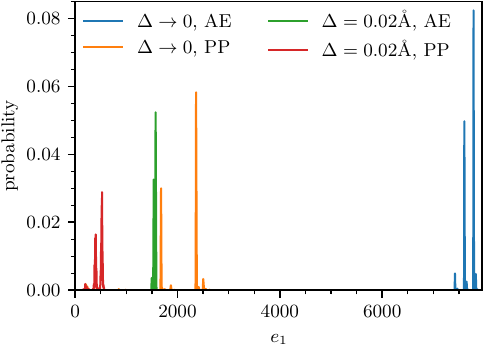}
	\includegraphics{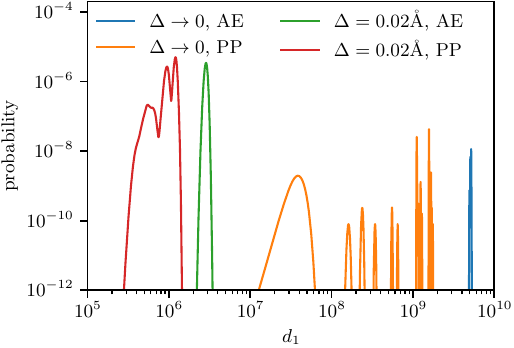}
    \caption{Distributions of the one-body $e_1$ (left) and $d_1$ (right) Trotter error terms over a thermal ensemble of mean-field Slater determinant states for different grid spacings $\Delta_r$ and when including all electrons (AE) or pseudizing the 1s electrons (PP).}
    \label{fig:e1_hist}
\end{figure}

\begin{figure}
    \centering
	\includegraphics{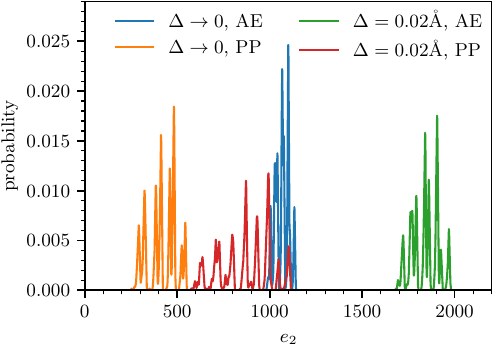}
	\includegraphics{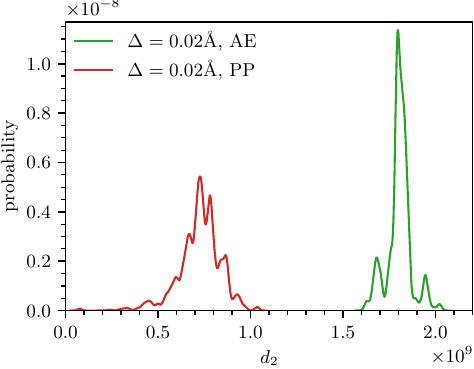}
    \caption{Distributions of the two-body $e_2$ (left) and $d_2$ (right) Trotter error terms over a thermal ensemble of mean-field Slater determinant states 
    when including all electrons (AE) or pseudizing the 1s electrons (PP).}
    \label{fig:e2_hist}
\end{figure}

\begin{figure}
    \centering
	\includegraphics{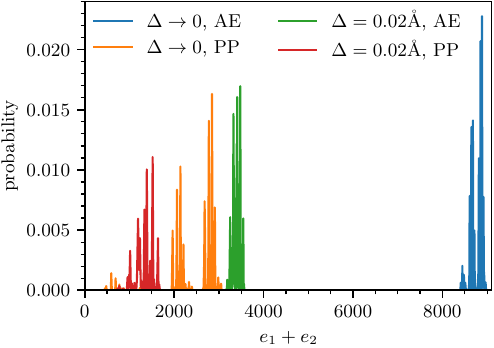}
	\includegraphics{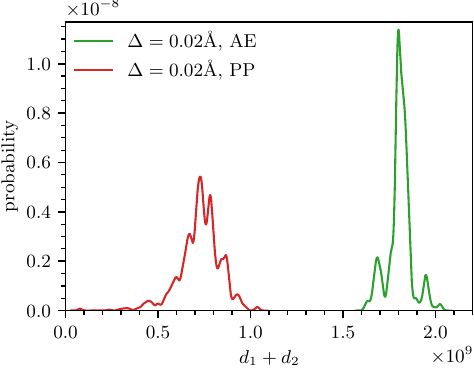}
    \caption{Distributions of the total Trotter error bounds $e_1+e_2$ and $d_1+d_2$ over a thermal ensemble of mean-field Slater determinant states for different grid spacings $\Delta_r$ and when including all electrons (AE) or pseudizing the 1s electrons (PP).}
    \label{fig:e1e2_hist}
\end{figure}

\subsection{Complete resource estimates \label{sec:resources-trotter}}
\begin{figure}
    \centering
    \includegraphics{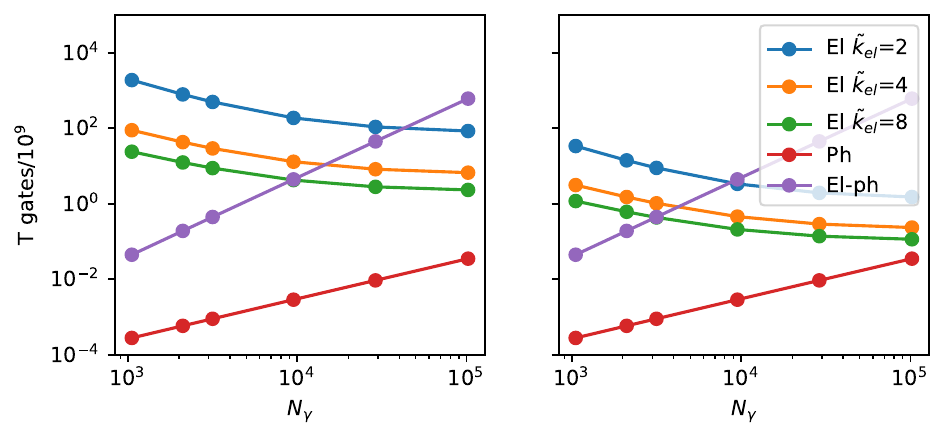}
    \caption{T-gate cost in billions as a function of $\ngridph$ for $\exp(-it\hel), \exp(-it\hph), \exp(-it\helph)$.
    Product formulae with $\tilde{k}_\textrm{el} = 2, 4, 8$ are considered  for implementing $\exp(-it\hel)$.
    The minimum value of $\ngridph = 10^3$ corresponds to the simulation of a single spectral feature in the iron opacity and $\ngridph = 10^5$ corresponds to the entire opacity spectrum. The left panel is for the all-electron simulation, the right panel is after pseudization of the 1s electrons.} 
    \label{fig:trotter_components}
\end{figure}

We begin by presenting the all-electron resource estimates for the components $\exp(-it\hel)$, $\exp(-it\hph)$, $\exp(-it\helph)$ in the left panel of Fig.~\ref{fig:trotter_components} as a function of $\ngridph$, from the minimum value corresponding to a single spectral feature of the opacity, to the maximum corresponding to the full opacity spectrum measured in experiment.
The costs of the photonic and electron-photon contributions scale as expected, linearly and quadractically respectively due to the fast-forwarding in the photonic Hamiltonian and the Fourier transform in the electron-photon Hamiltonian.
High-order Trotterization $\tilde{k}_\textrm{el} = 8$ for the electronic Hamiltonian simulation has the best performance, in agreement with prior studies \cite{rubin2024quantum}.

One also sees, at first glance, a counterintuitive reduction in the electronic simulation cost with increasing $\ngridph$.
This can be understood by looking at our expression for $\tee$, wherein we see that $\tee$ decreases as the number of features to simulate increases due to the uncertainty of the momentum of the wave packet increasing, and hence the uncertainty of the position decreasing.
Due to the shorter simulation time, the cost of the electronic simulation decreases with increasing $\ngridph$, in contrast to the photonic and electron-photon parts of the Hamiltonian.
As a consequence, there is a local minimum achieved in the sum of the costs of the three terms, which when multiplied by $r$ gives the total Hamiltonian simulation cost corresponding to 10 spectral features.

Qualitatively similar features are present after pseudization of the 1s electrons, however the total cost of the electronic part of the simulation is reduced by between 1 -- 2 orders of magnitude.
We find that the T gate cost of computing the pseudopotential term is consistently a factor of 5 -- 10 times cheaper than the cost of exponentiation of the electronic part of the Hamiltonian which is dominated by the cost of computing the 2-body Coulomb potential, and thereby has a negligible effect on total T gate costs.
Note also that the photonic and electron-photon terms are not affected by the pseudization, and so the pseudization shifts the gate-optimal value of $\ngridph$ down to $\ngridph \approx 3\times 10^{3}$ for $\tilde{k}_{el} = 8$ corresponding to three spectral features.

\begin{figure}
    \centering
    \includegraphics{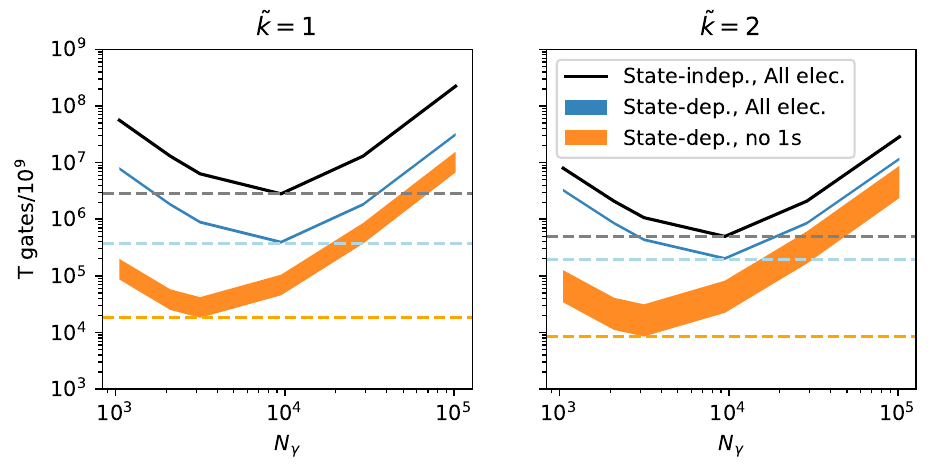}
    \caption{Simulation cost of a single shot required for computing iron opacity as function of $\ngridph$, includes the cost of state preparation and Hamiltonian simulation.
    Results using analytic first- and second-order Trotter formulae without pseudizing the 1s electrons are shown in the solid lines.
    State-dependent Trotter bounds without and with pseudization are shown as dashed lines and hatched regions.
    Note that additionally $\ns = 10$ repetitions of the Hamiltonian simulation would be required to resolve errors in the measured opacity.
    The dashed lines show the minimum configuration cost for each simulation type.}
    \label{fig:full_cost_trotter}
\end{figure}

In Fig.~\ref{fig:full_cost_trotter} we include the full cost of a single shot of the opacity simulation including state preparation and dynamics.
To construct these resource estimates, it was necessary to compute the Trotter bounds in Eqs.~\eqref{eq:trotter-formulae} and \ref{eq:trotter-formulae-statedep}, the details of which are given in App.~\ref{app:state-dep-trotter}.
State-independent bounds were computed analytically, and state-dependent bounds were computed using classical average atom simulation of a single iron nucleus.

The solid black lines show results with state-independent first- and second-order Trotter bounds without pseudization of the electrons and $\tilde{k}_\textrm{el} = 8$, then the blue and orange regions the state-dependent bounds with and without pseudization again with $\tilde{k}_\textrm{el} = 8$.
The dashed lines show the lowest cost simulation in each category: note that the minimum in $\ngridph$ shifts with pseudization as expected from Fig.~\ref{fig:trotter_components}.
The first notable point is that there is a significant improvement in resource estimates when optimizing Trotter bounds, in the case of first-order simulation an improvement of 2 orders of magnitude and for second-order simulation about 1.5 orders of magnitude improvement.
Most of the improvement comes from the combination of pseudization and state-dependent bounds, whereas computing state-dependent bounds without pseudization has a less dramatic effect on the Trotter bounds.

The second notable point is that the improvement moving from first to second-order Trotter formulas is very small, just a constant factor of about 2 when using the state-dependent bounds.
To understand why, we need to look at the asymptotic scaling of $\vec{\mathcal{E}}$ and $\vec{\mathcal{D}}$ with $\epsilon, \Delta_r$ near their maximal eigenvalue regimes.
These regimes are associated with close packing of electrons near each other, thereby increasing each term.
These asymptotics can be found in App.~\ref{app:state-dep-trotter}:
\begin{align}
& \vec{\mathcal{E}} \sim \Delta_r^{-2}, \ \vec{\mathcal{D}} \sim \ereg^{-3} \Delta_r^{-1}.
\end{align}
Importantly, we note that not only does $\vec{\mathcal{D}}$ scale with a different power of inverse grid spacing, as expected from the commutator formulas used to derive it, it also qualitatively scales differently because it is sensitive to the regularization at lowest order, whereas $\vec{\mathcal{E}}$ is not.
This is due to the form of these two as shown in Eq.~\eqref{eq:e-d-defn}, wherein $\vec{\mathcal{E}}$ vanishes as $\vec{r} \rightarrow 0$ while $\vec{\mathcal{D}}$ does not, specifically the third term where the momentum operator succeeds the position dependent term.
The result is that $\vec{\mathcal{D}}$ has much larger eigenvalues than $\vec{\mathcal{E}}$ as $\ereg = 10^{-3} \ll \Delta_r = 10^{-2}$, even when accounting for the additional square root corrections present in the expressions for $r$ in the second-order Trotter formula.
Higher-order product formulas should also be sensitive to $\ereg$ and should be qualitatively similar to the second-order Trotter bound, and thereby one expects that higher-order Trotter bounds $\tilde{k} \geq 4$ may yield significant improvements on the total simulation cost.

Given the higher cost of the Trotterized simulation to the interaction picture results shown in Fig.~\ref{fig:fig3-app}, we have opted to focus the main results on the interaction picture.
However, as noted above, with higher Trotter product formulas, e.g. 4th or 8th order, and with the corresponding state-dependent Trotter bounds, one may find that the fully Trotterized simulation outpaces the interaction picture simulation.
We leave this question open for future work.

\end{document}